\documentclass[12pt]{iopart}
\usepackage{graphicx}
\usepackage{epstopdf}
\usepackage{array,booktabs}
\usepackage{amsfonts}
\usepackage{hyperref}

\newcommand{\bfk}{{\bf k}}
\newcommand{\bfq}{{\bf q}}
\newcommand{\bfR}{{\bf R}}
\newcommand{\bfkR}{{\bf k}\cdot {\bf R}_\alpha}

\begin{document}

\topical{Chiral $d$-wave superconductivity in doped graphene}
\author{Annica M. Black-Schaffer$^1$ and Carsten Honerkamp$^2$}
\address{$^1$ Department of Physics and Astronomy, Uppsala University, Box 516, S-75120 Uppsala, Sweden} 
\address{$^2$ Institute for Theoretical Solid State Physics, RWTH Aachen University, D-52074 Aachen, Germany}
\ead{annica.black-schaffer@physics.uu.se, honerkamp@physik.rwth-aachen.de}
 
\begin{abstract}
A highly unconventional superconducting state with a spin-singlet $d_{x^2-y^2}\pm id_{xy}$-wave, or chiral $d$-wave, symmetry has recently been proposed to emerge from electron-electron interactions in doped graphene. Especially graphene doped to the van Hove singularity at $1/4$ doping, where the density of states diverges, has been argued to likely be a chiral $d$-wave superconductor. 
In this review we summarize the currently mounting theoretical evidence for the existence of a chiral $d$-wave superconducting state in graphene, obtained with methods ranging from mean-field studies of effective Hamiltonians to angle-resolved renormalization group calculations.
We further discuss multiple distinctive properties of the chiral $d$-wave superconducting state in graphene, as well as its stability in the presence of disorder. We also review means of enhancing the chiral $d$-wave state using proximity-induced superconductivity.
The appearance of chiral $d$-wave superconductivity is intimately linked to the hexagonal crystal lattice and we also offer a brief overview of other materials which have also been proposed to be chiral $d$-wave superconductors. 
\end{abstract}
\submitto{\JPCM}
\maketitle

%
% -------------------------------------------------- %
% INTRODUCTION
% -------------------------------------------------- %
\section{Introduction}
% Graphene:
Graphene is and has been one of the most exciting novel materials of the the new century \cite{Novoselov04}. While the true impact of graphene on technology is still to be determined, it has been a goldmine for fascinating physics. Among the physical properties which have generated most interest are the massless Dirac fermion spectrum of undoped graphene and the non-trivial structural physics of the two-dimensional carbon planes \cite{CastroNeto09}. 

% Interaction effects and SC:
Also interaction and correlations effects have been widely studied in graphene \cite{KotovRMP12}. A large number of exotic states of matter has been proposed theoretically, see e.g.~\cite{Tchougreeff92, Sorella92, Khevshchenko01,Herbut06, Hou_Chamon07, Honerkamp08, Raghu08, Liu_Li09, Drut09b, Herbut09, Gamayun10, Meng10, Ulybyshev13, Sorella12}, but most have not been experimentally observed as of yet. One exception is multi-layer graphene systems where there are now experimental reports of an energy gap opening at low temperatures, which has been ascribed to interactions effects \cite{Feldman09, Weitz10, Mayorov11, Bao12, Velasco12, vanElferenBi, vanElferenTri, FreitagBi}.
Obviously, the possibilities for superconductivity in graphene have also been explored. 
While the potential, and also realizations, of conventional phonon-mediated superconductivity in two-dimensional carbon materials is very actively pursued \cite{Xue12, Profeta12}, there are also a large number of theoretical works on how unconventional superconductivity can emerge in graphene and related systems. Most of these consider the possibility of chiral $d$-wave superconductivity in doped graphene, which is  exactly the focus of this review.

% Chiral SC:
Chiral superconductivity is characterized by breaking of both time-reversal and parity symmetries. In the usual context, it necessarily involves a complex linear combination of  two order parameters that often belong to a joint higher-dimensional representation of the point group of the crystal. The chiral $d$-wave superconducting state in graphene is a spin-singlet $d_{x^2-y^2} \pm id_{xy}$-wave state. Due to the sixfold symmetry of the honeycomb lattice, the two $d$-wave states carry an equal weight but have a relative $\pi/2$ phase shift. This results in a fully gapped bulk superconducting state at any finite doping levels in graphene. 
Furthermore, the phase of a $d_{x^2-y^2} \pm id_{xy}$-wave state winds around the Brillouin zone center twice with the direction, or equivalently chirality, set by the sign between the two $d$-wave components. This phase winding can be formalized into a non-zero topological invariant, such as a Chern or Skyrmion number. A non-zero topological invariant is directly related to the number of edge modes in a finite system, and the chiral $d$-wave state in graphene hosts two co-propagating, i.e.~chiral, edge modes crossing the bulk energy gap on every surface.

% Why is chiral interesting:
The time-reversal symmetry breaking and multi-component character naturally offers a very rich phenomenology for chiral superconductors. One very distinct property that could possibly be utilized in nanoscopic devices is the unidirectional transport at the edges of a chiral superconductor. 
Moreover, with some additional modifications, edge states or other localized states in a chiral superconductors can be turned into Majorana modes \cite{Black-Schaffer12PRL}, with possible application in robust quantum computing \cite{Nayak08}.
Related to the topological invariant, a chiral superconductor also represents a collective quantum state with a discrete degree of freedom that forms domains. 
The domains walls separating topologically distinct domains have been shown to have very interesting properties as electrical conduction channels \cite{Serban10}.

% p-wave chiral SC:
The known superconductor Sr$_2$RuO$_4$ is at present probably the most likely candidate for a chiral superconductor, here in the form of a proposed spin-triplet $p_x\pm i p_y$-pairing state \cite{KallinChiral, MacKenzieSruo}. This paired state is also present in the $A$-phase of superfluid $^3$He, see e.g.~\cite{VollhardtWoelfle}, as well as arises for composite fermions in the theory of the fractional quantum Hall effect \cite{MooreRead}. A spin-singlet chiral superconductor has, however, not yet been experimentally verified, and, at least in terms of a growing number of theoretical results, doped graphene seems to be a very promising candidate.
Moreover, chiral $d$-wave superconductivity in graphene might provide a natural link between graphene-based physics and the large correlated electron community that has developed around the high-temperature $d$-wave superconducting cuprates. The same pairing mechanisms proposed to generate the $d$-wave state in the cuprates would give a chiral $d$-wave state in graphene if they were to be present, simply due to the sixfold symmetric honeycomb lattice.

%Summary of contents:
We start this review with a brief summary of the different possible superconducting states in graphene. In Section \ref{sec:MFT} we then continue to discuss effective microscopic models, which include the effect of electron-electron interactions in graphene and yield chiral $d$-wave superconductivity as the leading superconducting state, both on the mean-field level and beyond. Following this we review in Section \ref{sec:RG} multiple renormalization group calculations using more realistic models for heavily doped graphene, which find graphene to be a chiral $d$-wave superconductor, especially at and around the van Hove singularity found at $1/4$ electron or hole doping. After having reviewed the currently mounting theoretical evidence for the existence a chiral $d$-wave state in doped graphene, we discuss in Section \ref{sec:prop} numerous distinctive properties of chiral $d$-wave superconductors. We also review in Section \ref{sec:robustness} works which have studied the robustness of the chiral $d$-wave state in graphene, both in terms of its stability in the presence of disorder and impurities, and also the possibility of enhancement using external proximity-induced superconductivity.
In Section \ref{sec:relsys} we provide a brief overview of other potential chiral $d$-wave superconductors where very similar physics as in graphene might be present. Finally, in Section 8 we provide a short summary and offer a brief outlook towards the future.

% -------------------------------------------------- %
% D6h SYMMETRY GROUP.
% -------------------------------------------------- %
\section{Superconducting symmetries in graphene}
\label{sec:SC_D6h}
Superconductivity is an ordered state appearing below the transition temperature $T_c$ in materials with an effective attractive interaction between the electrons. At the superconducting transition global U(1) symmetry is always broken, but it is possible that the superconducting state also breaks additional symmetries present in the normal-state Hamiltonian above $T_c$. The term {\em unconventional superconductivity} is often used to label such states with additional broken symmetries. These other broken symmetries can include crystal lattice symmetry, spin-rotation symmetry, and also time-reversal symmetry. A general symmetry analysis of the possible superconducting states and their symmetries can be performed using group theory. 
Quite generally (see e.g.~\cite{Sigrist91}) we can start in momentum space by writing an effective Hamiltonian with attractive interaction between pairs of electrons with zero total momentum as \footnote{For simplicity we ignore band indices here.}:
%EQUATION:
\begin{eqnarray}
\label{eq:Heff}
H = \sum_{\bfk \sigma} \varepsilon(\bfk) c^\dagger_{\bfk \sigma} c_{\bfk \sigma} + \frac{1}{4} \sum_{\bfk \bfk' \sigma_1  \sigma_2  \sigma_3\sigma_4} V_{\sigma_1  \sigma_2 \sigma_3\sigma_4}(\bfk,\bfk') c^\dagger_{-\bfk\sigma_1}c^\dagger_{\bfk \sigma_2} c_{\bfk'\sigma_3} c_{-\bfk'\sigma_4}.
\end{eqnarray}
Here $\varepsilon(\bfk)$ is the band energy and $V_{\sigma_1  \sigma_2 \sigma_3\sigma_4}(\bfk,\bfk') $ is the effective attractive electron-electron interaction matrix element with the antisymmetry properties as given in Ref.~\cite{Sigrist91}. 
To treat the Hamiltonian in (\ref{eq:Heff}) in mean-field theory we define a superconducting order parameter, often also called the gap function:
%EQUATION:
\begin{eqnarray}
\label{eq:Deltaeff}
\Delta_{\sigma \sigma'}(\bfk) = -  \frac{1}{2}  \sum_{\bfk' \sigma_3 \sigma_4} V_{\sigma'\sigma \sigma_3 \sigma_4}(\bfk,\bfk')\langle c_{\bfk'\sigma_3} c_{-\bfk'\sigma_4} \rangle.
\end{eqnarray}
Ignoring fluctuations around this mean-field value we can write down a general quadratic BCS Hamiltonian, which can be solved using a Bogoliubov transformation. At temperatures very close to $T_c$ the gap function is small and its self-consistency equation can be linearized to:
%EQUATION:
\begin{eqnarray}
\label{eq:BCSlinear}
\nu \Delta_{\sigma_1 \sigma_2}(\bfk) = -  \frac{1}{2}  \sum_{\sigma_3 \sigma_4} \langle
V_{\sigma_2 \sigma_1 \sigma_3 \sigma_4}(\bfk,\bfk')\Delta_{\sigma_3 \sigma_4}(\bfk')\rangle_{\bfk'},
\end{eqnarray}
where the  the average is taken over the Fermi surface and the prefactor $\nu = (\ln (1.14 \varepsilon_c/(k_BT_c))^{-1}$, with $\varepsilon_c$ being the cut-off energy for the attractive interaction.
Equation (\ref{eq:BCSlinear}) is an eigenvalue problem, where the largest eigenvalue $\nu$ determines $T_c$ and the symmetry of the order parameter $\Delta$.  At lower $T$, subsequent transitions may occur in special cases, adding (subdominant) order parameters with other symmetries.   

Even without knowing the details of the attractive interaction $V$ it is possible to perform a general symmetry analysis. For this purpose, we expand the momentum dependence of the gap function, either around the Fermi surface or in the whole Brillouin zone, with respect to a set of basis functions that can be classified according to the irreducible representations of the symmetry group of the normal-state Hamiltonian.\footnote{See e.g.~Ref.~\cite{Sigrist91} for a more thorough symmetry treatment of unconventional superconductivity.} Normally, apart from accidental degeneracies, the superconducting state belongs to a single irreducible representation, or is an unequal mixture of a dominate symmetry and subdominant ones if multiple superconducting transitions take place as the temperature is lowered. It is thus possible  and useful to classify all potential superconducting states by looking at the irreducible representations of the symmetry group of the normal-state Hamiltonian. If the material is already superconducting, any possible subdominant order parameter belongs to an irreducible representation of the symmetry group of the original superconducting state, which can have a reduced symmetry compared to the normal-state Hamiltonian.

For graphene, with its two-dimensional honeycomb lattice, the crystal symmetry group is the hexagonal group $D_{6h}$, with $k_z = 0$, equivalent to evenness under in-plane reflection.\footnote{Alternatively, $C_{6v} $ can also be used with no addition of $k_z = 0$.}
In terms of spin-space symmetries, we are here going to limit ourselves to $s_z = 0$ spin-pairing, i.e.~spin-singlet and spin-triplet $s_z = 0$ pairing, since the normal state in graphene is not spin-polarized. From the $s_z=0$ spin-triplet case we can also reconstruct the $s_z = \pm 1$ unitary triplet states by spin rotation.
We can thus classify all possible zero-momentum pairing states in graphene using the irreducible representations of $D_{6h}$. Since the superconducting order parameter is fermionic in nature, the even-parity representations ($g$) correspond to spin-singlet pairing, whereas the odd-parity representations ($u$) are spin-triplet states.
In table \ref{tab:irreps} we write down the simplest basis functions for the irreducible representations present in graphene along with figures showing their symmetries in the graphene Brillouin zone. 
%
% TABLE:
\begin{table}[h]
 \caption{\label{tab:irreps} Irreducible representations (irreps) for the hexagonal $D_{6h}$ symmetry group with $k_z = 0$. The basis functions are the simplest possible basis functions satisfying the symmetry requirements of the irrep. 
These basis functions are valid at small $k$, but do not necessarily obey the translation symmetry of the reciprocal space.
The simplest basis functions which also obey translation symmetry are schematically displayed in the graphene Brillouin zone (thick black lines) in the last column, with green circles indicating the Fermi surface for lightly doped graphene at the two inequivalent corners of the Brillouin zone. Graphene doped to and beyond the van Hove singularity has a Fermi surface centered around $\Gamma$.}
\begin{indented}
\item[] \begin{tabular}{lll}
 \br
 Irreps & Basis function & Brillouin zone symmetry \\
 \mr
 $A_{1g}$ & 1, $k_x^2 + k_y^2$ & 
 \parbox[c]{0em}{\includegraphics[height =2cm]{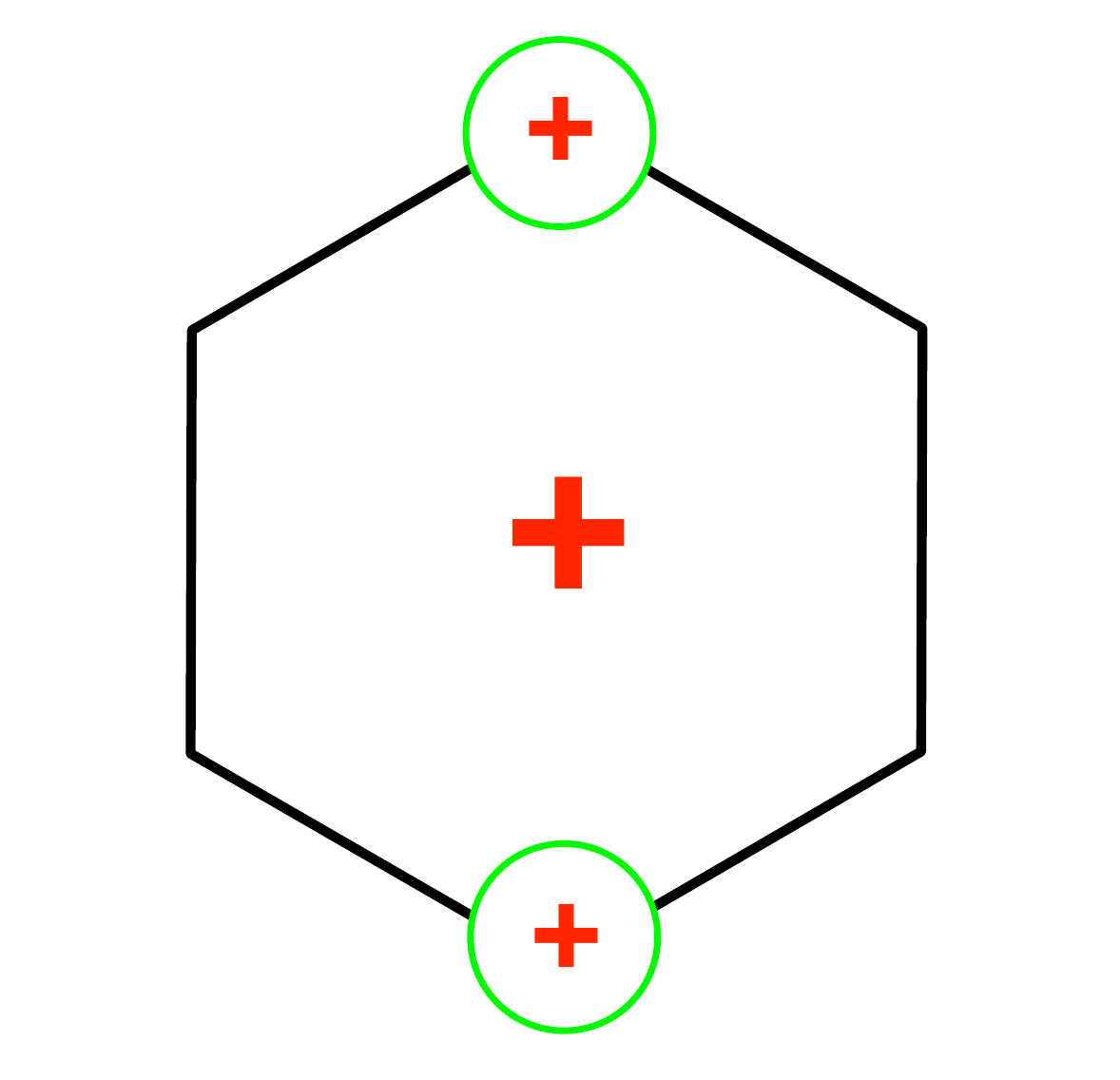} }\\
 $A_{2g}$ & $k_x k_y (k_x^2 - 3k_y^2)(k_y^2 - 3k_x^2)$ & 
 \parbox[c]{0em}{\includegraphics[height =2cm]{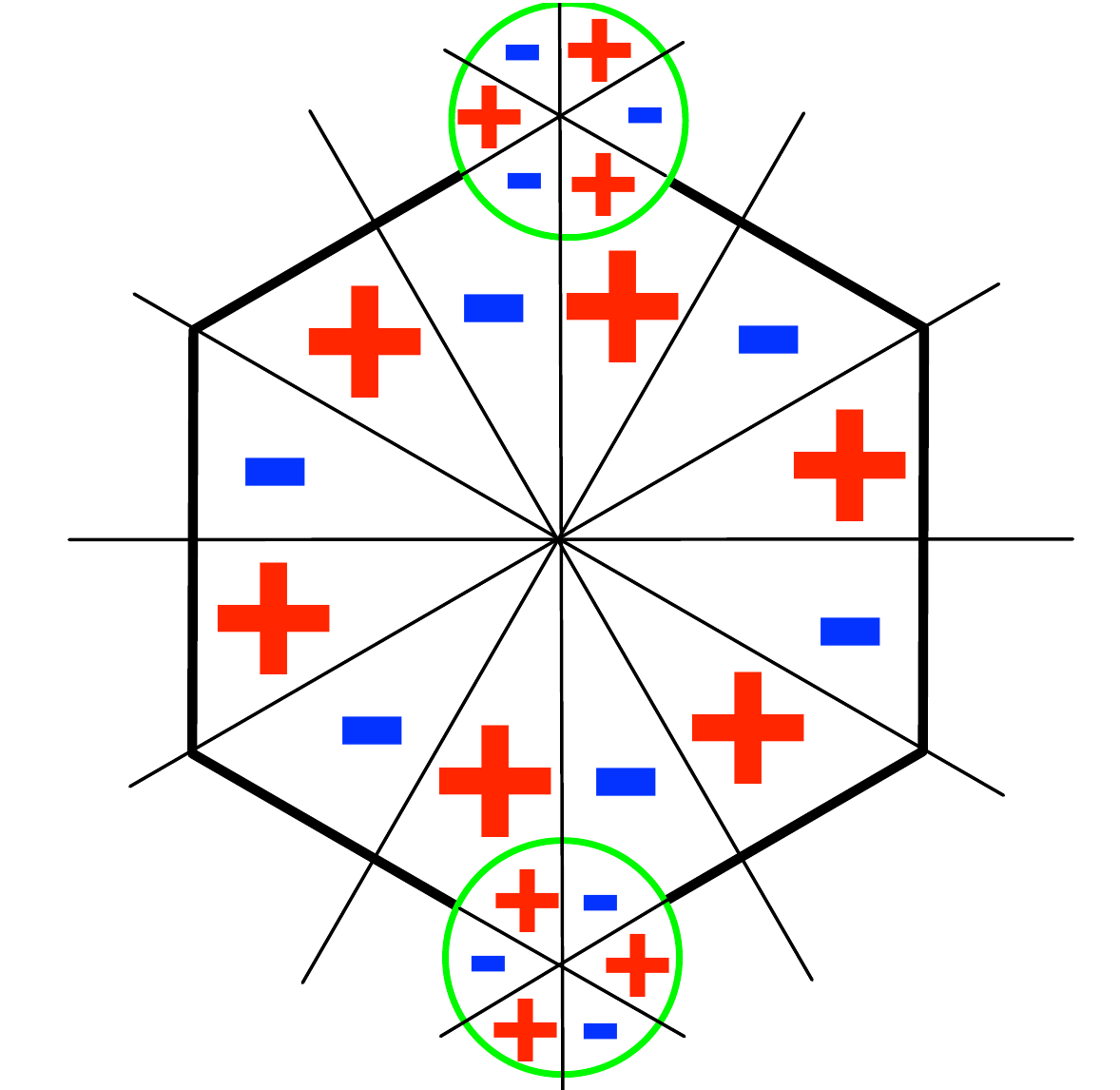} }\\
 $E_{2g}$ & $(k_x^2-k_y^2,2k_x k_y)$ & 
 \parbox[c]{0em}{\includegraphics[height =2cm]{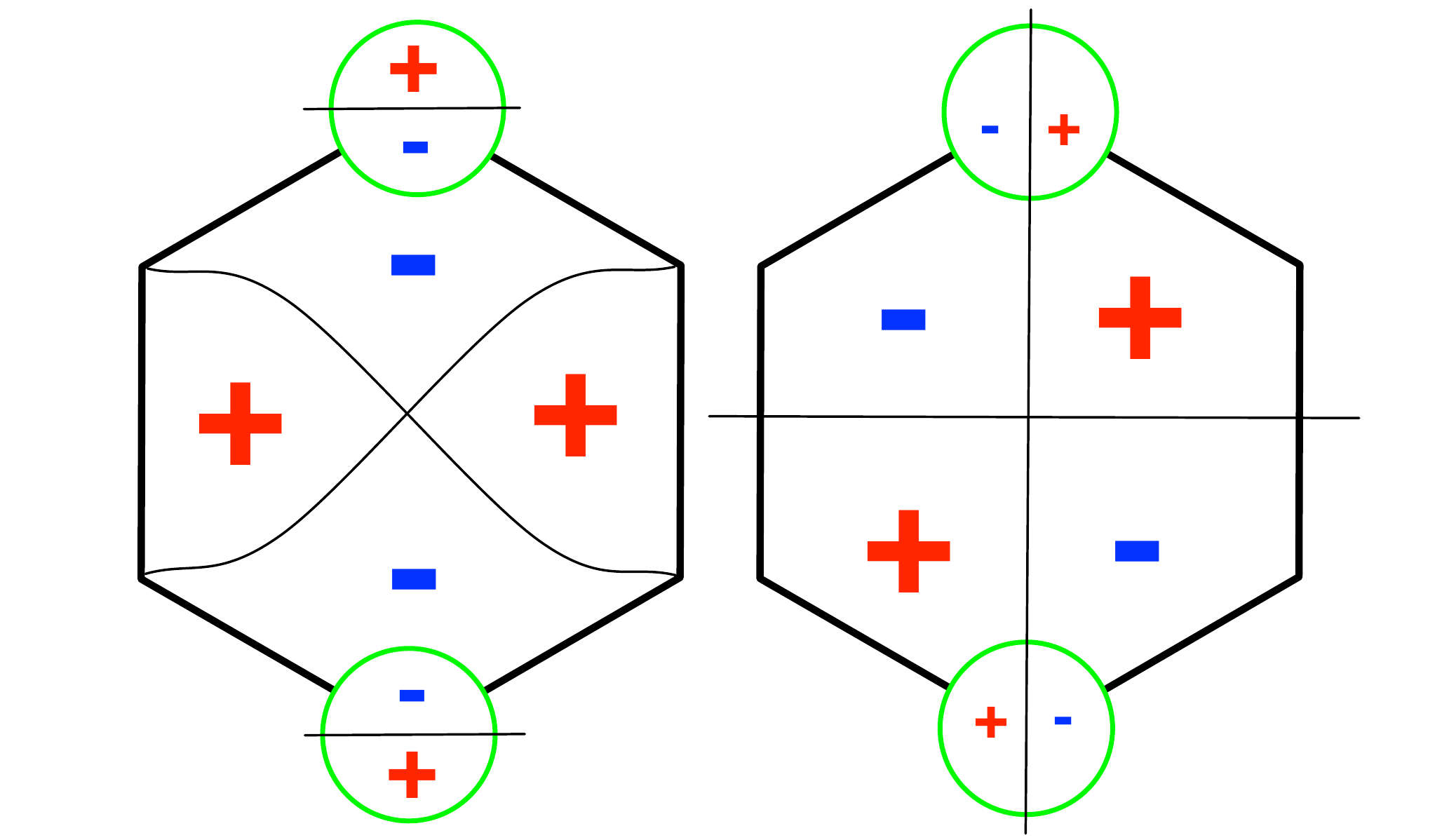} }\\
\mr
$B_{1u}$ & $k_x(k_x^2-3k_y^2)$ & 
 \parbox[c]{0em}{\includegraphics[height =2cm]{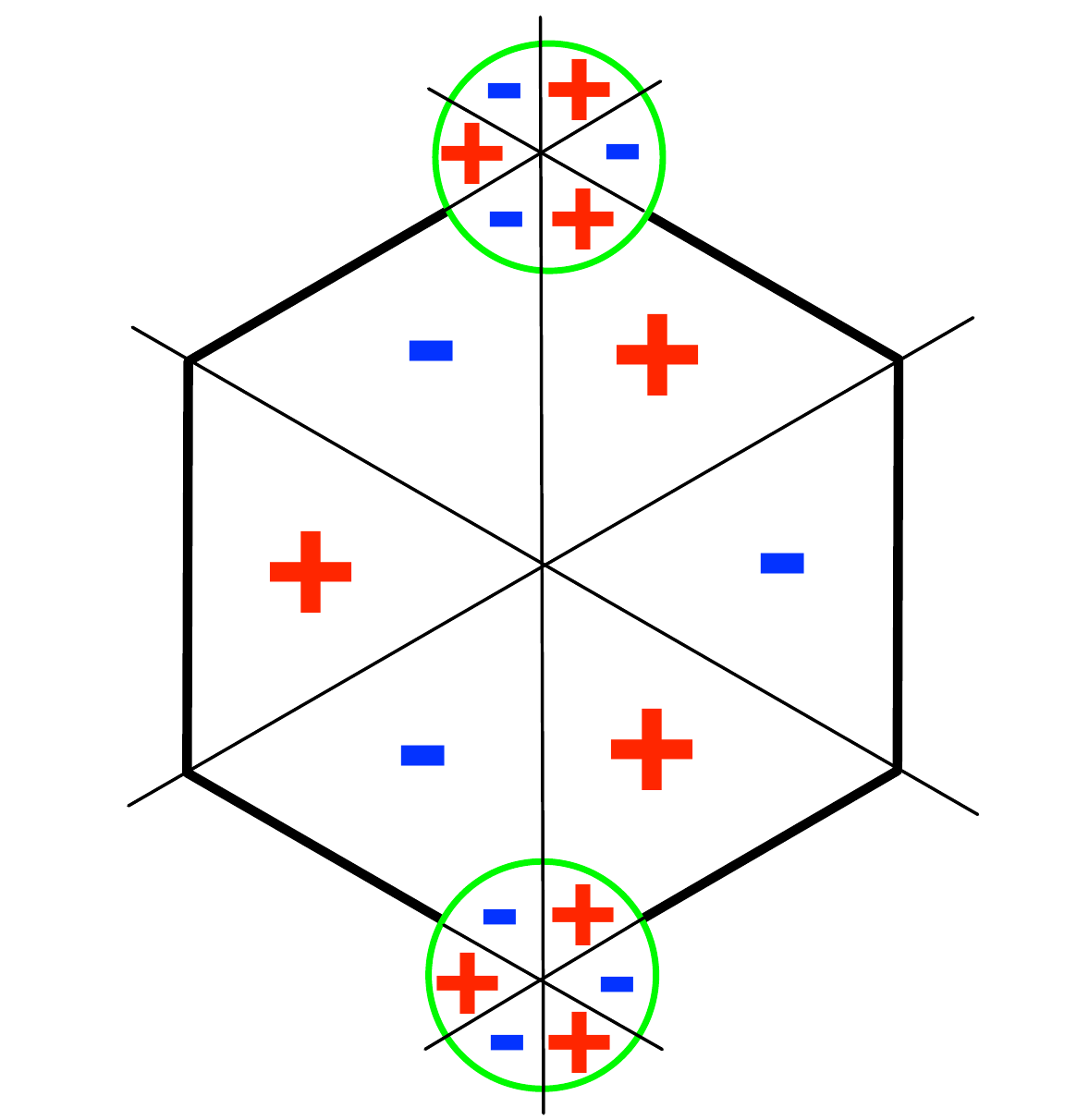} }\\
$B_{2u}$ & $k_y(k_y^2-3k_x^2)$ & 
 \parbox[c]{0em}{\includegraphics[height =2cm]{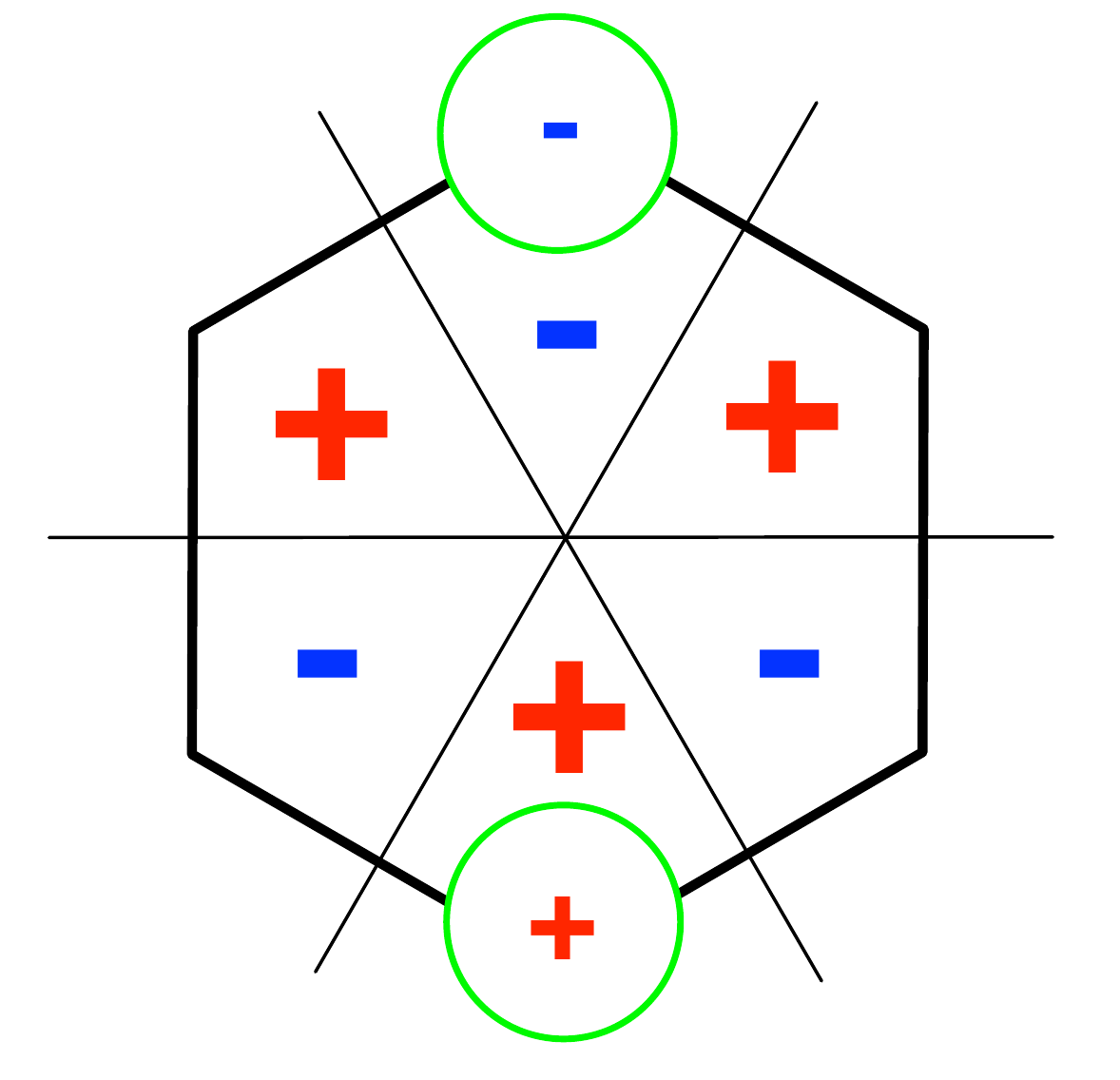} }\\
$E_{1u}$ & $(k_x, k_y)$ & 
 \parbox[c]{0em}{\includegraphics[height =2cm]{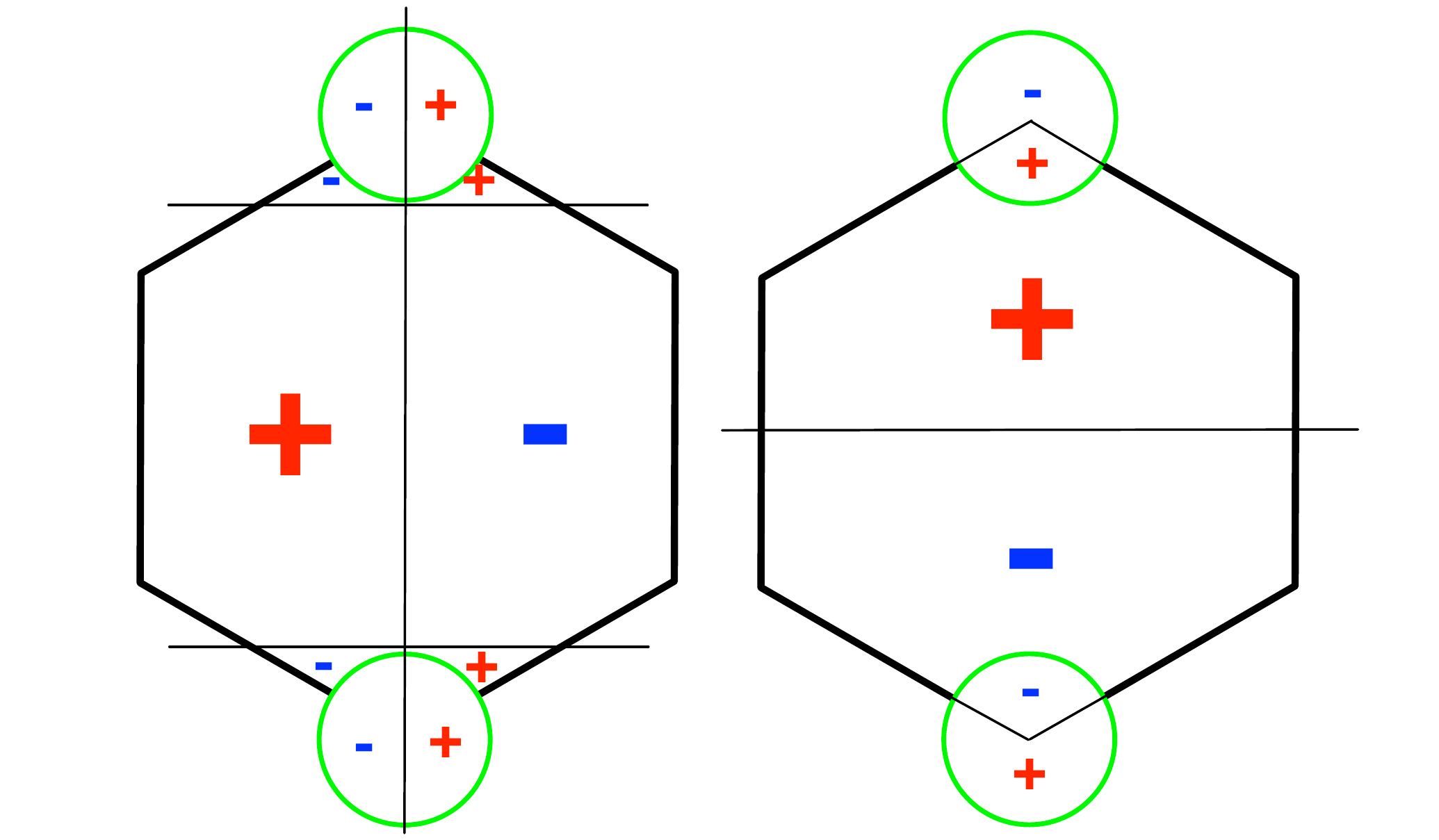} }\\
 \br
 \end{tabular}
 \end{indented}
\end{table}
The Fermi surface of lightly doped graphene is also indicated with green circles in order to easily deduce the symmetry in the low doping regime. These Fermi circles should be understood to be present at all corners of the Brillouin zone. For graphene doped to and above the van Hove singularity, the Fermi surface is centered around the zone center at $\Gamma$ and the available superconducting symmetries can likewise be easily deduced in the figures.

The basis functions given in table \ref{tab:irreps} are the simplest possible basis functions satisfying the symmetry requirements of the representation in question. Of course, infinitely many other basis functions with higher powers in the $k$-components also satisfy the same symmetry requirements. 
However, the quasiparticle energy (for a one-band system)  with unitary spin pairing is $E(\bfk) = \sqrt{\varepsilon(\bfk)^2 + |\Delta(\bfk)|^2}$ and it is thus energetically favorable for $\Delta(\bfk)$ to have as few nodes as possible on the normal-state Fermi surface $\varepsilon(\bfk) = 0$. This naturally leads to the listed lowest order polynomial basis functions being by far the most likely to occur. 
Furthermore, the basis functions listed in table \ref{tab:irreps} should be understood as representatives for the transformation behavior. However, they do not necessarily obey the translational symmetry in reciprocal space of the lattice system. Taking this into account leads to minor modifications of the nodal structure of the superconducting state for irreducible representations with two- and fourfold symmetries, as can be seen in the figures for the $k_x^2-k_y^2$ and $k_x$ symmetries. In these cases the modifications are chosen such as to minimize the number of nodes in graphene doped below the van Hove singularity.
Another way to obtain basis functions that respect the reciprocal space translational symmetry would be to create bond form factors for the pairing, and to superimpose them in a way to get the corresponding symmetries. Then the basis functions in table  \ref{tab:irreps} would come out as expansions of these form factors around $\Gamma$.  

Again using the above energy argument for the number of nodes in $\Delta(\bfk)$, the most likely superconducting states are those with a minimum number of nodes, or even preferably, with a fully gapped quasiparticle spectrum. 
Naturally, the fully isotropic $A_{1g}$, or $s$-wave symmetry state, is a fully gapped superconducting state, also referred to as the conventional superconducting state since it does not break any additional symmetries. However, a sixfold lattice also allows for several other fully gapped states not present in materials with two- or fourfold symmetry. 
The $B_{2u}$ state is a spin-triplet $f$-wave state with multiple nodes for Fermi surfaces centered around $\Gamma$, but it  is fully gapped for lightly doped graphene as it has no nodes on the Fermi surface for low doping levels. The Hubbard model in the weak coupling regime on the honeycomb lattice close to half-filling has been proven (asymptotically exact) to give this $f$-wave state \cite{Raghu10}. It has also been found to be the leading pairing instability for dominant nearest-neighbor repulsions \cite{Honerkamp08}.
Such a Kohn-Luttinger mechanism \cite{Kohn65}, where pairing is generated from weak repulsive interactions, has also recently been shown to give this $f$-wave pairing state at low doping levels for more spatially extended repulsion \cite{Nandkishore14}.

Next let us address the the two-dimensional representations $E_{2g}$ and $E_{1u}$. The linearized BCS equation (\ref{eq:BCSlinear}) gives the same $T_c$ for any basis function belonging a two-dimensional irreducible representation. However, below $T_c$ higher order terms will favor a specific symmetry combination of these basis functions. Quite generally and shown explicitly for $E_{2g}$ in the next section, a complex combination of the basis functions are usually favored as it fully gaps the quasiparticle spectrum (for Fermi surfaces avoiding $K,K'$, and $\Gamma$) and thus minimizes the free energy. For $E_{2g}$ the symmetry takes the form $(k_x \pm ik_y)^2 = (k_x^2-k_y^2 \pm 2ik_x k_y)$, which is exactly the $d_{x^2-y^2} + id_{xy}$-wave or chiral $d$-wave state. The chiral $d$-wave state has been found to appear in the $t$-$J$ model on the honeycomb lattice at low but finite doping levels both at the mean-field level \cite{Black-Schaffer07,Wu13} and in quantum Monte Carlo simulations \cite{Pathak10,Ma11}. It has also been found to be the leading instability around the van Hove singularity in renormalization group calculations for weak repulsive interactions \cite{Nandkishore11, Wang11, Kiesel12}, as well as for an explicit Kohn-Luttinger mechanism \cite{Gonzalez08}.
The complex and equal weight combination of the two fourfold symmetric $d$-wave solutions in graphene is dictated by the crystal lattice. This is distinctly different from tetragonal and square lattices, where the two different $d$-wave solutions belong to different irreducible representations, and can thus in general never be of equal weight. One example are the high-temperature cuprates superconductors, which are known to be $d_{x^2-y^2}$-wave superconductors. A subdominant $id_{xy}$ symmetry has been proposed to exist in the cuprates as a second subdominant superconducting state, for example at surfaces \cite{Fogelstrom97,Covington97}, magnetic impurities \cite{Balatsky97}, or in a magnetic field \cite{Krishana97,Elhalel07}. However, the $d_{xy}$ component is here both generated by extrinsic effects and subdominant to the original $d_{x^2-y^2}$-wave order parameter.
Finally, the $E_{1u}$ solutions lead in a similar manner to the chiral $p$-wave combination $k_x \pm ik_y$. This order parameter combination is well-known from the square lattice, and is e.g.~very likely realized in Sr$_2$RuO$_4$ \cite{Maeno03, Kallin12}. 
Due to the intrinsically complex order parameter both of these chiral states also break time-reversal symmetry $K$, since $K\Delta(\bfk) =  \Delta^\ast(-\bfk) \not= e^{i\theta } \Delta(\bfk)$ for some fixed phase $\theta$.

To summarize, we conclude that the chiral $d$-wave superconducting state in graphene is one of several fully gapped states, and should thus be an energetically favorable state. Furthermore, we know from the studies of the high-temperature cuprate superconductors that $d$-wave superconducting symmetry is generally favored for systems with strong Coulomb repulsion. For strong on-site Coulomb repulsion the superconducting state needs to avoid same-site pairing which corresponds to isotropic $\bfk$-space pairing. At the same time, the Coulomb repulsion generally favors antiferromagnetic tendencies and thus a spin-singlet superconducting state. The spin-singlet state with the lowest number of nodes but still avoiding same-site pairing is exactly the $d$-wave state. Based on these arguments it seems natural to expect to find the chiral $d$-wave state in materials with a two-dimensional hexagonal lattice and strong Coulomb interactions. In the next section we will show that this is the case for a simple effective model. 
Moreover, for graphene doped at and beyond the van Hove singularity, the chiral $d$-wave state is still fully gapped, whereas both spin-triplet $f$-wave states now host multiple nodes. Therefore one might expect that the chiral $d$-wave state  becomes the favored state even for weak Coulomb interactions in this doping regime.

Before closing this general discussion it is worth mentioning that more exotic superconducting states, not included in the classification in table \ref{tab:irreps}, have also been proposed for the honeycomb lattice. These are states with finite momentum pairing, or so-called Fulde-Ferrell-Larkin-Ovchinikov (FFLO) pairing \cite{Fulde64,Larkin64}. For such pairing the crystal symmetry group does not have to be that of the full Brillouin zone.
For attractive nearest neighbor interaction a FFLO spin-triplet state breaking translation invariance through a Kekule pattern has been proposed \cite{Roy10}. Moreover, a uniform pairing state with intrapocket pairing (pairing within one single Dirac cone) has very recently been found from weak long-range repulsive interactions in perturbative renormalization group calculations for bilayer graphene \cite{Murray13}.

% -------------------------------------------------- %
% NN BOND PAIRING
% -------------------------------------------------- %
\section{Chiral $d$-wave superconductivity in effective models for graphene}
\label{sec:MFT}
As briefly discussed in the last section, $d$-wave superconductivity is very often the favored superconducting state in the presence of strong Coulomb repulsion, as exemplified most prominently by the high-temperature cuprate superconductors. On the honeycomb lattice, the two $d$-wave solutions, $d_{x^2-y^2}$ and $d_{xy}$,  are dictated by group theory to be degenerate at $T_c$, but are allowed to develop into a fully gapped superconducting chiral $d_{x^2-y^2}+id_{xy}$-wave combination below $T_c$. In this section we will explicitly show that this chiral $d$-wave state is the preferred superconducting state in effective models proposed to capture the low-energy physics of interacting electrons on the graphene honeycomb lattice. We will mainly focus on a mean-field solution to an effective $t$-$J$ model, but also review results from accurate numerical many-body techniques applied to simple models.

\subsection{Electron interactions in graphene}
While many properties of graphene are well captured by a non-interacting electron picture, the role of Coulomb interactions in graphene has also received a lot of attention \cite{KotovRMP12}. For example, magnetism has experimentally been reported both in nanographene \cite{Shibayama00, Enoki09, Tao11},  and in graphite in the presence of disorder \cite{Esquinazi03} or grain boundaries \cite{Cervenka09}, although pristine graphene has not been found to be either magnetic \cite{Sepioni10} or gapped \cite{Elias11, Mayorov11}. Theoretically, on-site Coulomb repulsion exceeding $U>3.9 t$ has been found to give an antiferromagnetic state in undoped graphene in quantum Monte Carlo simulations \cite{Meng10, Sorella12}. This relatively large value of $U$ is a consequence of undoped graphene being a semimetal with only a point-like Fermi surface. The strength of the on-site Coulomb interaction in graphene was recently estimated to be $U = 3.3t$ from first-principles \cite{Wehling11U}. Due to  limited screening in pristine graphene, also longer range repulsion was found to be important, with the nearest neighbor repulsion $V = 2.0t$ and then further diminishing with distance, resulting in a dielectric constant $\epsilon = 2.5$ \cite{Wehling11U}. The effective fine structure constant then becomes $\alpha = \frac{e^2}{\epsilon \hbar v_F} \approx 0.9$. This is, according to hybrid Monte Carlo simulations, in close proximity to the $\alpha$ needed for an insulating transition in graphene due to the long-range Coulomb interaction \cite{Drut09,Drut09b, Ulybyshev13}.
Taken together, these results show that both the short-range and long-range Coulomb interactions in graphene are likely sufficiently strong to, at the very least, put pristine graphene relatively close to a transition to an ordered ground state, which, somewhat depending on the exact nature of the short-range interaction, will have an antiferromagnetic spin ordering. 
Furthermore, with doping, the density of states at the Fermi level in graphene increases and can thus enhance instabilities towards an ordered ground state.

\subsection{Effective $t$-$J$ model}
\begin{figure}[htb]
\begin{center}
\includegraphics[scale = 1]{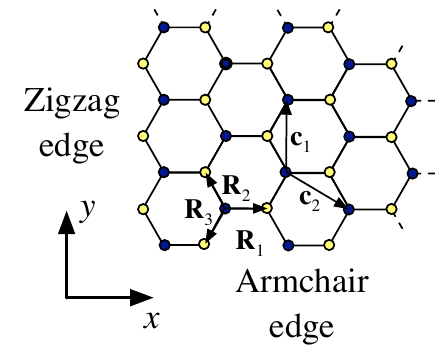}
\caption{\label{fig:honeycomb} (Color online). The graphene honeycomb lattice with sublattice A (blue) and B (yellow) with lattice vectors $\{ {\bf c}_1, {\bf c}_2\}$ and lattice constant $|{\bf c}| = 2.46$~\AA, nearest neighbor bonds $\bfR_\alpha$ ($\alpha = 1,2,3$), and zigzag and armchair edges indicated. 
}
\end{center}
\end{figure}

We will here not further dwell on the exact strength of the Coulomb interaction, but we are instead interested in deriving and studying a simple effective model which gives chiral $d$-wave superconductivity as the leading superconducting instability.
Unbiased theoretical evidence that chiral $d$-wave superconductivity appears and even wins over other ordered states in heavily doped graphene is presented in the next section.
For this purpose we start with a simple Hubbard model for the honeycomb lattice:
%EQUATION:
\begin{eqnarray}
\label{eq:HU}
H_U = -t \sum_{\langle i,j\rangle,\sigma} a^\dagger_{i\sigma}b_{j\sigma} + {\rm H.c.} + \mu \sum_{i,\sigma} a^\dagger_{i\sigma}a_{i\sigma} + b^\dagger_{i\sigma}b_{i\sigma}
 + U\sum_{i,\sigma}n_{i\uparrow} n_{i\downarrow}.
\end{eqnarray}
Here $t \approx 2.8$~eV is the nearest neighbor hopping amplitude, $\mu$ is the chemical potential, and $U$ is  the on-site repulsion. Furthermore, $a^\dagger_i$ ($b^\dagger_i$) is the creation operator on site $i$ in the $A$ (B) sublattice, as shown in Figure \ref{fig:honeycomb}, $\langle i,j\rangle$ denotes summation over nearest neighbors, and $n_i$ is the number operator on site $i$. We are mainly going to be concerned with finite doping which also provides additional screening of the long-range tail of the Coulomb interaction. Thus only including on-site repulsion should be a reasonable first-order approximation of the electron interactions in doped graphene. 
At finite doping the superconducting instabilities are also largely immune to any potential gap generation at the Dirac points produced by additional sublattice symmetry breaking or spin-orbit coupling terms.
Based on recent first-principles calculations $U = 3.3t$ \cite{Wehling11U} and cannot simply be seen as a small perturbation. 
%
% FIGURE:
In the opposite, large-$U$ limit, the Hubbard model can be rewritten as a $t$-$J$ Hamiltonian \cite{Hirsch85, GrosJoyntRice87, ZhangRice88, Choy96}, where the effective interaction to lowest order is $J = 2t^2/U$ and between nearest-neighbor spins:
% EQUATION:
\begin{eqnarray}
\label{eq:HtJ}
\fl H_{t-J} = -t \sum_{\langle i,j\rangle,\sigma} a^\dagger_{i\sigma}b_{j\sigma} + {\rm H.c.} + \mu \sum_{i,\sigma} a^\dagger_{i\sigma}a_{i\sigma} + b^\dagger_{i\sigma}b_{i\sigma}
 + J\sum_{\langle i,j\rangle,\sigma}{\bf S}_i \cdot {\bf S}_j - \frac{1}{4}n_i n_j.
\end{eqnarray}
Due to the very large on-site repulsion in this limit, the Hilbert space for $H_{t-J}$ is reduced and excludes doubly occupied sites. This explicit many-body effect significantly complicates accurate treatment, but within mean-field theory we can replace the strict prohibition of doubly occupied sites with statistical weighting factors. This  is known as renormalized mean-field theory \cite{Zhang88, Lederer89, Anderson04, Edegger07, Ogata08}, which includes the rescaling $t \rightarrow 2t\delta/(1+\delta)$ \cite{Vollhardt84} and $J \rightarrow 4J/(1+\delta)^2$ \cite{Zhang88}, with $\delta = 1-n$ denoting the doping away from half-filling per site, with $\delta = 0$ in pristine graphene.
 
In the form $H_{t-J}$ is written in (\ref{eq:HtJ}) it is clear that a mean-field antiferromagnetic state should emerge for strong enough $J$. But, we can also rewrite the interaction term as
%EQUATION:
\begin{eqnarray}
\label{eq:RVB}
{\bf S}_i \cdot {\bf S}_j - \frac{1}{4}n_i n_j = -Jh^\dagger_{ij}h_{ij}, \quad {\rm where} \quad h^\dagger_{ij} = \frac{1}{\sqrt{2}}(a_{i\uparrow}^\dagger b_{j\downarrow}^\dagger - a_{i\downarrow}^\dagger b_{j\uparrow}^\dagger)
\end{eqnarray}
is the nearest-neighbor spin-singlet creation operator for unit cell $i$ when $i$ belongs to the A sublattice, with a corresponding term existing for $i$ belonging to the B sublattice. Thus $H_{t-J}$ in (\ref{eq:HtJ}) in fact consists of a (renormalized) band structure and an effective resonance valence bond (RVB) interaction term \cite{Anderson73, Anderson87}.   
Interestingly, already early treatments by Pauling and others \cite{Paulingbook, CastroNeto09view} of $p\pi$-bonded planar organic molecules such as benzene, of which graphene is the infinite extension, rested heavily on the the idea of RVB interactions, favoring spin-singlet nearest neighbor bonds over polar configurations with single or double occupancy of the orbitals. For example, good estimates were achieved using RVB interactions for the C-C bond distance, cohesive energy, and some excited state properties.
In fact, since $U= 3.3t$ is not necessarily large enough to warrant a strong-coupling treatment, phenomenologically introducing an effective RVB term has been proposed as a viable approach to the intermediate coupling regime for graphene \cite{Baskaran02, Black-Schaffer07}. The effective RVB coupling can in this case be estimated as the energy gain for a two-electron state in the lowest singlet configuration: $J = \frac{1}{2}\sqrt{U^2 + 16t^2}-U/2 \sim t$ in graphene \cite{Baskaran02}.
Thus, both a strong-coupling approach and a chemistry-based phenomenological argument result in a Hamiltonian of the type  (\ref{eq:HtJ}), i.e.~a simple band structure with effective nearest-neighbor spin-singlet bond correlations. For simplicity, we are therefore here going to use an electron-electron interaction term of the form (\ref{eq:RVB}) and work with generic values (renormalized or not) for $t$ and $J$.

\subsubsection{Mean-field treatment}
The Hamiltonian in~(\ref{eq:HtJ}) with RVB interactions can be treated within mean-field theory using a complete Hartree-Fock-Bogoliubov factorization, yielding order parameters for both nearest-neighbor spin-singlet pairing (particle-particle channel) and hopping (particle-hole channel) \cite{Black-Schaffer07, Wu13}. However, if we assume that the band structure is, at most, rescaled isotropically, we only have to work with an (again) renormalized $t$ and a superconducting mean-field order parameter $\Delta_{ij} = -J\langle a_{i\downarrow} b_{j\uparrow} - a_{i\uparrow}b_{j\downarrow}\rangle$.
The resulting BCS mean-field Hamiltonian can be written after a Fourier transform to reciprocal space as:
%EQUATION:
\begin{eqnarray}
\label{eq:HMF}
\fl H_{\rm MF} = -t \sum_{\bfk,\alpha,\sigma} e^{i\bfkR} a^\dagger_{\bfk\sigma}b_{\bfk\sigma} + {\rm H.c.}
+ \mu \sum_{\bfk,\sigma} a^\dagger_{\bfk\sigma}a_{\bfk\sigma} + b^\dagger_{\bfk\sigma}b_{\bfk\sigma} \nonumber \\
- \sum_{\bfk,\alpha} \Delta_\alpha e^{i\bfkR} (a^\dagger_{\bfk\uparrow}b^\dagger_{-\bfk\downarrow} - a^\dagger_{\bfk\downarrow}b^\dagger_{-\bfk\uparrow} + {\rm H.c.}) + \frac{N}{J}\sum_\alpha 2|\Delta_\alpha|^2.
\end{eqnarray}
Here $\bfR_\alpha$, with $\alpha = 1,2,3$, labels the three nearest neighbor vectors on the honeycomb lattice, see figure \ref{fig:honeycomb}. 
We have assumed translational invariance but we allow for the order parameters to be independent on the three nearest neighbor bonds, such th at $\Delta_\alpha = \Delta_{i,j=i+\bfR_\alpha}$. In the last term in (\ref{eq:HMF}) $N$ is the number of $\bfk$-points in the first Brillouin zone or equivalently the number of unit cells. This term is irrelevant unless the the total free energy is to be calculated and will thus be mostly ignored below.

In order to be able to find the normal-state band structure and its Fermi surface we need to diagonalize the kinetic energy term in (\ref{eq:HMF}), which is done using the following basis transformation:
% EQUATION:
\begin{eqnarray}
\label{eq:banddiag}
\left( \begin{array}{c}
a_{{\bf k}\sigma}  \\ b_{{\bf k}\sigma}
\end{array} \right) = \frac{1}{\sqrt{2}}
\left( \begin{array}{c}
c_{{\bf k}\sigma} + d_{{\bf k}\sigma} \\ e^{-i\varphi_{\bf k}}(c_{{\bf k}\sigma} - d_{{\bf k}\sigma})
\end{array} \right),
\end{eqnarray}
where $c_{\bfk\sigma}^\dagger$ ($d_{\bfk\sigma}^\dagger$) now creates an electron in the lower (upper) $\pi$-band with energy $\mu -(+) \varepsilon(\bfk)$, where $\varepsilon(\bfk) = t|\sum_\alpha e^{i\bfkR}|$ and $\varphi(\bfk) = {\rm arg}(\sum_\alpha e^{i\bfkR})$. This transformation results in a BCS Hamiltonian with both inter and intra-band pairing \cite{Black-Schaffer07}:
%EQUATION:
\begin{eqnarray}
\label{eq:Hband}
\fl H_{\rm MF} = \sum_{\bfk,\sigma} (\mu - \varepsilon(\bfk))c_{\bfk \sigma}^\dagger c_{\bfk \sigma} + (\mu + \varepsilon(\bfk)) d_{\bfk \sigma}^\dagger d_{\bfk \sigma} \nonumber \\
+ \sum_{\bfk} \Delta_i(\bfk) (c^\dagger_{\bfk \uparrow}c^\dagger_{-\bfk\downarrow} - d^\dagger_{\bfk \uparrow}d^\dagger_{-\bfk\downarrow})
+ \sum_{\bfk} \Delta_I(\bfk) (d^\dagger_{\bfk \uparrow}c^\dagger_{-\bfk\downarrow} - c^\dagger_{\bfk \uparrow}d^\dagger_{-\bfk\downarrow}).
\end{eqnarray}
The intraband pairing $\Delta_i(\bfk) = \sum_{\alpha} \Delta_\alpha \cos(\bfkR -\varphi(\bfk))$ is a spin-singlet pairing of even-parity in the Brillouin zone. The interband pairing $\Delta_I(\bfk) = i \sum_{\alpha}\Delta_\alpha \sin(\bfkR -\varphi(\bfk))$ is also a spin-singlet pairing term but of odd-parity and also odd in band index ($c,d$), as to still satisfy the overall fermionic nature required for a superconducting order parameter.
The Hamiltonian in (\ref{eq:Hband}) can be diagonalized using a Bogoliubov-Valatin transformation with a total of four different quasiparticle operators, since we have a two-band system. The resulting quasiparticle energies are given by
%EQUATION:
\begin{eqnarray}
\label{eq:EQP}
\fl E_{QP} =\pm \sqrt{\varepsilon^2 + \mu^2 +|\Delta_i|^2+|\Delta_I|^2  \! \pm \! \sqrt{4\varepsilon^2\mu^2 + 2|\Delta_I|^2(2\varepsilon^2  + |\Delta_i|^2) + \Delta_i^2\Delta_I^{\dagger 2} + \Delta_i^{\dagger 2}\Delta_I^2}},
\end{eqnarray}
where we have kept the $\bfk$-dependence implicit. 
We directly see that the presence of the interband order parameter make the quasiparticle energies not following the standard $E = \sqrt{\varepsilon^2 + |\Delta|^2}$ BCS form. Thus, even though the order parameters are expressed in the band basis, where we have direct access to the Fermi surface and its zero-energy excitations in the normal state, the presence of the interband pairing makes the quasiparticle energy spectrum highly non-trivial. Thus the nodal structure, i.e.~where quasiparticles appear at zero energy, is also not straightforwardly identified at this stage.

\subsubsection{Order parameter symmetries}
The order parameters $\Delta_\alpha$ can be calculated self-consistently once we know the quasiparticle spectrum and its eigenstates. Close to $T_c$ the order parameters are small and after some straightforward algebra \cite{Black-Schaffer07} we arrive at linear self-consistency equations for $\Delta_\alpha$ as function of the inverse transition temperature $\beta_c = (k_BT_c)^{-1}$:
%EQUATION:
\begin{eqnarray}
\label{eq:selfcons}
\fl \Delta_\alpha = & \frac{J}{N}  \sum_{\bfk,\beta} \left[\cos(\bfkR-\varphi) \cos(\bfk \cdot \bfR_\beta-\varphi) 
\left( \frac{\tanh(\frac{\beta_c}{2} (\varepsilon+\mu))}{2(\varepsilon +\mu)} + \frac{\tanh( \frac{\beta_c}{2}(\varepsilon-\mu))}{2(\varepsilon-\mu)} \right) \right. \nonumber \\
\fl & \left. + \sin(\bfkR-\varphi)\sin({\bfk\cdot \bfR_\beta}-\varphi) \left( \frac{\sinh(\beta_c\mu)}{2\mu \cosh( \frac{\beta_c}{2} (\varepsilon+\mu)) \cosh( \frac{\beta_c}{2}(\varepsilon-\mu))} \right) \right] \Delta_\beta.
\end{eqnarray}
The first part has the regular $\tanh(\beta_cE/2)/(2E)$ BCS form, but here doubled because of the presence of two bands. The second part is non-standard and due entirely to the interband pairing between the upper and lower $\pi$-band. At finite doping the interband pairing can be expected to be very small because it pairs electrons at notably different energy levels. The second part is consequently largely insignificant at higher doping levels. It turns out to also be very small for reasonably low temperatures, and we therefore often simply ignore it. 

\begin{figure}[htb]
\begin{center}
\includegraphics[scale = 0.6]{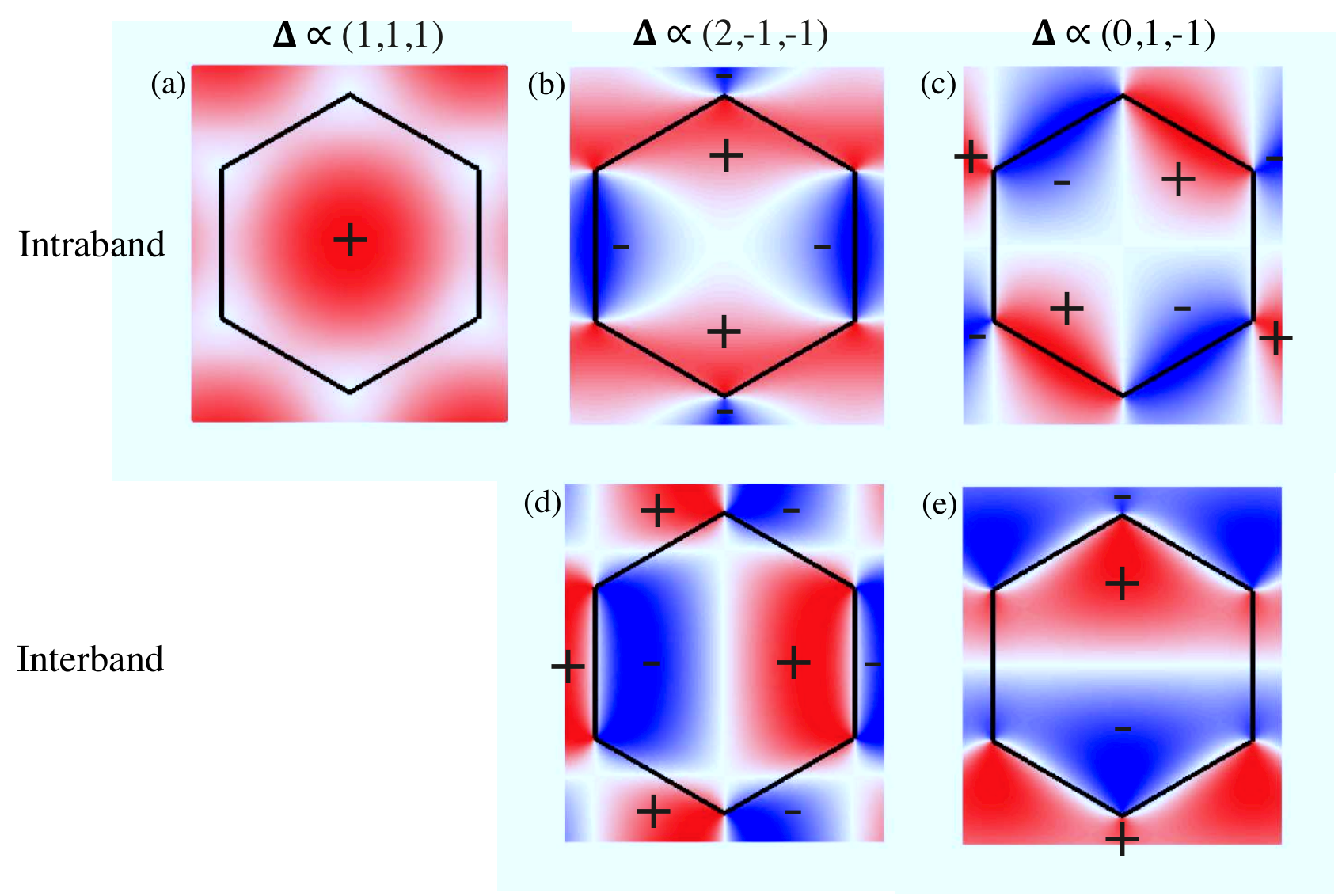}
\caption{\label{fig:bandOPs} (Color online). The $\bfk$-dependence of the intra- (upper figures) and interband (lower figures) order parameters for the ${\bf \Delta} = \frac{1}{\sqrt{3}}(1,1,1)^T$ extended $s$-wave solution (a) and the ${\bf \Delta} = \{ \frac{1}{\sqrt{6}}(2,-1,-1),\frac{1}{\sqrt{2}}(0,1,-1)\}$ solutions (b,c,d,e). The intraband $d$-wave and interband $p$-wave characters are clearly evident in the latter plots. Red indicates positive values, blue indicate negative values, black lines show the first Brillouin zone. In order to produce real values, the interband order parameters have been divided with $i$.
}
\end{center}
\end{figure}

The self-consistency equations (\ref{eq:selfcons}) can be written in matrix form using the vector ${\bf \Delta} = (\Delta_1,\Delta_2,\Delta_3)^T$:
%
% EQUATION:
\begin{eqnarray}
\label{eq:gapmatrix}
\frac{1}{J}{\bf \Delta} = 
\left( \begin{array}{ccc}
A & B & B \\ B & A & B \\ B & B & A \\
\end{array} \right) {\bf \Delta},
\end{eqnarray}
where $B=B(\beta_c)$ is the right-hand side of (\ref{eq:selfcons}) divided by $J$ when $\alpha \neq \beta$ and $A=A(\beta_c)$ is the right-hand side divided by $J$ when $\alpha = \beta$. The eigenvalues  to the above matrix is easily found to be
%
% EQUATION:
\begin{eqnarray}
\label{eq:eigenvalues}
\frac{1}{J} =  \left\{
\begin{array}{ll}
A+2B, & {\rm extended} \ s{\rm-wave} \\
A-B, & d{\rm -wave} \ (p{\rm-wave}).
\end{array} 
\right.
\end{eqnarray}
The first solution has eigenvector ${\bf \Delta} = \frac{1}{\sqrt{3}}(1,1,1)^T$ and is thus an isotropic bond order parameter. It results in the $\bfk$-dependence of the intraband order parameter $\Delta_i$ being directly proportional to $\varepsilon(\bfk)$. This is thus an extended $s$-wave state, belonging to the $A_{1g}$ irreducible representation and plotted in figure \ref{fig:bandOPs}(a). The interband pairing $\Delta_I = 0$ for the extended $s$-wave solution. The second solution is twofold degenerate and spanned by the vectors ${\bf \Delta} = \{ \frac{1}{\sqrt{6}}(2,-1,-1)^T,\frac{1}{\sqrt{2}}(0,1,-1)^T\}$. The intraband pairing $\Delta_i$ has for both of these solutions a fourfold, or $d$-wave symmetry, in $\bfk$-space as illustrated in Figures \ref{fig:bandOPs}(b,c), and they are thus the basis functions spanning up the $E_{2g}$ irreducible representation. The interband pairing $\Delta_I$ has a twofold, or $p$-wave symmetry, displayed in Figures \ref{fig:bandOPs}(d,e) and they thus belongs to the $E_{1u}$ irreducible representation. 
% FIGURE:
Since the interband pairing does not significantly influence the self-consistency equation (\ref{eq:selfcons}) at finite doping levels, we will call the two different solutions extended $s$-wave and $d$-wave, respectively.
The twofold degeneracy at $T_c$ of the second solution in (\ref{eq:eigenvalues}) is a consequence of the eigensolutions belonging to two-dimensional irreducible representations and can only be lifted by higher order corrections present at temperatures below $T_c$. 
Despite the complications from interband pairing, we can conclude by studying the quasiparticle energies (\ref{eq:EQP}) that at finite doping the $s$-wave solution will aways be fully gapped, whereas the $d$-wave solution, taken as any real combination of its two basis vectors, will have nodal quasiparticles down to zero energy. It is however possible to also fully gap the $d$-wave solutions by creating a complex combination of the type $d_{x^2-y^2} \pm id_{xy}$.

Before continuing, we point out again that the order parameter symmetries derived above are valid in the band basis, i.e.~in the basis with a defined normal-state Fermi surface. If we instead study the symmetry of the order parameters in the atomic basis, i.e.~before the band diagonalization, the different superconducting states have significantly more complicated symmetries. Expanded around the $K_{\pm} = (K,K')$ Dirac points the extended $s$-wave symmetry takes the form $ik_x\pm k_y$, i.e.~an effective $p+ip$ order, but with a sign change between the two Dirac points as to still preserve time-reversal symmetry \cite{Uchoa07, Linder09}. The two $d$-wave orders, on the other hand, have both a constan, $s$-wave part and a $p+ip$ part, again with sign changes between the two Dirac points \cite{Linder09}. For the complex combination $d_{x^2-y^2} \pm id_{xy}$, the order parameters have constant $s$-wave pairing around one of the Dirac points and a $ik_x+k_y$ state around the other Dirac point \cite{Jiang08}. 
As seen, the order parameter symmetries are thus significantly simplified in the band structure basis, the only drawback is that an additional interband pairing instead appears. However, sufficiently away from the Dirac point this order parameter should not be important.

\begin{figure}[htb]
\begin{center}
\includegraphics[scale = 0.5]{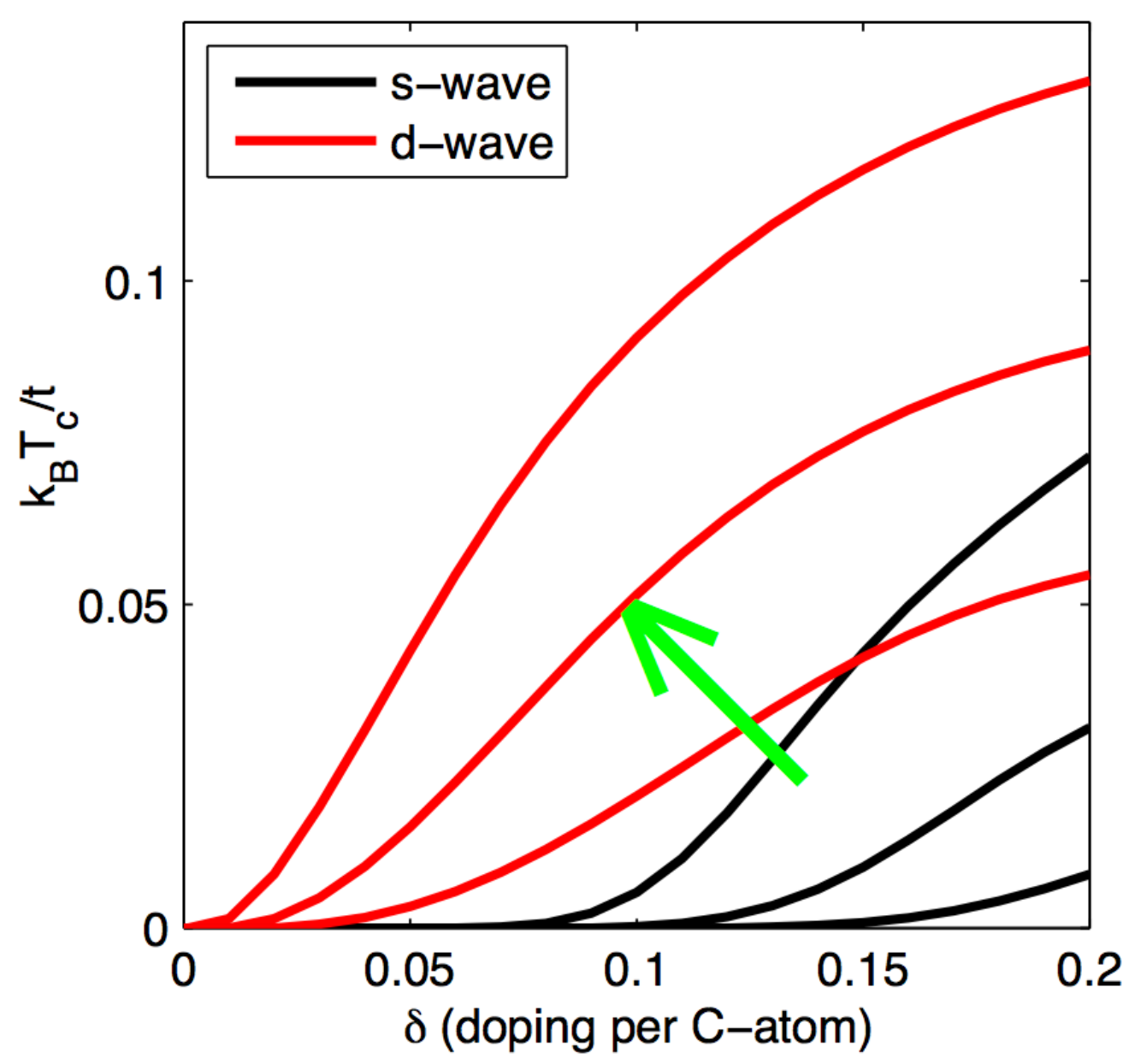}
\caption{\label{fig:Tcvsdoping} (Color online). Superconducting transition temperature $T_c$ for the extended $s$-wave (black) and $d$-wave (red) solutions as function of doping level $\delta$ in graphene for $J/t = 0.8, 1, 1.2$ (increasing values indicated by green arrow) in (\ref{eq:HMF}).
}
\end{center}
\end{figure}

To determine the actual transition temperatures for the different solutions, we proceed by solving (\ref{eq:selfcons}) numerically. Figure \ref{fig:Tcvsdoping} shows the transition temperatures for the $s$- and $d$-wave solutions as function of doping $\delta$ for doping levels below the van Hove singularity (at $\delta = 0.25$) and for several different coupling constants $J$. 
%
% FIGURE:
In pristine graphene there is a quantum critical point at $J/t = 1.91$ for achieving superconductivity, since the Fermi surface then only consists of two points. At zero doping the $s$- and $d$-wave solutions are also degenerate. For finite doping there is no quantum critical point, but always a finite $T_c$ for a non-zero coupling constant $J$. Also, the $d$-wave solution has a much higher $T_c$ than the $s$-wave solution and it is thus the preferred superconducting instability for Fermi surfaces centered around $K,K'$ \cite{Black-Schaffer07}. If we reinstate the statistical weight factors present in the strong coupling $t$-$J$ model, the $d$-wave solution is still the preferred solution at finite doping \cite{Wu13}. Also, solving the strong coupling $t$-$J$ model using slave-boson theory produces results in agreement with these mean-field results \cite{Wu13}.
If we go beyond the weakly doped regime to close and beyond the van Hove singularity, the $d$-wave solution is still dominating, until at extremely large doping levels the leading superconducting instability finally turn into an $s$-wave state for the effective $t$-$J$ model (\ref{eq:HMF}) \cite{Lothman14}.

From the self-consistency equation at $T_c$ (\ref{eq:selfcons}) we cannot determine which linear combination of the two different $d$-wave solutions appear below $T_c$. It was early established numerically that only solutions which leave $E_{QP}$ sixfold symmetric will be allowed by the self-consistency condition below $T_c$ \cite{Black-Schaffer07}. This limits the $d$-wave solutions to combinations ${\bf \Delta} = \frac{1}{\sqrt{3}}(1,e^{i2\pi/3},e^{i4\pi/3})^T$, and permutations thereof, i.e.~the three complex cube roots of 1. Decomposing this particular solution into its real and imaginary part we see directly that this is the chiral $d_{x^2-y^2}- id_{xy}$-wave solution. This solution, and its permutations of course, has a fully gapped quasiparticle spectrum for $\bfk \neq K,K',\Gamma$. It also breaks time-reversal symmetry since it is an intrinsically complex order parameter, where the complex phase cannot be gauged away. 
The preference for the $d_{x^2-y^2}\pm id_{xy}$-wave solutions can also very generally be obtained using a Ginzburg-Landau expansion of the free energy. Calling the (possibly complex) coefficients in front the two different $d$-wave basis functions $\Delta_1$ and $\Delta_2$, the fourth-order expansion of the free energy on the hexagonal lattice has the form \cite{Sigrist91}:
%
% EQUATION:
\begin{eqnarray}
\label{eq:FGL}
\fl F = F_0(T) + \alpha(T-T_c) (|\Delta_{1}|^2 + |\Delta_2|^2) + \beta_1(|\Delta_{1}|^2 + |\Delta_2|^2)^2 + \beta_2 (\Delta_1^\ast \Delta_2 - \Delta_1 \Delta_2^\ast)^2.
\end{eqnarray}
Here $F_0$ is the normal state free energy and $\alpha, \beta_1, \beta_2$ are coefficients. For a superconducting transition at $T_c$ we need $\alpha, \beta_1 >0$. If further $\beta_2 <0$ then a $d_{x^2-y^2}$- or a $d_{xy}$-wave state would arise, whereas for $\beta_2> 0$ the complex combinations $d_{x^2-y^2} \pm id_{xy}$ is energetically favored \cite{Sigrist91}. For both graphene around the van Hove singularity \cite{Nandkishore11} and for generic circular Fermi surfaces on the hexagonal lattice \cite{Kuznetsova05}, it has been shown that $\beta_2>0$, and thus the chiral  $d_{x^2-y^2} \pm id_{xy}$-wave combination is the widely favored solution. This also agrees with the much simpler energy argument which only aims to minimize the number of nodes is the quasiparticle spectrum.

The two disjoint Fermi surfaces at low to moderate doping in graphene have also been proposed to open up for the possibility of a non-chiral, time-reversal invariant $d_{x^2-y^2}+id_{xy}$-wave state \cite{Tran11,Wu13}. Here the order parameter on the Fermi surface around $K$ has, say, $d_{x^2-y^2}+id_{xy}$-wave symmetry, whereas the order parameter on the Fermi surface around $K'=-K$ has the opposite chirality, i.e.~$d_{x^2-y^2}-id_{xy}$-wave symmetry, thus effectively canceling the overall chirality. However, assuming zero-momentum pairing, such that the electrons in a Cooper pair come from the $K$ and $K'$ valleys, and a spin-singlet state this mixed chirality state has been shown to not be physically possible \cite{Black-Schaffer14tJ}. The argument boils down to the fact that the $d_{xy}$-wave component, which changes signs between the two valleys, ends up with odd spatial parity in the Brillouin zone and must thus be a spin-triplet state, whereas the $d_{x^2-y^2}$-wave component is still in a spin-singlet state, which is inconsistent with assuming only spin-singlet pairing. Thus a $d$-wave state from e.g.~an effective $t$-$J$ model, which explicitly gives zero-momentum spin-singlet pairing, cannot have such mixed chirality. Within the $t$-$J$ model real space modulations of the chiral $d$-wave state was also investigated but a net zero sum chirality state was never found \cite{Black-Schaffer14tJ}.
However, going beyond this model and allowing for spin-triplet/spin-singlet mixing, while keeping the coupling between the two Fermi surfaces weak, could possibly still produce a mixed chirality state.

Finally, we also briefly comment on how more spatially extended interactions have shown to not, in any significant manner, change the symmetry of the superconducting state. A mean-field study of a model with $J_2$ spin-singlet bond pairing on next-nearest-neighbor bonds instead of on nearest-neighbor bonds also yields $d_{x^2-y^2}+id_{xy}$-wave symmetry for the intraband pairing \cite{Black-Schaffer12}. Moreover, including both $J$ and a smaller $J_2$ has been shown to further enhance the $d$-wave state over the extended $s$-wave state \cite{Wu13}. 

\subsection{Accurate numerical many-body results}
The Hubbard Hamiltonian (\ref{eq:HU}) has also directly been studied on the doped honeycomb lattice using quantum Monte Carlo (QMC) techniques. These results are of special interest since the estimated Hubbard-$U$ parameter in graphene is of intermediate strength, where strictly speaking neither a weak-coupling nor a strong-coupling approach is justifiable. 
The variational Monte Carlo method applied to the Hubbard Hamiltonian with  $U = 2.4t$, starting with the mean-field solution derived above with an added Gutzwiller-Jastrow factor, has been shown to yield a chiral $d$-wave superconducting state as the ground state for a wide range of finite doping values \cite{Pathak10}. The superconducting phase was shown to form a dome structure in the temperature-doping phase diagram, similar to the high-temperature cuprate superconductors. 
Both determinant quantum Monte Carlo and constrained path Monte Carlo methods have also been applied to the Hubbard model for $U = 3t$ \cite{Ma11}. Also here the chiral $d$-wave state was found close to charge neutrality. However, the long-range part of the $d$-wave pairing correlations was found to vanish in the thermodynamic limit, indicating that electron correlations might not be strong enough to give an intrinsic superconducting state in lightly doped graphene.

Furthermore, the $t$-$J$ model has also recently been studied on the honeycomb lattice using a Grassman tensor product state variational calculation \cite{Gu13}. These calculations targeted the possibility of superconductivity arising when doping a Mott insulating phase at charge neutrality. The chiral $d$-wave superconducting state was found to appear at finite doping. A finite co-existence region between superconductivity and antiferromagnetism was also found at low doping levels. 
These results agree with functional renormalization group calculations on the lightly doped honeycomb lattice, where chiral $d$-wave superconductivity was found to develop from a spin-density wave state at charge neutrality generated by a finite Heisenberg interaction \cite{Honerkamp08}.

A combination of exact diagonalization, density matrix renormalization group and variational Monte Carlo methods on finite clusters have also very recently been applied to both the Hubbard model and the $t$-$J$ model at the van Hove singularity at $1/4$ doping levels \cite{JiangMesaros14}. For the $t$-$J$ model chiral $d$-wave superconductivity was found to always be a very competitive state, and the leading instability at $J/t>0.8$, whereas for the Hubbard model, a chiral spin-density-wave state (and possibly a novel spin-charge-Chern liquid state) was found in the ground state. Moreover, with any slight doping away from the van Hove singularity, the extra charge was argued to very generally favor chiral $d$-wave superconductivity.

Both mean-field calculations and these numerical many-body results thus show that chiral $d$-wave superconductivity appears quite generally at finite doping levels in models which effectively tries to capture the electron-electron interactions in graphene. If graphene finally becomes superconducting or not depends according to these models on the strength and spatial extent of the electron-electron interactions. Based on the lack of experimental detection of ordered states, electron-electron interactions are probably a little bit too weak to cause both an antiferromagnetic state at charge neutrality and chiral $d$-wave superconductivity in lightly doped graphene.
In the next section we will go beyond the lightly doped regime and review multiple calculations which show that graphene doped close to the van Hove singularity, where the density of states diverges, are likely in a chiral $d$-wave superconducting state at very low temperatures. Here the divergent density of state significantly help to enhance the effect of electron-electron interactions.ÊÊ

% -------------------------------------------------- %
% RG RESULTS
% -------------------------------------------------- %
\section{Renormalization group results for graphene}
\label{sec:RG}
Here we turn to discuss how the renormalization group (RG) framework has been used in order to obtain an unbiased picture of the possible ground states in models for graphene. Compared to the studies mentioned so far which focus only on $d$-wave superconductivity, the RG studies aim at comparing the relative strengths of all different ordering possibilities and can thus give a more definitive answer on the possibility for chiral $d$-wave superconductivity in graphene.

To start the discussion let us quickly repeat some important basic aspects of graphene that were already introduced in Sections \ref{sec:SC_D6h} and \ref{sec:MFT}.
For the electrons near the Fermi level in graphene, the simplest model is a honeycomb lattice with lattice sites connected by nearest neighbor hopping, as  written, e.g., in (\ref{eq:HU}). Then, at charge neutrality or zero doping, the Fermi surface consists of just two points on the corners of the Brillouin zone, the famous Dirac-points $K$ and $K'$.  Away from these points, the absolute value of the band energy rises linearly. 
This implies a density of states that grows proportionally to the energy distance from the Dirac point. While the vanishing of the density of states at the Fermi level weakens the effect of electronic interactions, the particle-hole symmetry at charge neutrality and the valley degeneracy imply a rich picture of potential interaction effects. Namely, for each state with wave vector  ${\bf k}$ with band energy $\varepsilon$ in the band above the Fermi level near one Dirac point $K$, there is a state with wave vector  ${\bf k}$ with the opposite band energy $-\varepsilon$, a state with the same band energy $\varepsilon$ near $K'$ with wave vector ${\bf k}+ {\bf K'}- {\bf K}$,  and another state near $K'$ also with wave vector ${\bf k}+ {\bf K'}- {\bf K}$ but the opposite energy $- \varepsilon$. 

The presence of partner states with the same or opposite band energy for every wavevector, also very close to the Fermi level, is called {\em nesting}. If the partners have the same energy, the nesting is said to be in the particle-particle channel, whereas if they have opposite band energies, the nesting is in the particle-hole channel.
If the density of states in a given model remains nonzero down to $\varepsilon \to 0$, any nesting results in logarithmically divergent one-loop diagrams in perturbation theory for temperature $T \to 0$. In higher orders in perturbation theory, e.g.~in ladder- or RPA-type summations, the logarithmical divergences for $T \to 0$ pile up to power-law divergences at nonzero $T$, which is interpreted as the onset of symmetry breaking or long-range ordering in a particular channel. In two spatial dimensions, long-range ordering at $T>0$ is not straightforward due to the importance of collective fluctuations, but very likely a divergence in perturbation theory can still be interpreted as the onset of relevant strong correlations of the type indicated by the most divergent channel.  The fact that there are several nesting possibilities for graphene already indicates that there is likely an interesting interplay between different particle-hole and particle-particle channels competing for dominance at low $T$. However, the vanishing of the density of states at charge neutrality means that none of the one-loop diagrams actually diverges, but they all saturate at finite values for $T \to 0$. Then, perturbation theory can actually have a non-zero radius of convergence, for sufficiently small interactions. Consequently, the semi-metallic dispersion of the honeycomb lattice remains robust for weak interactions and no long-range order occurs even at $T = 0$. It is currently believed that monolayer pristine graphene is in this parameter regime, as no indications for spontaneous long-range ordering or energy gap opening have so far been found experimentally \cite{Sepioni10,Elias11,Mayorov11}.

It is usually expected, and also found in many theories, that there is in this case still a nonzero threshold value for the interaction strength above which the instability occurs anyways. In a simplified picture, this can be discussed using the Stoner criterion for ordering:
\begin{equation}
\label{stoner}
g \chi (\bfq) \ge 1 \, , 
\end{equation} 
with a coupling constant $g$ and a wave vector dependent susceptibility (i.e.~a one-loop diagram) $\chi (\bfq)$ which depends on the type of ordering, particle-particle, particle-hole symmetry etc., being considered. Since the density of states vanishes in monolayer graphene, all possible $\chi (\bfq)$ remain finite for $T\to 0$. Hence $g$ has to become larger than $1/\chi$ in order to cause an instability and thus a ground state change. As $g$ will be some average of the bare interaction, increasing the bare coupling will usually cause several of the mentioned nested channels to become 'critical' in (\ref{stoner}). Therefore, the ordering criterion (\ref{stoner}) and also mean-field theories usually indicate many different ordering possibilities. 
Upon doping, Fermi circles open around the $K$ and $K'$-points, with increasing aspherical deformations upon further doping. Then the particle-hole instabilities become weakened, and pairing instabilities might arise as the dominating ones at sufficiently low temperatures. Yet, the particle-hole channels still play an important role in deciding about the symmetry of the pairing, as is for example known from the study of spin- or charge-fluctuation mechanisms for pairing \cite{Scalapino,Onari}. 
Finally reaching the van Hove doping at $\delta = 1/4$, both the particle-hole channel at the wave vectors connecting the van Hove points, as well as the particle-particle bubble at zero total moment diverge at low $T$, which results in multiple ordered state possibilities.  

Hence, the study of symmetry breaking in models like the one on the honeycomb lattice can become  a rather delicate task as there are different singular channels in perturbation theory.
Here, RG techniques represent a powerful tool to unravel these ambiguities. Their advantage is that they sum up all possible one-loop contributions, usually corresponding to specific ordering tendencies, to the perturbation theory in an unbiased way. This allows for both including the interplay between different channels and seeing which channel wins the competition towards an ordered state.

Here we do not give a full-fledged formulae or a derivation of the renormalization group equations. For this, the reader is referred to recent reviews by e.g.~Metzner {\it et al.} \cite{Metzner12} and Platt {\it et al.} \cite{Platt13}. We instead just sketch the main ingredients. The starting point for the RG is a bare Hamiltonian, given in terms of a dispersion and interactions. Some works make explicit reference to ab-initio derived Wannier representations of the conduction and valence $\pi$-bands of graphene and use cRPA values for the interactions, while other works remain more abstract and define effective coupling constants without an explicit algorithm for how they could be computed from first principles. Most works employ the functional integral formalism, where the Hamiltonian defines the initial action of the system. The dispersion goes into the quadratic part of the action, while the interaction is quartic in the fermion fields.
The RG then consists of integrating out the quadratic part of the action by defining a cutoff, or RG scale $\Lambda$, above which the modes are considered and below which the modes are left untouched. Most studies, see e.g. the reviews \cite{Metzner12} and \cite{Platt13}, use a band energy cutoff, such that modes with $|\varepsilon_b(\bfk)| \ge \Lambda$ in band $b$ are integrated out for cutoff $\Lambda$. However, at least one work on the honeycomb lattice has employed a cutoff in Matsubara frequency space \cite{Wang11}. These different cutoff variants have been used in other contexts too \cite{Metzner12}, and it is the general understanding that, besides some known caveats, the choice of the cutoff is mainly a technical question, without any practical importance for the physical results.

When a certain part of the modes are integrated over, the perturbation expansion for the vertex functions of the theory changes. The usual object of study is the quartic interaction vertex, determined by a coupling function $V_\Lambda$. This change can be cast in the form of an ordinary differential equation for the coupling function, schematically of the structure
\begin{equation}
\label{RGflow}
\frac{dV_{\Lambda}}{d \Lambda} = V_{\Lambda} \circ \dot \chi_\Lambda \circ V_{\Lambda} \, . 
\end{equation}
The right hand side is the condensed notation for five one-loop diagrams, which either contain particle-particle diagrams or particle-hole diagrams. With respect to the diagram rules known from normal many-body perturbation theory in the Matsubara formalism, there are cutoff functions on the two internal lines. On one line, there is also a cutoff function differentiation with respect to the RG scale $\Lambda$, $\frac{d}{d\Lambda} \chi_\Lambda (k)$,Êwhile on the other line there is a normal cutoff function $\chi_\Lambda (k')$. Here, $k$ and $k'$ denote the quantum numbers of the propagator lines and $\Lambda$ is the RG scale. In the full formalism, the propagators should be the renormalized ones, but in the usual approximation, the self-energy is neglected such that the internal lines are given by the bare propagator multiplied by the cutoffs. In the momentum-shell RG the cutoff function is typically chosen as a function of the band energy $\varepsilon(\bfk)$ such that it is zero for $|\varepsilon(\bfk)| \le \Lambda$ and unity for $|\varepsilon(\bfk)| > \Lambda$. For the numerical treatment, this step-function-like cutoff function is usually somewhat smeared out. Another variant of RG flows are temperature flows, where by a rescaling of the terms in the action, the RG describes the evolution of the interactions when the temperature is lowered. Then, the right hand side of the flow equation (\ref{RGflow}) instead has temperature-derivatives of the one-loop diagrams.

The object of interest, i.e.~the coupling function $V_\Lambda$, is usually a function of orbital and band indices and of the incoming and outgoing wave vectors. Thus one actually has a differential equation for a function of many variables; a functional differential equation. Treating this functional dependence of the flowing object in a reasonable way is often called {\it functional renormalization group} or fRG in short. In the current literature the frequency dependence of $V_\Lambda$ is usually neglected, but there is a wider span of ways to describe the dependence of $V_\Lambda$ on the wave vectors. As most of the action happens either near the $K$ and $K'$ or the mid-edge points $M_1$, $M_2$ and $M_3$ of the Brillouin zone for the honeycomb lattice, one approach is to approximate the general wave vector dependence by the interaction values $g_i$ for processes $i$ acting between these points. This is in the spirit of the $g$-ology that was successfully used to understand qualitative physics of one-dimensional interacting Fermi systems \cite{Solyom79}. This procedure is best used in the orbital representation, when the projection or contraction of the interactions on the mentioned points in the Brillouin zone is well-defined. If the interaction is instead expressed in the band representation, the orbital-to-band transformation matrix elements have to be taken into account. Then the windings of the Bloch functions, e.g.~around the $K$ and $K'$-points, complicate the contraction on single points and the angular dependence of the coupling function has to be resolved. This can be done by using so-called $N$-patch discretizations \cite{Zanchi98, Halboth00, Honerkamp01}, where the regions around the Brillouin zone points of interest are split up into angular sectors, and the coupling function is held constant within these patches. Another variant to capture more of the wave vector dependence is channel decompositions of the vertex \cite{Husemann09, Maier13}, in particular, the so-called singular-mode fRG \cite{Wang11}, which is described in a bit more detail below.

The initial condition for the RG flow (\ref{RGflow}) is given by the bare interactions, typically parameterized by on-site repulsion $U$, nearest and second-nearest neighbor density-density repulsions $V_1$ and $V_2$ and a Heisenberg spin-exchange between nearest neighbors $J$. Upon integrating out modes, different components of the coupling function develop differently. In most cases, one observes a flow to strong coupling at some critical scale $\Lambda_c$, where at least one class of coupling function components becomes very large in absolute values. This phenomenon is basically a more sophisticated version of the textbook Cooper instability, i.e.~it strongly signals a change of the ground state. Here the flow has to be stopped as the approximations, like the neglect of self-energy corrections (discussed in more detail in e.g.~\cite{Metzner12}), eventually renders the scheme invalid. Nevertheless, important information can be obtained form analyzing the diverging coupling function. The class of coupling functions diverging most strongly signals which type of new ground state occurs. For example, in doped graphene it is mainly the pair scattering processes with total momentum equal to zero that diverge, and in generalization of Cooper instabilities the new ground state should be Cooper-paired. From the precise angular dependence of the pair scattering, the gap symmetry can also be determined, for example by considering the linearized BCS gap equations with the effective RG coupling.
 Furthermore, a comparison between BCS theory and the RG flow for the reduced BCS model shows that the critical scale $\Lambda_c$ should, up to a factor of order 1, coincide with the pairing gap or the critical temperature $T_c$. For a more accurate value of $T_c$, flows at nonzero $T$ or a temperature-flow \cite{Honerkamp01b} may be used as well.

\subsection{Undoped and weakly doped graphene}
Coming back to graphene and regarding the effects of interactions and possible long-range ordering at low scales, there is a wealth of literature for the situation near charge neutrality using a number of methods, see e.g.~\cite{Tchougreeff92, Sorella92, Khevshchenko01,Herbut06, Hou_Chamon07, Honerkamp08, Raghu08, Liu_Li09, Drut09b, Herbut09, Gamayun10, Meng10, Sorella12, Ulybyshev13}). There, the predominant ordering tendencies are in the particle-hole channel, and usually superconductivity is not among the leading candidates. Hence we do not discuss the charge-neutral situation any further here.

For the weakly doped regime in graphene, various RG works have, however, found pairing instabilities. For example, in an $N$-patch study, Honerkamp \cite{Honerkamp08} found a spin-triplet $f$-wave instability, with a sign change between $K$ and $K'$, when the nearest neighbor repulsion $V_1$ is sufficiently strong. The $f$-wave symmetry form a one-dimensional representation under the crystal point group and there is thus no chance to get a chiral state in this case. Nevertheless, this odd-parity pairing state may host sub gap edge states under appropriate conditions.
A $d_{x^2-y^2}+id_{xy}$ instability was also found possible when including an antiferromagnetic Heisenberg interaction $J$.  In this case, the found $d_{x^2-y^2}+id_{xy}$-pairing is a chiral $d$-wave state. 
The same instability has also been obtained in the context of the $t$-$J$-$J'$ model (without occupation number constraint for the fRG part) on the honeycomb lattice \cite{Wu13}. However, it is somewhat unclear how a very large Heisenberg coupling can appear in graphene, as briefly discussed in Section \ref{sec:MFT}.
%Here it was also pointed out that since the coupling between the two Fermi circles is relatively weak, the two Fermi circles could potentially adopt pairing states with opposite chirality. Then the resulting $d\pm id'$-state was argued to be fully gapped, time-reversal invariant, and non-chiral \cite{Tran11}. However, this study was not directly aimed at graphene, as  it is somewhat unclear how a very large Heisenberg coupling can appear in graphene, as briefly discussed in Section \ref{sec:MFT}.

\subsection{Graphene doped to the van Hove singularity}
The most natural stage where chiral superconductivity can occur in graphene is near the van Hove singularity at $1/4$ electron or hole doping. If the dispersion is modeled with just a nearest-neighbor hopping, the resulting Fermi surface for $1/4$-doping is a hexagon with corners at the $M$-points. At these points the dispersion exhibits a quadratic saddle point, leading to a logarithmic divergence in the density of states. 
The $M$-points are located at the end points of the flat sides of the Fermi surface hexagon. Hence these van Hove points enhance the particle-hole nesting between parallel Fermi surface sides. The nesting has the dual effect of both boosting particle-hole fluctuations at these wave vectors, most naturally in the form of a spin-density-wave instability, and of creating an attractive interaction for unconventional pairing channels. 

This strongly doped van Hove situation has recently been analyzed by three different groups using different RG approaches. Nandkishore {\it et al.} \cite{Nandkishore11} used a $g$-ology-like approach that is strictly valid only exactly at the van Hove doping, although much of the physics for doping levels nearby can be extracted from this as well. They found that directly at the van Hove filling and very close to the instability, the two $d$-wave pairing components wins over the spin-density wave tendencies that dominate a distance away from the van Hove point. They further argued, based on Ginzburg-Landau arguments, that the time-reversal symmetry breaking chiral $d_{x^2-y^2}+id_{xy}$ combination gives the best energy gain in the paired state. This work established the qualitative picture of possible chiral $d$-wave superconductivity in graphene doped to the van Hove singularity by analyzing the minimal model for this situation. The picture has been worked out further in subsequent publications, showing that there should be a first-order transition between the potential spin-density-wave and chiral $d$-wave superconducting states  \cite{Nandkishore12}, and also embedding the special case of graphene into a broader picture for fermions on hexagonal lattices \cite{Nandkishore14}.

\begin{figure}
 \centering \resizebox{0.48\columnwidth}{!}{\includegraphics{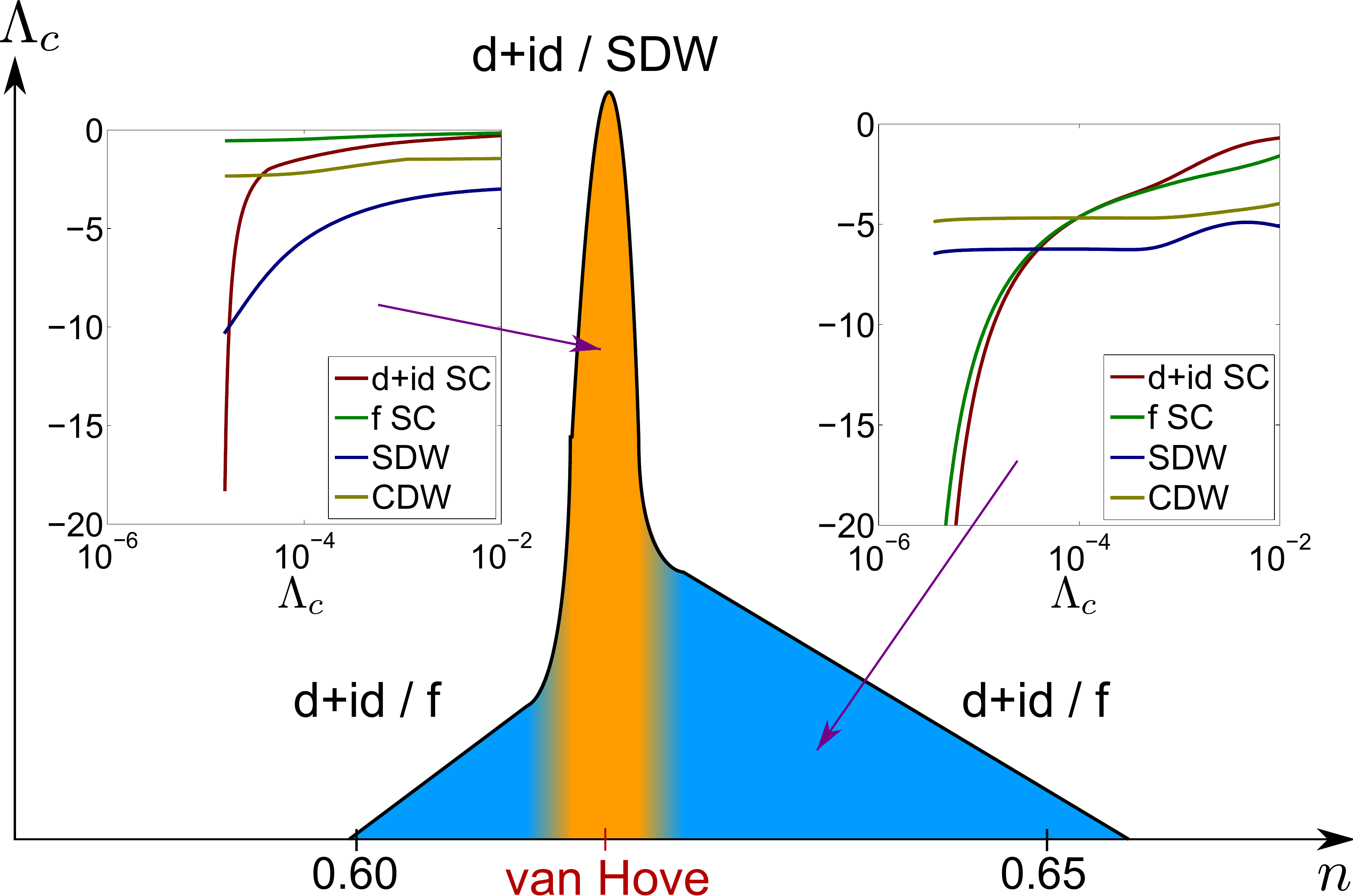} }
 \caption{(Color online). fRG phase diagram from Ref. \cite{Kiesel12}. The main plot shows the instability scale as a function of doping level with additional longer-ranged hoppings included.
At and around the van Hove singularity (orange area), chiral $d$-wave pairing competes with a spin-density-wave (SDW). The left inset picture shows the flow of various interaction channels, indicating a dominant chiral $d$-wave instability at the van Hove singularity. 
Away from the van Hove singularity (blue area), the critical scale drops and whether the chiral $d$-wave or $f$-wave superconductivity instability is preferred depends on the precise decay profile of the interaction.  The right inset picture shows the flow of the interaction channels in such a case.
(Reprinted figure by permission from M.~L.~Kiesel {\it et al.}, Phys. Rev. B {\bf 86}, 020507 (2012), \cite{Kiesel12}.
\href{http://link.aps.org/abstract/PRB/v86/p020507}{Copyright \copyright~(2012) by the American Physical Society.})
}
 \label{KieselPhaseDiagram}
\end{figure}

\begin{figure} 
\centering \resizebox{0.48\columnwidth}{!}{\includegraphics{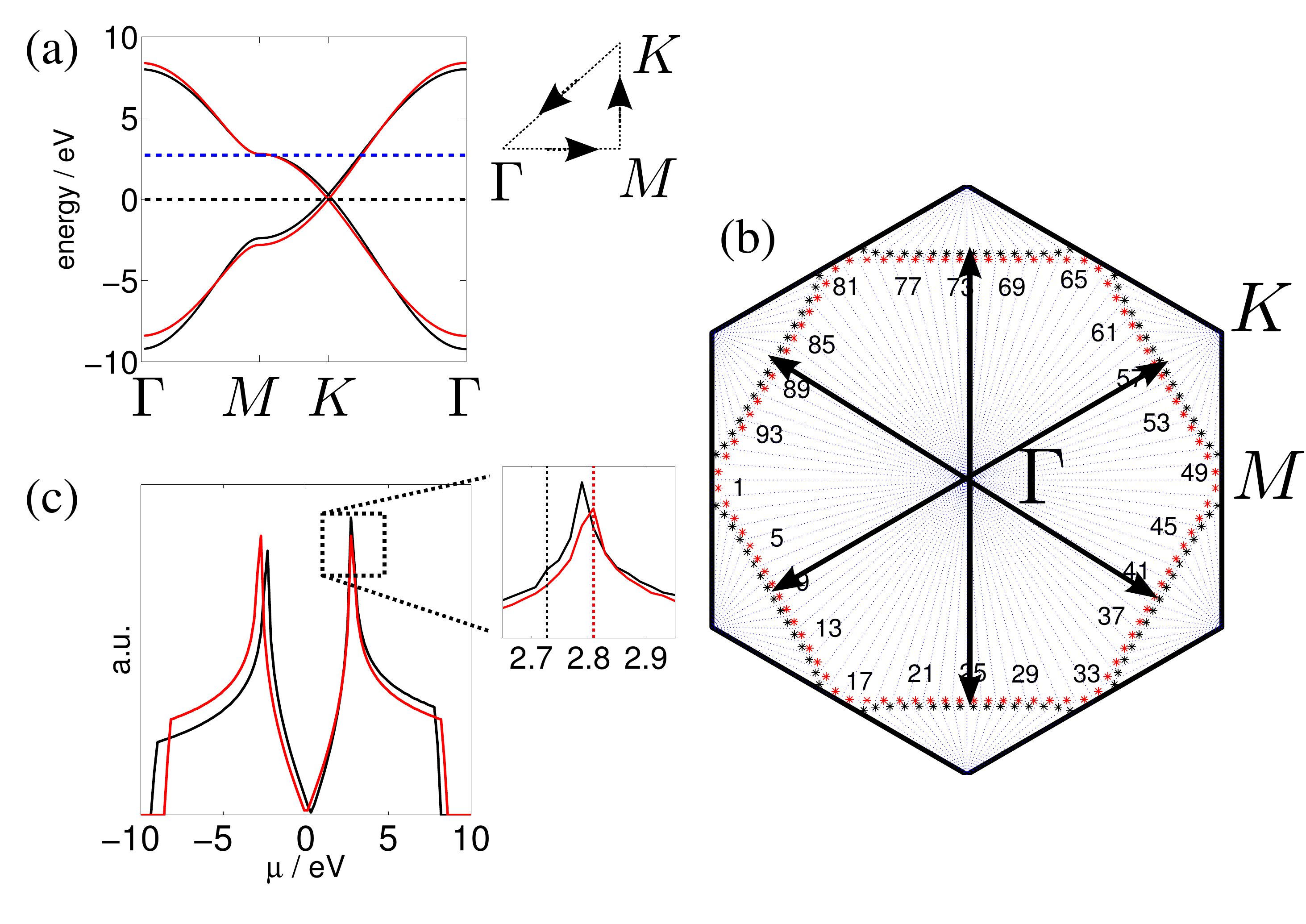} }
 \caption{(Color online). (a) Band structure of the honeycomb lattice with nearest-neighbor hopping only (red) and additional smaller second- and third-nearest neighbor hoppings (black). (b) The Brillouin zone displaying the Fermi surface near the van Hove point (the dashed blue horizontal line in (a) indicates the chemical potential at the van Hove point). The 96 patches used in the fRG and the nesting vectors are indicated as well. (c) Density of states for both band structures in (a). 
(Reprinted figure by permission from M.~L.~Kiesel {\it et al.}, Phys. Rev. B {\bf 86}, 020507 (2012), \cite{Kiesel12}. \href{http://link.aps.org/abstract/PRB/v86/p020507}{Copyright \copyright~(2012) by the American Physical Society.})
}
\label{KieselBandstructure} 
\end{figure}

Kiesel {\it et al.} \cite{Kiesel12} very recently analyzed the same situation using $N$-patch fRG. This approach is more flexible and also allows the study of doping levels away from the van Hove point, as well as a systematically investigation of how changes in the interaction profile affect the result. Furthermore, the model parameters were here adapted from ab-initio results and both longer-ranged hopping parameters and longer-ranged interactions were considered as well. A well-rounded qualitative picture should thus be obtainable.
 Again, chiral $d$-wave pairing was found to be the dominant pairing instability near the van Hove filling, as shown in the phase diagram in Figure \ref{KieselPhaseDiagram}, taken from their work. 
At the van Hove filling chiral $d$-wave superconductivity was found to win over the spin-density-wave state, especially for 'realistic' model parameters, at least if one dares to flow long enough to get close to the instability. Moreover, finite longer range hopping parameters somewhat distort the perfect hexagon of the Fermi surface at the van Hove filling. This decreases the degree of nesting and hence the spin-density-wave tendencies, as visible in Figure \ref{KieselBandstructure}. This further strengthens the chiral $d$-wave pairing in its competition with the spin-density-wave state.
The pairing scales obtained in this work was in the range $10^{-4}$ of the hopping parameter, i.e.~compatible with a transition temperature of a few Kelvins. Of course impurities may very well become a determining factor for the true experimental $T_c$, an effect not considered in this work. 

A third very recent fRG study looking for chiral pairing on the honeycomb lattice is from Wang {\it et al.} \cite{Wang11}. They used the so-called singular-mode fRG, which is based on the same flow equations as in the previously mentioned $N$-patch fRG, but uses a different representation for the electronic interactions. Rather than discretizing the wave vector dependence around the Fermi surface and working with a coupling function that depends on three wave vectors, singular-mode fRG uses a channel decomposition (see also \cite{Husemann09}) and form factors to express the wave vector dependence of the coupling function. This way a better resolution of the modes away from the Fermi surface and of the long-wavelength ordering tendencies is obtained. Even with this approach, the competition between spin-density-wave and chiral $d$-wave pairing was clearly reproduced. The authors interpret their results at the van Hove filling as dominance of the spin-density-wave order, and only away from the van Hove filling did they find a dominating chiral $d$-wave state. Regarding the order of the leading instabilities for this situation one has to say, however, that these differences are well in the uncertainty range of data interpretations. In addition, as mentioned above, the details of the competition will definitely depend on model details such as distance-dependence of the hopping, fine-tuning the degree of the nesting, and the interaction parameters. Still, all RG studies of the van Hove situation share the same features that the spin-density-wave tendency grow first and at larger scales, while chiral $d$-wave pairing develops later in the flow, but rises more steeply in the end. So, drawing distinction lines in tentative phase diagrams heavily depends on how long one trusts the RG flows, and also on the values of the initial interactions.

\begin{figure}
\centering \resizebox{0.48\columnwidth}{!}{\includegraphics{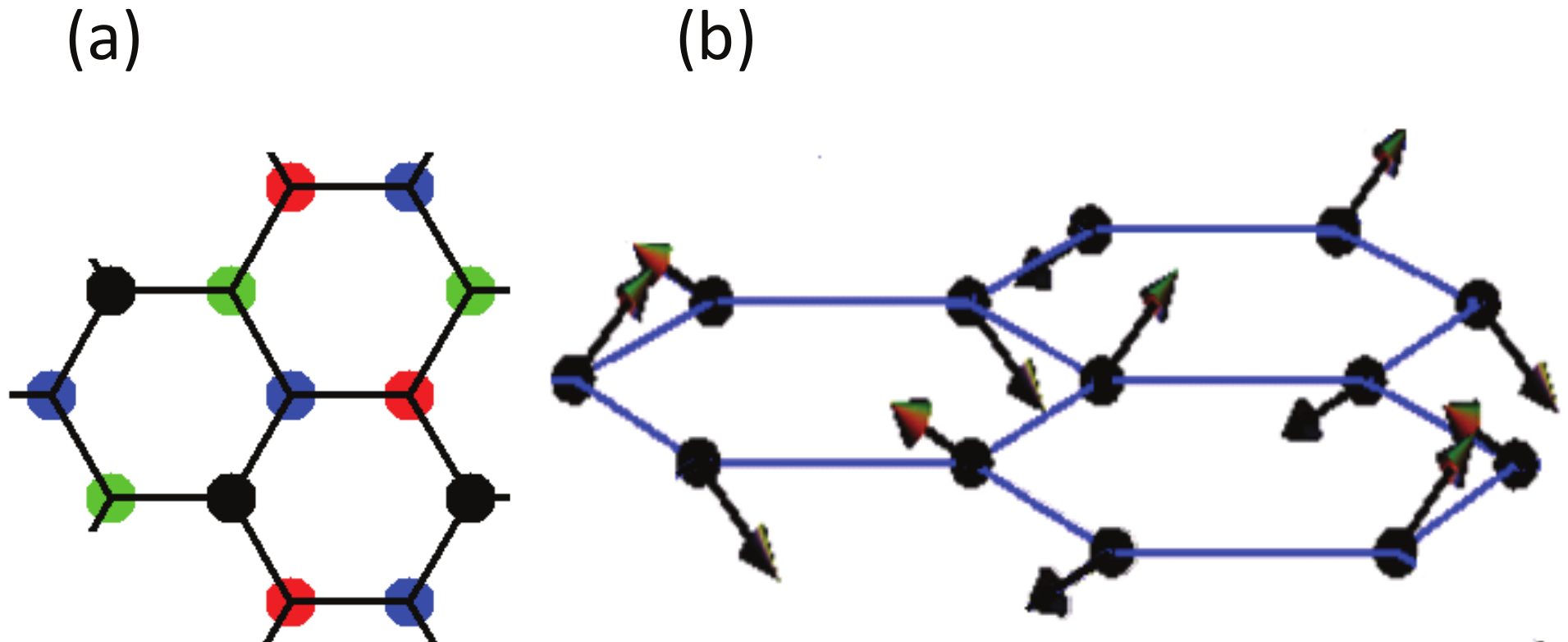} }
 \caption{(Color online). Chiral spin-density-wave order on the honeycomb lattice. (a) The spin expectation values on the black, red, green, and blue sublattices (different gray scales) point in different directions for each color. (b) A three-dimensional view of the chiral spin-density-wave order.
(Reprinted figure by permission from W.-S.~Wang {\it et al.}, Phys. Rev. B {\bf 85}, 035414 (2012), \cite{Wang11}. \href{http://link.aps.org/abstract/PRB/v85/p035414}{Copyright \copyright~(2012) by the American Physical Society.})
}
 \label{WangSpinOrder}      
\end{figure}

Wang {\it et al.} \cite{Wang11} also discuss the spin order in the spin-density-wave phase. The three nesting vectors appear equally strong the fRG flow, as required by symmetry. The energetically most favorable spin order can then be studied best in a mean-field picture. The upshot of these considerations is that the spin order should be chiral, i.e.~non-collinear, with a tripling of the unit cell and four spin directions on the sites within the enlarged unit cell, as shown in Figure \ref{WangSpinOrder}. The spin order breaks time-reversal and the reflection symmetry of the lattice, and is hence appropriately called chiral. This is the same chiral spin-density-wave state found in finite clusters at the van Hove filling using exact diagonalization and density matrix renormalization group calculations on the Hubbard and $t$-$J$ models \cite{JiangMesaros14}.
According to \cite{Nandkishore12b}, the chiral order is only present at the lowest temperatures and gives way for an uniaxial half-metal phase at higher temperatures.

Summarizing these works, we can conclude that according to the RG approach, doping levels approaching the van Hove singularity in graphene is indeed a promising place to look for the occurrence of chiral spin-singlet $d$-wave superconducting pairing. Different independent RG calculations find a strong tendency toward chiral $d$-wave pairing in the proximity to the van Hove filling on the honeycomb lattice. Theoretically, there are still some question marks regarding the validity of these RG studies, not only just concerning the above-mentioned possible smallness of the energy scale for pairing. One fine point is that the van Hove situation itself is known to create non-Fermi-liquid features in the electronic spectral function \cite{Katanin04, Feldman08}. This and additional instabilities, e.g.~due to the coupling to the lattice, might substantially change the low-energy spectrum near the Fermi level and might in bad cases invalidate RG studies without self-energy effects. 
On the other hand, treating these effects in RG increases the effort considerably, and there are very few examples where self-energy effects have been taken into account in two-dimensional problems. Furthermore, the van Hove situation is difficult to study with other, possibly more controlled theoretical approaches like (cluster-)dynamical mean-field theory (DMFT) or QMC, due to the finite-size nature of these approaches. Another difficulty for the RG technique is collective fluctuations. For example, the Mermin-Wagner theorem on the absence of continuous symmetry breaking in two dimensions at nonzero $T$ is not fulfilled in the RG schemes discussed here and strictly speaking all finite-$T$ phase diagrams for the honeycomb lattice have to be viewed as diagrams indicating the dominating ordering tendency, but not true phases. At the same time, in real graphene, the substrate or other couplings to the environment resolve the issue of strict infinite two-dimensionality. 
While all these question marks should be kept in mind, it may be difficult to get to more conclusive statements from theory in the near future. 
In experimental realizations the sample quality and impurities may also be decisive factors that need to be controlled in order to find the pairing state. This is briefly further discussed in Section~\ref{sec:robustness}. It is also possible that the strong doping necessary to reach the van Hove point cannot be treated anymore in a rigid band picture and that additional bands due to the dopant atoms also have to be considered.

\subsection{Bilayer graphene}
The theoretical search for chiral superconducting pairing has also been extended to bilayer graphene systems. For layered systems, the general picture from fRG studies \cite{Scherer12a,Scherer12b} is that the order of relevance of the possible instabilities is not changed  by the adding more layers, but that the the stacking affects the ordering scale in a significant way. While in the monolayer system at charge neutrality, an instability requires a non-zero interaction strength due to the vanishing density of states, in the AB bilayer system the density of states remains non-zero even at the lowest scales and thus an exponential behavior is obtained for the critical scale: $\Lambda_c \sim e^{-1/g}$, with the appropriate dimensionless coupling constant $g$. 

More specifically regarding possible superconducting pairing for bilayer graphene, only the lightly doped situation has so far been studied using the RG technique. The general picture was recently studied in a $g$-ology-like RG work by Murray and Vafek \cite{Murray13}. Besides a phase with non-zero pair momentum and a $f$-wave paired phase without time-reversal symmetry breaking, the authors also detected chiral  $d$-wave pairing instabilities. They further tried to estimate the energy scale for pairing by combining their scale for particle-hole instabilities with the experimentally observed gap scales in bilayer graphene systems and ended up with $T_c$ around 1~K.  However, as they point out, impurities might suppress this scale significantly.
As expected, the nature of the pairing state depends on the nature of the gapped state at charge neutrality. In particular, the chiral $d$-wave superconducting state was predicted to occur upon doping the spin-density-wave state, which is obtained for Hubbard-like initial interactions, much in analogy with other systems with strong antiferromagnetic fluctuations. This way, characterizing the superconducting state would constrain any potential order at charge neutrality and vice versa.

% -------------------------------------------------- %
% PROPERTIES
% -------------------------------------------------- %
\section{Properties of the chiral $d$-wave state}
\label{sec:prop}
In this section we will review some of the, often rather exotic, properties of the chiral $d$-wave superconducting state in graphene. The honeycomb lattice is special insofar that it forces the $d_{x^2-y^2}$ and $d_{xy}$ components to be of equal size in the chiral $d$-wave state, but most of the properties we list below is also present in superconductors where one of the two $d$-wave orders are clearly subdominant. The key thing is not the equal weight but simply the presence of the additional order parameter and the relative $\pi/2$ phase shift between the two order parameters. 
A subdominant complex $d_{xy}$ order has been proposed to appear in the high-temperature $d_{x^2-y^2}$-wave cuprate superconductors in the presence of, for example, magnetic fields \cite{Krishana97, Laughlin98, Elhalel07}, impurities \cite{Balatsky98}, and surfaces \cite{Fogelstrom97, Covington97, Gustafsson13}.

% ENERGY GAP
\subsection{Quasiparticle energy gap}
Probably the most defining property of a superconducting state is its quasiparticle energy spectrum. The BCS quasiparticle energy for a one-band model with band structure $\varepsilon(\bfk)$ is given by
% EQUATION:
\begin{eqnarray}
\label{eq:EQPk}
E_{QP}(\bfk) = \sqrt{\varepsilon(\bfk)^2 + |\Delta(\bfk)|^2}.
\end{eqnarray}
A conventional $s$-wave superconductor has a constant order parameter $\Delta$ and thus have a finite energy gap equal to $|\Delta|$. A $d_{x^2-y^2}$-wave superconductor, on the other hand, has quasiparticles at arbitrary low energies for the intersections of the lines $k_x = \pm k_y$ and the normal state Fermi surface, $\varepsilon(\bfk) = 0$. Such points/lines are referred to as nodal points/lines. 
Any added complex (subdominant) $d_{xy}$ order to the nodal $d_{x^2-y^2}$-wave superconducting state will, however, in general fully gap the quasiparticle energy spectrum.
For example, on the square lattice $\Delta(\bfk) = \Delta_0(\cos k_x - \cos k_y + i \delta \sin k_x \sin k_y)$, with $\delta$ being the fraction of the $d_{xy}$ to the $d_{x^2-y^2}$ component. Then $\Delta$ is only zero at the $\Gamma$-point and at the Brillouin zone corners $\bfk = (\pm \pi,\pm \pi)$. Thus, as long as the Fermi surface ($\varepsilon(\bfk) = 0$) does not pass through these high-symmetry points, the superconducting state has a fully gapped quasiparticle spectrum. For the graphene honeycomb lattice, the chiral $d$-wave pairing  equivalently produces a fully gapped state for doping levels away from charge neutrality ($\epsilon(\bfk) = 0$ at $K,K'$) or a completely empty or full band ($\epsilon(\bfk) = 0$ at $\Gamma$).

% TOPOLOGY
\subsection{Non-trivial topology}
The spin-singlet $d_{x^2-y^2} + id_{xy}$-wave superconducting state also has a non-trivial topology.
It preserves full SU(2) spin-rotation symmetry, as long as the normal state is spin-degenerate. However, it breaks time-reversal symmetry since the time-reversal operator $K$ acts as $K \Delta (\bfk) = \Delta^\ast (-\bfk)$ on a spin-singlet order parameter \cite{Sigrist91}. It thus belongs to class C in the Altland-Zirnbauer classification of Bogoliubov-de Gennes systems \cite{AltlandZirnbauer97, Schnyder08}. The form of the order parameter, where only $k_x$ and $k_y$ is present, clearly indicates two-dimensionality, which is obvious in graphene but is also effectively present in layered systems such as the the cuprate superconductors. Two-dimensional class C superconductors can be classified by an integer-valued ($\mathbb{Z}$) topological invariant \cite{Schnyder08}. This means that the space of quantum ground states is partitioned into topologically distinct sectors which each can be labeled by an integer number.

 It is possible to classify two-dimensional class C superconductors \cite{Sato10} using the integer TKNN number \cite{TKNN}, i.e.~the first Chern number developed for quantum Hall systems. However, the non-trivial topology is more easily visualized by instead using a Skyrmion number formula, which gives the Chern number as \cite{Volovik89,Volovikbook92,Volovik97,Read00}:
% EQUATION:
\begin{eqnarray}
\label{eq:winding}
\mathcal{N} = \frac{1}{4\pi} \int_{\mathrm{BZ}} d^2k \, \hat{\bf m} \cdot \left( \frac{\partial  \hat{\bf m}}{\partial k_x}  \times \frac{\partial  \hat{\bf m}}{\partial k_y} \right) ,
\end{eqnarray}
with the unit vector
\begin{eqnarray}
\label{eq:windind2}
\hat{\bf m} = \frac{1}{\sqrt{\varepsilon(\bfk) ^2 + |\Delta (\bfk)|^2} }\left( \begin{array}{c}
     \mbox{Re} \, \Delta (\bfk)    \\ \mbox{Im} \, \Delta (\bfk)   \\Ê\varepsilon(\bfk)     
\end{array} \right).
\end{eqnarray} 
The scalar triple product under the integral is the directed area on the unit sphere that is spanned by the $\hat{\bf m}$ unit vector when it moves an infinitesimal area $d^2k$. Hence the integral counts how much space on the unit sphere is covered when $\bfk=(k_x,k_y)$ covers the full Brillouin zone. At the bottom of the band $\hat{\bf m}$ points roughly to the south pole, while at the top of the band $\hat{\bf m}$ should point close to the north pole of the unit sphere. Whether the integral over the full Brillouin zone  gives a nonzero contribution or not now depends on if $\hat{\bf m}$ has a finite winding along the lines of constant $\varepsilon$. If is has, then $\hat{\bf m}$  will visit the full sphere at least once and give a nonzero $\mathcal{N}$. If not, then its motion will lead to contributions that cancel in the integral resulting in $\mathcal{N} = 0$, which denotes a topologically trivial state.

For a $m_z = 0$ spin-triplet $p_x\pm ip_y$-wave superconductor state, where $\Delta (\bfk) = \Delta_0 (k_x\pm ik_y) \propto \cos \phi \pm i \sin \phi $ for small $\bfk$ and with $\phi = \arctan(k_y/k_x)$, we get a winding of $2\pi$ for the $\hat{\bf m}$-vector around constant energy lines. The unit sphere will in this case be covered once leading to $\mathcal{N} = \pm 1$, since the $\hat{\bf m}$-vector describes a $xy$ in-plane vortex combined with a reversal of the third component between the bottom and the top of the band. For the spin-singlet $d_{x^2-y^2} +i d_{xy}$ state with $\Delta (\bfk) \propto \cos 2 \phi \pm i \sin 2 \phi $ the winding is double that and hence the sphere is covered twice, leading to $\mathcal{N} = \pm 2$. 
Reversely, a nonzero $\mathcal{N}$ implies a winding by $2\pi \mathcal{N}$ of the superconducting phase at the Fermi level. This is because the mapping of the Brillouin zone to $\mathcal{N}$-times the unit sphere of the $\hat{\bf m}$-vector maps the Fermi level to the equator, and an $\mathcal{N}$-fold coverage requires an $\mathcal{N}$-fold winding of the $xy$-content of the $\hat{\bf m}$-vector in (\ref{eq:windind2}).
If the superconducting phase winds around a closed Fermi surface $\mathcal{N}$-times, the mapping covers the equator $\mathcal{N}$-times. Below the Fermi surface, i.e.~for $\varepsilon <0$, the vector is the in the southern hemisphere, whereas for $\varepsilon >0$ it visits the northern hemisphere. This is topologically identical to the Skyrmion number $\mathcal{N}$. Thus a $\mathcal{N}$ Skyrmion number implies an  $\mathcal{N}$-fold winding of the superconducting state and vice versa.

The $m_z = 0$ spin-triplet $p_x\pm ip_y$-wave and the spin-singlet $d_{x^2-y^2} +i d_{xy}$-wave states are the two prime examples of chiral superconductors with full SU(2) spin-rotation invariance. Chiral here refers to the fact that the order parameter has a certain ``handedness" set by the sign of the Skyrmion number. For a superconducting state chirality implies that the state has to break time-reversal symmetry. Many time-reversal symmetry breaking superconducting states are, however, not chiral. For example, in the other often discussed fully gapped time-reversal breaking state in the cuprates, $d_{x^2-y^2}+is$, the $\hat{\bf m}$-vector just oscillates between two extrema when moving around the ${\bf k}$-point origin and does not cover the full circle, thus giving $\mathcal{N} = 0$ \cite{Volovik97}.
In a multiband superconductor, the Skyrmion winding numbers have to be calculated for each individual band crossing the Fermi level and can then add up to zero \cite{Raghu10bands}. 
In the case of graphene, this chirality of a superconducting state should not be confused with the chirality of the honeycomb band structure, which is related to the winding in sublattice space. 

% QHE
\subsubsection{Quantized Hall effects}
Integer quantum Hall states classified by a TKNN integer have a quantized Hall conductance directly proportional to the TKNN number \cite{TKNN}. However, in a superconductor the condensate consists of spinless charge $2e$ Cooper pairs, whereas the spin is carried by the quasiparticle excitations, which on the other hand do not have definite charge. As a consequence, the usual charge Hall conductance cannot be quantized in a superconductor. Still, there exist other quantized Hall effects in chiral superconductors with a nonzero $\mathcal{N}$; both the spin Hall conductance $\sigma^s$ and the thermal Hall conductance $\kappa$ are quantized. The spin Hall conductance generates a finite spin current $j^z$ in a direction transverse to the direction of the variation of an external Zeeman field:
%
% EQUATION:
\begin{eqnarray}
\label{eq:sqH}
j_x^z  = \sigma_{xy}^s \left(  -\frac{dB^z(y)}{dy}\right).
\end{eqnarray} 
It was shown in \cite{Senthil99} that the spin Hall conductance is quantized in a chiral $d$-wave state according to $\sigma_{xy}^s = 2 {\rm sgn}(\Delta_{d_{xy}}\Delta_{d_{x^2-y^2}})  = \mathcal{N}$ for small $d_{xy}$ components. With the quantized value of the spin Hall conductance being a universal property, which cannot change within the same topological phase, the result must valid for any finite size of the $d_{xy}$ component \cite{Volovik97}.
The thermal Hall conductance similarly measures the transverse heat current as a function of a temperature gradient. The thermal Hall conductance is given by $\kappa_{xy} =2\pi^2Tk_B^2/(3h){\rm sgn}(\Delta_{d_{xy}}\Delta_{d_{x^2-y^2}})$ in a chiral $d$-wave superconductor and thus $\kappa_{xy}/T$ is a quantized quantity \cite{Senthil99,Horovitz03}.  

% EDGE STATES
\subsubsection{Edge states}
The non-trivial topology of the chiral $d$-wave state gives rise to edge states crossing the bulk energy gap. This is a direct consequence of the deeply rooted bulk-boundary correspondence, which states that materials with non-trivial topology in the bulk necessarily have boundary, or edge, states at zero energy, see e.g.~\cite{Hasan10}. Very generally speaking, at the edge of a material any possible non-trivial topology of the bulk band structure has to transform into the trivial structure of the vacuum. Such a topological phase transition is only possible if the energy gap is closing at at least some point in reciprocal space. Thus, somewhere in the region between the interior bulk of the material and the vacuum outside, there needs to be states at zero energy, otherwise it is impossible for the topological invariant to change. This interplay between topology and  edge modes crossing zero-energy was originally found in one-dimensional systems by Jackiw and Rebbi \cite{Jackiw76}, and is very well-established in quantum Hall systems \cite{Halperin82} and the more recently discovered topological insulators and superconductors \cite{Hasan10, Qi11}. 
For systems classified with a Chern number the difference between the number of right and left moving edge modes is equivalent to the change in Chern number across the interface \cite{Hasan10}. For a chiral $d$-wave superconductors this means two co-propagating, or chiral, edge modes crossing the bulk gap \cite{Volovik97}.

The existence of two zero-energy edge states can also be seen from a qualitative quasiclassical argument that only requires knowledge of the $\bfk$-dependence of the order parameter. For Bogoliubov quasiparticle excitations at energy $E$, the amplitudes of the two single-particle states that get hybridized by the gap function are described by 
$u(\bfk) = \sqrt{\frac{1}{2}[1+ (1- |\Delta (\bfk)|^2/E^2)^{1/2}]}$ and $v(\bfk)= \sqrt{\frac{1}{2}[1- (1- |\Delta (\bfk)|^2/E^2)^{1/2}]}$, with the gap phase $\eta (\bfk) = \Delta (\bfk)/|\Delta (\bfk) |$ attached to either of the two amplitudes, usually such that the second  component becomes $\eta (\bfk) v(\bfk)$. 
Now, if we assume that the momentum $\bfk$ forms an angle $\theta$ with the edge, then the trajectory that gets (specularly) reflected at the edge has the angle $\pi - \theta$. The condition for a sub-gap state is then \cite{Hu94, Kashiwaya00}
%
%EQUATION:
\begin{eqnarray}
\label{eq:bscond}
\eta (\theta ) \eta^* (\pi - \theta) = \frac{u(\theta)  u(\pi - \theta) }{ v(\theta )  v(\pi - \theta )}.
\end{eqnarray}
For $E$ smaller than all gaps $|\DeltaÊ|$ occurring in (\ref{eq:bscond}), the right hand side is an energy-dependent complex number $c(E)$ of modulus 1, i.e.~$|c(E)|=1$.
Now, if the phase of the gap winds like $e^{im\theta }$ around a Fermi surface, we get $\eta (\theta ) \eta^* (\pi - \theta)  = - e^{2im \theta}$, leading to (\ref{eq:bscond}) taking the simple form $e^{2im\theta} = - c$. This shows that for $m=1$, there exists an angle $\theta$ for which we find a bound state at every sub-gap energy. For the chiral $d$-wave state we have $m=2$ and thus we instead expect to find two different angles for any given sub-gap energy, resulting in two edge states in the bulk energy gap. This is of course the same conclusion as we arrived at above using the topological invariant.

\begin{figure}[htb]
\begin{center}
\includegraphics[scale = 1]{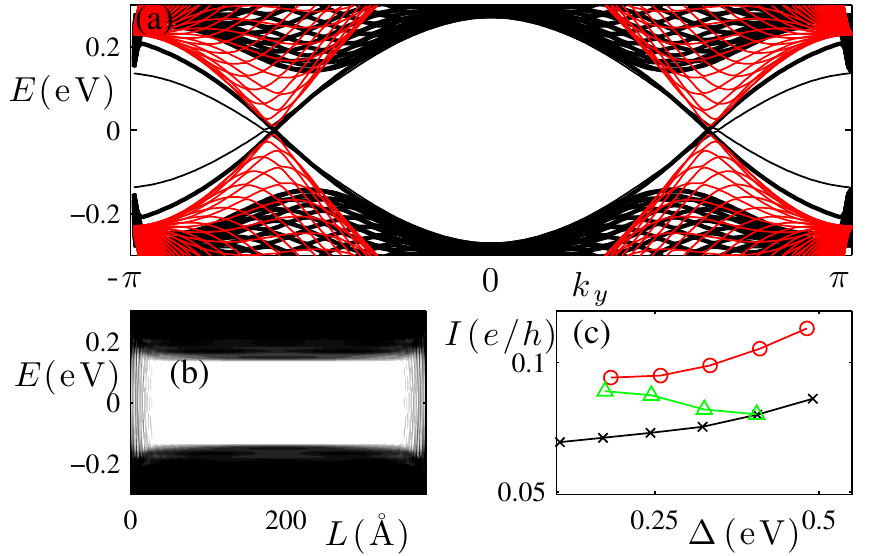}
\caption{\label{fig:edge} (Color online). Zigzag edge ribbon for chiral $d$-wave superconducting graphene close to the van Hove singularity. (a) Band structure for the self-consistent state which have a pure $d_{x^2-y^2}$-wave state on the edge (thick black line), a non-self-consistent constant $d_{x^2-y^2}+id_{xy}$ state (thin black line), and a non-selfconsistent constant $d_{x^2-y^2}$-wave state (red lines), all of equivalent amplitudes. (b) Local density of states across the ribbon for the self-consistent solution in (a) interpolating between 0.2 (black) to 0 (white) states/(eV unit cell), with parameters chosen to give a 0.18~eV bulk energy gap. (c) Quasiparticle edge current as function of the superconducting bulk order parameter amplitude for a zigzag edge at the van Hove singularity at $\mu = t$ (black crosses), at $\mu = 0.8t$ (red circles), and for the armchair edge at $\mu = t$ (green triangles). 
(Reprinted figures by permission from A.~M.~Black-Schaffer, Phys. Rev. Lett. {\bf 109}, 197001 (2012), \cite{Black-Schaffer12PRL}. \href{http://link.aps.org/abstract/PRL/v109/p197001}{Copyright \copyright~(2012) by the American Physical Society.})
}
\end{center}
\end{figure}

A detailed study of the $d$-wave superconducting state in graphene close to different edges was preformed in \cite{Black-Schaffer12PRL}. Both the zigzag and armchair edge terminations were shown to be strongly pair breaking for the $d_{xy}$ component, whereas they both enhanced the $d_{x^2-y^2}$ component. The $d_{xy}$ component was found to only recover inside the material with the recovery length inversely proportional to the strength of the superconducting state.
Any edge of a chiral $d$-wave superconducting graphene sheet can thus be expected to be in a $d_{x^2-y^2}$ state. Even though the very edge region does not break time-reversal symmetry, the chiral $d$-wave state in the bulk still guarantees the existence of chiral states crossing the bulk energy gap somewhere in the, now extended, edge region. Figure \ref{fig:edge}(a) shows the band structure of a zigzag edge ribbon with two co-propagating chiral edge states spanning the bulk energy gap per edge. The self-consistent solution resulting in a pure $d_{x^2-y^2}$ edge (thick black) is compared to the non-selfconsistent solution with bulk $d+id'$ character throughout the ribbon (thin black). Clearly there is very little change in the band edge structure, despite the dramatic effect of the edge on the order parameter itself. This is due to to the fact that the pure $d_{x^2-y^2}$ bulk solution (red lines) has its nodal quasiparticles reaching zero energy at essentially the same momentum as the chiral edge states. 
%
% FIGURE:
Figure \ref{fig:edge}(b) further shows how the chiral edge states are also still well-localized to the edge, despite the strong pair breaking effect on the $d_{xy}$ component.

The two edge states in a chiral $d$-wave superconductor consist of regular fermionic Bogoliubov excitations. However, if the  spin-rotation symmetry is broken by a Rashba spin-orbit term and a Zeeman field, it is possible to transform the chiral edge modes into Majorana fermion edge modes \cite{Black-Schaffer12PRL}. Majorana edge modes only appear in superconductors where the spin degeneracy is broken and one can roughly see a regular fermionic mode as two fully overlapping Majorana modes due to spin degeneracy.
The change in edge modes marks a topological phase transition and can thus only occur if the bulk energy gap closes. In graphene doped close to the van Hove singularity such a topological phase transition takes place in the chiral $d$-wave state when the Zeeman field $h_z = \pm2\Delta$ \cite{Black-Schaffer12PRL}. For larger Zeeman fields each edge carries three Majorana modes, of which two can be combined into a remnant of one of the original chiral modes.

% SPONTANEOUS CURRENTS
\subsubsection{Spontaneous edge currents}
The two co-propagating, or chiral, edge states of a $d_{x^2-y^2} + id_{xy}$-wave superconductor carry a spontaneous edge current \cite{Fogelstrom97,Volovik97,Laughlin98}. For chiral $d$-wave superconducting graphene the edge current has been calculated using the charge continuity equation in combination with the Heisenberg equation for the particle number \cite{Black-Schaffer12PRL}. As seen in figure \ref{fig:edge}(c) the current depends not only on the size of the superconducting order parameter, but also on the doping of the system and the type of edge. The overall current behavior can be understood by studying the evolution of the zero-energy crossings $k_0$ of the chiral edge modes, as changes in the spontaneous current are proportional to changes in $k_0$ \cite{Volovik97,Black-Schaffer12PRL}.
Edge currents in chiral $d$-wave superconductors have also been shown to have a surprising size and direction dependence on the distance from the edge. In \cite{Braunecker05} these were interpreted as Friedel oscillations at two frequencies, $2k_F$ and $\sqrt{2}k_F$. The former is the usual Friedel oscillation of continuum states, whereas the latter is due to the zero-energy edge states. These two oscillations were found to be of equal amplitude and also comparable to the non-oscillating part of the current, leading to a reversal of the current direction in some regions close to the edge. 
The spontaneous edge currents give rise to a finite magnetization which is screened from the bulk of the superconductor by counter-propagating super currents. The spontaneous magnetization and its temperature and spatial dependence has been studied for a $d_{x^2-y^2}+id_{xy}$ state in the cuprate superconductors \cite{Horovitz03}. 

% DOMAIN WALLS, VORTICES
\subsubsection{Domain walls and vortices}
There is a twofold ground state degeneracy in the bulk between the two different chiral states $d_{x^2-y^2}+id_{xy}$ and $d_{x^2-y^2}-id_{xy}$, as evident by the topological invariant $\mathcal{N} = \pm 2$. There can thus exist domain walls in a chiral $d$-wave superconductors, which separates domains with different chiralities.
Following the same argument leading to two chiral edge states for a $d_{x^2-y^2}+id_{xy}$-wave superconductor, a domain wall between these two degenerate bulk solutions hosts four co-propagating, or chiral, states \cite{Volovik97}. This is because at the domain wall the topological invariant changes from $\mathcal{N} = 2$ to $-2$, which necessarily results in four chiral modes in the domain wall. 

A superconducting vortex can be seen as yet another example of a type of edge in a superconductor. A vortex core in a conventional spin-singlet $s$-wave superconductor hosts a spectrum of so-called Caroli-de Gennes-Matricon bound states \cite{Caroli64}, located at finite subgap energies.
In chiral superconductors this spectrum can sometimes be shifted downwards in energy, such that zero-energy core states appears \cite{Kopnin91,Volovik97,Volovik99}. %This has been proposed also for chiral $d$-wave superconductors, which would then be predicted to have $|\mathcal{N}|N$ zero-energy states, where $N$ is the winding number of the vortex \cite{Volovik97}.
Following recent classifications of topological insulators and superconductors, zero-energy states only exist at a defect in a particular topological class if that class is non-trivial in dimensions $r+1$, where $r$ is the dimensionality of the defect \cite{Ryu10,Teo10}. Since a vortex has $r= 0$ and class C for the chiral $d$-wave state is trivial in one dimension \cite{Schnyder08,Ryu10,Teo10}, there should not exist any zero-energy states for zero-dimensional defects in the chiral $d$-wave state, although subgap states are still allowed. 
%A superconducting vortex can be seen as yet another example of a type of edge in a superconductor, in this case circular, and thus vortex cores also host zero-energy states in the chiral $d$-wave state. A vortex core in a conventional $s$-wave superconductor contains a spectrum of so-called Caroli-de Gennes-Matricon bound states \cite{Caroli64}, located at finite subgap energies. In a chiral superconductor this spectrum is shifted downwards in energy, resulting in $|\mathcal{N}|N$ zero-energy core states, where $N$ is the winding number of the vortex \cite{Kopnin91,Volovik97,Volovik99}. Thus a singly quantized vortex in a chiral $d$-wave superconductor contains two zero-energy states.

% PROBES
\subsection{Unique experimental signatures}
Among the many properties listed above, there are plenty of distinctive experimental signatures for the chiral $d$-wave superconducting state. These include quantized spin and thermal Hall conductance, spontaneous edge currents and magnetization, a fully gapped bulk, and zero energy states at edges, domain walls, and in vortices. 

The quantized spin and thermal Hall effects and the spontaneous edge currents are direct consequences of  the non-trivial topology of chiral $d$-wave state and thus not present in the non-chiral $d_{x^2-y2}$- and $d_{xy}$-wave states. 
Moreover, the chiral $d$-wave state has a fully gapped bulk, whereas the non-chiral $d$-wave states have nodal quasiparticles in the bulk. All these properties can thus be used to clearly distinguish between a chiral $d$-wave state and the non-chiral $d_{x^2-y2}$- and $d_{xy}$-wave states. 
In terms of edge states, however, the $d_{x^2-y2}$- and $d_{xy}$-wave states can also have zero-energy edge states, but only on surfaces where the order parameter changes sign between the incoming scattering angle $\theta$ and the reflected angle $\pi-\theta$. This result follows from a similar quasiclassical argument as the one given above for the chiral edge states, see e.g.~\cite{Kashiwaya00}. This means that the $d_{x^2-y^2}$-wave state has zero-energy states on the armchair edge, whereas the $d_{xy}$-wave state has zero-energy states on the zigzag edge \cite{Linder09, Black-Schaffer09}.
However, the zero-energy states in non-chiral $d$-wave superconductors are bound states at zero-energy, in contrast to the propagating chiral states in a chiral $d$-wave superconductor. %These bound zero-energy states can be viewed as bound states formed in a normal layer on top of the superconductor in the limit where the width of the normal layer goes to zero. The bound states are formed in the normal layer due to retro-reflectivity of the Andreev reflection, causing electrons which travel in the normal layer to form closed trajectories. For order parameters with different signs at angles $\theta$ and $\pi-\theta$ these bound states end up at zero energy, see e.g.~\cite{Kashiwaya00}. 

In a comparison to an (extended) $s$-wave superconducting state the chiral $d$-wave state can be identified by essentially all properties discussed above, the only common property of the ones listed is the fully gapped energy spectrum in the bulk. Notably, the existence of zero-energy states has been shown to give the chiral $d$-wave state a distinctive Andreev conductance spectrum, with a pronounce zero-bias conductance peak, through a normal/superconducting graphene junction compared to an $s$-wave superconductor \cite{Jiang08}.
Electronic Raman scattering has also recently been proposed to distinguish between the chiral $d$-wave state and the extended $s$-wave state \cite{Lu13}.

\newpage
% SECTION: 
\section{Robustness of the chiral $d$-wave state} 
\label{sec:robustness}
In this section we will briefly review studies of the robustness of the chiral $d$-wave superconducting state. We will primarily focus on how it might be suppressed, but also enhanced, by materials properties and engineering. In terms of suppressing the superconducting state we discuss the stability of the chiral $d$-wave state in the presence of disorder and impurities. Apart from externally applied perturbations, this is the prime candidate for suppressing a superconducting state in a real material. We also review proposals on how to enhance the chiral $d$-wave state by proximity effect to external superconductors. If chiral $d$-wave superconductivity in graphene is present but very weak, such proximity effects could boost the changes of experimental discovery.

% IMPURITIES
\subsection{Impurity effects}
As with any real material, the role of disorder and impurities is always important in superconductors, see e.g.~\cite{Balatsky06}. Conventional $s$-wave superconductors are well-known to be robust against non-magnetic disorder. This result is known as the Anderson's theorem \cite{Anderson59} and is a consequence of how, even in the presence of disorder, it is possible to pair time-reversed electron states into a Cooper pair with a uniform order parameter. For an unconventional superconductor with a $\bfk$-dependent order parameter, the same protection is not present and both magnetic and non-magnetic impurities are in general pair breaking. This influences the local properties around an impurity, resulting for example in low-energy impurity-induced resonance states, as extensively reviewed in \cite{Balatsky06}.
However, the very presence of superconductivity at high temperatures in the cuprate superconductors makes it evident that some amount of disorder is clearly not debilitating in unconventional superconductors. 
In fact, it has even been shown that impurities can help to produce additional subdominant orders. For example, magnetic impurities in a $d_{x^2-y^2}$-wave superconductor has been shown to locally induce an $id_{xy}$ component \cite{Balatsky98}. This is a consequence of the coupling of the impurity spin to the orbital moment of the condensate. The resulting time-reversal broken state near the impurity has even been shown to carry an induced charge current, a clear characteristic of the non-trivial topology \cite{Graf00}. Also disorder generated by surfaces has been proposed to induce additional complex order parameters in the $d_{x^2-y^2}$-wave superconducting state \cite{Fogelstrom97, Covington97}. Very recent experimental data on small grains of YB$_2$C$_3$O$_{7-\delta}$ \cite{Gustafsson13} agree with the presence of a $d_{x^2-y^2}+is$ superconducting state \cite{Black-Schaffer13cuprate}.

In terms of the intrinsic chiral $d$-wave state in heavily doped graphene, disorder effects have been investigated using a random fluctuating chemical potential \cite{Black-Schaffer12PRL}. Even for moderately strong disorder the chiral $d$-wave state, with equal weight of the two $d$-wave components, was found to survive essentially unchanged. At very strong disorder the superconducting state is weakened and a substantial extended $s$-wave component emerges, as it is more robust to disorder than the chiral $d$-wave state. 

\begin{figure}[htb]
\begin{center}
\includegraphics[scale = 0.4]{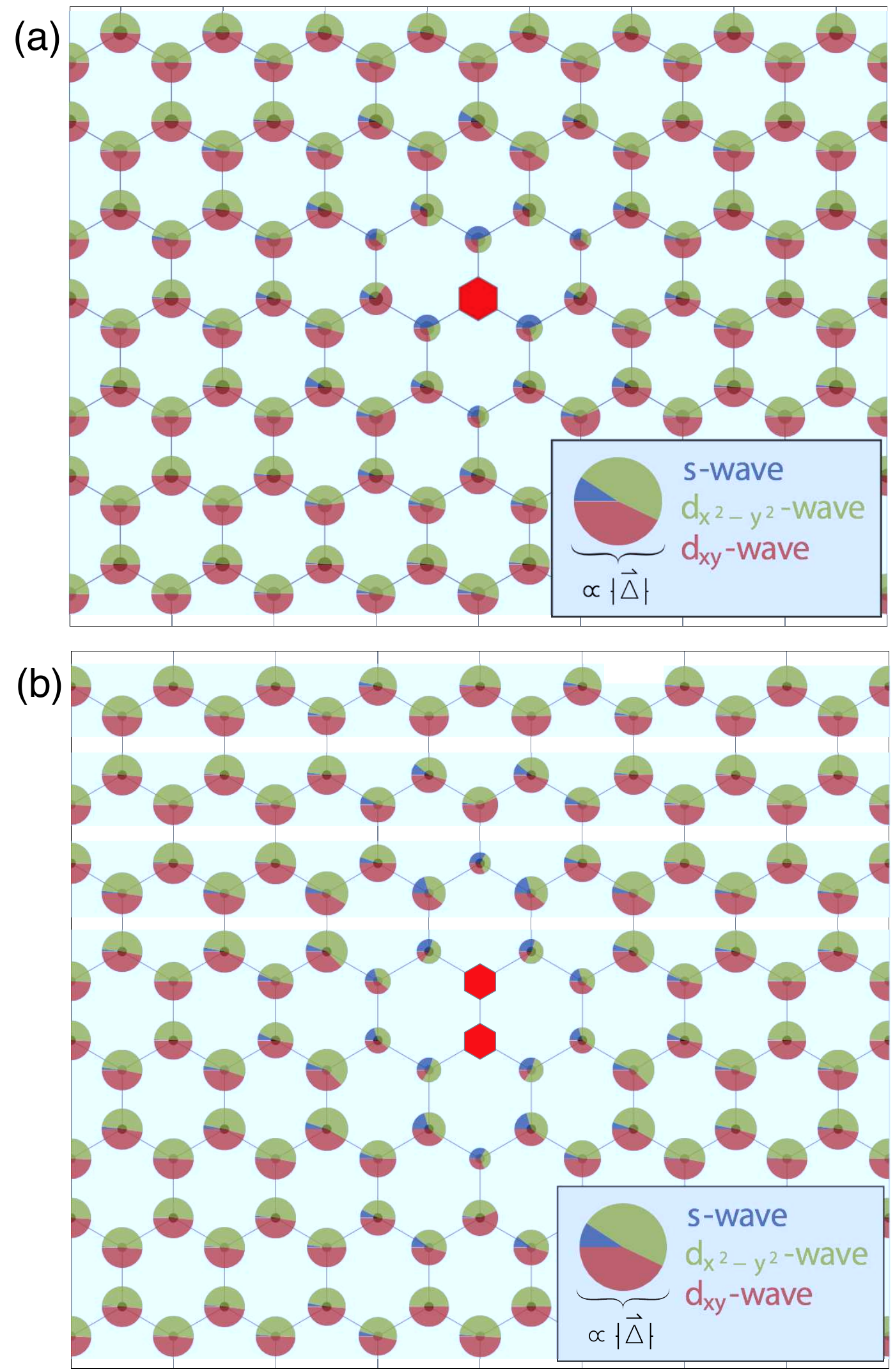}
\caption{\label{fig:vacancies} (Color online). A qualitative view of the order parameter near a single (a) and bivacancy (b) in  heavily doped superconducting graphene. The local wave-character of each site is shown by a pie chart with the radius proportional to the magnitude of the order parameter. The vacancy sites are indicated by red polygons.
(Reprinted figures by permission from T.~L\"{o}thman {\it et al.}, arXiv:1402.3195), \cite{Lothman14}.}
\end{center}
\end{figure}

Also the effect of individual non-magnetic impurities has been studied in chiral $d$-wave superconducting graphene \cite{Lothman14}. Since the $d_{x^2-y^2}+id_{xy}$ symmetry in graphene is dictated by the symmetry of the lattice, it is an interesting question how the order parameter symmetry is influenced in the presence of translation symmetry breaking impurities. For individual point-like vacancies (unitary scattering limit) the chiral $d$-wave symmetry was found to be only  locally perturbed by the appearance of a small extended $s$-wave component, as seen in figure \ref{fig:vacancies}(a). The chiral $d$-wave state was found to heal exponentially fast away from the impurity, with a recovery length of approximately one lattice constance even for a weak superconducting state \cite{Lothman14}. 
%
% FIGURE:
The equivalent recovery length for a conventional $s$-wave state at the same doping level in graphene was found to be 0.4 lattice constants and thus the chiral $d$-wave state should be quite resilient to impurities despite its unconventional and exotic nature. Also bivacancies, which explicitly break the point group symmetry of the lattice, have been investigated \cite{Lothman14}. Despite the symmetry breaking, the results for bivacancies was found to be very similar to those of a single impurity, as can be seen in figure \ref{fig:vacancies}(b).
The local density of states around single impurities have also been studied, showing how even non-magnetic defects induce localized subgap states at finite energy in the fully gapped chiral $d$-wave state \cite{Pellegrino10, Lothman14}.

In the existing proposals for doping graphene close or to the van Hove point \cite{McChesney10,Efetov10}, the deposition of charge-donating adatoms can lead to a significant amount of disorder. Hence the studies that show that the chiral $d$-wave pairing state is only perturbed  locally provide important indications that the $d$-wave state can be experimentally realized. However, if the doping gets large, the dopants may strongly change the electronic structure such that the pairing mechanism gets affected as well, e.g.~by deformations of the Fermi surface. Such effects go beyond the single-impurity studies conducted so far \cite{Pellegrino10, Black-Schaffer12PRL, Lothman14}.

% ENHANCING THE D-WAVE PAIRING
\subsection{Superconducting proximity effects}
While numerical results are promising for the appearance of chiral $d$-wave superconductivity at doping levels approaching the van Hove singularity in graphene, no experimental confirmation exist yet. Even if the chiral $d$-wave state seems to be quite resilient to disorder and impurities as discussed above, heavy doping of graphene still poses significant materials challenges for a superconducting state to be viable. Here we review a few approaches which have been shown to enhance the chiral $d$-wave superconducting state in graphene, with the aim of increasing the prospects of experimental discovery.
The primary candidate for enhancing superconducting correlations in a material is the superconducting proximity effect from an external superconductor. By depositing a superconductor on top of the graphene sheet, proximity-induced superconductivity is induced in the graphene, both directly under the superconductor and in an exponential tail around its edges. For example, a graphene Josephson junction can be created by depositing two superconductors close to each other on a graphene sheet and graphene becomes superconducting even inside the junction. The central idea here is that if the chiral $d$-wave state is too weak to be experimental detected on its own, then graphene-superconductor heterostructures can enhance the chiral $d$-wave superconducting correlations, which in turn give rise to unique experimental signatures.

Experimentally, superconducting proximity effect in graphene was reported using conventional $s$-wave superconductors already rather early \cite{Heersche07,Shailos07,Du08}. 
Unfortunately, there is no direct coupling between an $s$-wave symmetric state and $d$-wave states, since they have different orbital symmetries, with zero averaged overlap \cite{Black-Schaffer07,Black-Schaffer09}. It is, however, still possible to achieve a non-zero Josephson coupling between an external $s$-wave superconductor and the intrinsic $d$-wave state for junctions with disordered interfaces. The Josephson current has in this case been shown to increase with as much as $50\%$  do to the presence of an intrinsic $d$-wave state \cite{Black-Schaffer09}.
Using a high-temperature cuprate superconductor naturally avoids the mismatched orbital symmetries. The superconducting decay length in a cuprate-graphene junction has been shown to have a $1/(T-T_c)$ functional dependence, where $T_c$ is the intrinsic $d$-wave transition temperature in graphene \cite{Black-Schaffer10}. Thus, even if $T_c$ is too small to be directly experimentally detected, its signature is present in the superconducting decay length at much higher temperatures. This also leads to significantly enhanced Josephson currents in cuprate-graphene Josepshon junctions, even far above the intrinsic chiral $d$-wave transition temperature in the graphene \cite{Black-Schaffer10}.

The different orbital momentum between an external $s$-wave superconductor and the intrinsic chiral $d$-wave state in graphene can also be overcome in a doubly quantized vortex in the external $s$-wave superconductor. Here the $2\hbar$ center-of-mass angular momentum of the vortex can be transferred into the orbital angular momentum of the chiral $d$-wave state and the $d$-wave state can thus appear in the vortex core \cite{Volovik88, Salomaa89, Sauls09}. By this process the intrinsic chiral $d$-wave state in graphene has been shown to be significantly strengthened in the core of a doubly quantized $s$-wave superconducting vortex \cite{Black-Schaffer13vortex}. Due to the circular geometry of the vortex, the proximity effect is in this case significantly enhanced compared to linear Josephson junctions. Furthermore, it has also been shown that the interplay between the chiral edge states of the $d$-wave superconducting core and the original vortex core states gives rise to sudden radial changes in spatial profile of the lowest-energy core states as function of temperature, in a temperature range ten times higher than the intrinsic $T_c$ for the $d$-wave state \cite{Black-Schaffer13vortex}. These spatial changes produce favorable experimental conditions for discovering the chiral $d$-wave state using scanning tunnel spectroscopy.

%
% -------------------------------------------------- %
% RELATED SYSTEMS
% -------------------------------------------------- %
\section{Related systems}
\label{sec:relsys}
We have so far almost exclusively focused on the possibility of achieving chiral $d$-wave superconductivity in graphene. The two main ingredients for bulk chiral $d$-wave superconductivity -- a lattice where the two $d$-wave symmetries belong to the same irreducible representation and electron-electron driven superconductivity -- are, however, likely to be present in other materials as well. 
Lattices where the two $d$-wave symmetries belong to the same irreducible representation have a three- or sixfold symmetry axis, as these rotations forces fourfold symmetries to be degenerate. It is thus in general rather straightforward to determine if a material has the required lattice symmetry. 
It is, however, another issue altogether to know if the electron-electron interactions can give rise to superconductivity. If the electron interactions are very strong, the same physics as in the cuprates might be present. Of course, this limit is still highly non-trivial since we do not yet fully understand the underlying mechanism for superconductivity in the high-temperature cuprate superconductors. If interactions are more moderate, and even weak, then the effect of the interactions can still be heavily enhanced if the density of states at the Fermi level is very large. This latter scenario is particularly interesting in the presence of van Hove singularities in the band structure, where the density of states is divergent.
We will in this section review a few other materials which have the necessary lattice symmetry and are, or have been proposed to be, superconducting. We do not attempt to provide a full list, but want to merely illustrate that chiral $d$-wave superconductivity is likely present in many, widely different, materials.

\subsection{Graphene-like systems}
Very reasonably candidates for searching for chiral $d$-wave superconductivity are graphene derivatives, which naturally have a sixfold point group symmetry.

\subsubsection{Bilayer graphene }
Bilayer graphene has received a lot of attention, mainly because of the enhanced density of states at zero doping due to parabolically touching bands. Such band structures have been shown to be unstable towards spontaneous symmetry breaking \cite{Sun09} and experimental results have revealed signatures of interaction-driven excitonic states in suspended and undoped graphene bilayers \cite{Feldman09, Weitz10, Mayorov11, Bao12, Velasco12, vanElferenBi, vanElferenTri, FreitagBi}. 
Very recently weak-coupling renormalization group calculations have shown that superconductivity emerges upon doping away from these excitonic states \cite{Vafek14}. For long-range interactions a fully gapped $f$-wave state or a pair-density-wave state emerges, whereas for a short-range Hubbard interaction the superconducting state either has a chiral $d$-wave or a pair-density-wave symmetry \cite{Murray13}. The pair-density-wave is a paired state at finite momentum, i.e.~a FFLO state, with the two paired electrons belonging to the same Fermi surface (around $K$ or $K'$). The pair-density-wave state is very sensitive to trigonal warping and is destroyed by large further-neighbor hopping in favor of the $f$- or $d$-wave states. The appearance of the chiral $d$-wave state is in agreement with earlier strong-coupling mean-field results on the doped graphene bilayer lattice \cite{Milovanovic12}. Interestingly, the chiral $d$-wave state has a $4\pi$ winding around each Fermi pocket on the graphene bilayer, compared to the $2\pi$ winding for monolayer graphene \cite{Milovanovic12, Murray13}.
A classification of all fully gapped superconducting states in bilayer graphene has also recently been carried out \cite{Roy13}.

\subsubsection{Bilayer silicene }
Going beyond carbon-based honeycomb systems, silicene, the silicon equivalent to graphene, has also been proposed to host chiral $d$-wave superconductivity in bilayer systems \cite{Liu13}. The structure of bilayer silicene is slightly buckled and the band structure has sizable Fermi pockets. When including a realistic Hubbard interaction within the random-phase approximation (RPA) a chiral $d$-wave state has been shown to arise, mediated by antiferromagnetic spin fluctuations. The Fermi pocket area be manipulated by strain in bilayer silicene, thus opening the possibility to easily tune the superconducting transition temperature.

\subsubsection{Intercalated graphites and similar materials }
When discussing superconductivity in graphene-like systems, we should probably also mention that superconductivity has been long established in alkali metal intercalated graphites \cite{Hannay65, Belash87, Weller05} and, more recently, also in the few-layer versions of such systems \cite{Xue12}. The alkali atom intercalated graphites are phonon-driven $s$-wave superconductors with CaC$_6$ reaching the highest transition temperature at 11.6~K. 
The pairing has been associated with the interlayer bands that mainly reside on the intercalated metals and thus this is likely not related to intrinsic superconductivity in graphite \cite{Csanyi05}.
Recent theoretical results have also shown that graphene decorated by both Ca and Li can be phonon-mediated superconductors, but now with LiC$_6$ reaching the highest temperatures \cite{Profeta12}. Because of the mismatch between $s$- and $d$-wave symmetries, there is no coupling between the phonon-driven $s$-wave superconducting state and a potential interaction-driven $d$-wave state \cite{Black-Schaffer07}. 
Beyond the intercalated graphites, experimental superconducting signals have also been reported in graphite \cite{Kopelevich00} and graphite mixed with both sulfur \cite{daSilva01,Moehlecke04} and pure water \cite{Scheike12}. The properties of the superconducting state in these cases are, however, still largely unknown.

% Non-graphene materials:
\subsection{Other hexagonal lattice systems}
Two-dimensional lattices with a sixfold rotational axis perpendicular to the plane can either have the a triangular or a honeycomb structure. The band structure is in both cases very similar for a single orbital system. For the honeycomb lattice the Fermi surface at half-filling is centered around $K$ and $K'$ and only transforms into a $\Gamma$-centered surface for very heavy electron or hole doping. The triangular lattice at half-filling instead has its Fermi surface around $\Gamma$, with heavy electron doping causing a change to two separated Fermi surfaces at $K$ and $K'$. Thus both the honeycomb and the triangular lattices for single orbitals do not only have the required sixfold symmetry but also have a van Hove singularity reachable upon doping.
Similar to the honeycomb lattice, the triangular lattice near half-filling has been shown to host a chiral $d$-wave superconducting state both as a consequence of RVB interactions \cite{LeeFeng, Koretsune02, BaskaranTria, KumarShastry, WangLeeLee} and antiferromagnetic exchange interactions \cite{HonerkampTria}.
Below we list several different known superconductors which all have an (effective) hexagonal lattice and are considered to be candidate chiral $d$-wave superconductors.

\subsubsection{SrPtAs }
The pnictide SrPtAs is a material with honeycomb layers, which has recently been discovered to be superconducting with $T_c = 2.4$~K \cite{Nishikubo11}. Even if this temperature is markedly lower than for the iron-pnictides, its hexagonal structure, with weakly coupled PtAs honeycomb layers alternated by Sr triangular layers, adds the possibility of chiral $d$-wave superconductivity. 
In fact, recent muon spin-rotation ($\mu$SR) experiments have revealed that time-reversal symmetry is broken at $T_c$  \cite{Biswas13} and several experiments have reported a lack of  nodes in the quasiparticle spectrum \cite{Biswas13,Bruckner13}, all consistent with a chiral $d$-wave state.
Due to layer stacking, the point group is reduced to $D_{3d}$ in SrPtAs and the crystal structure also lacks local inversion symmetry, although the unit cell possesses a global inversion center \cite{Nishikubo11, Goryo12}. 
The Fermi surface in SrPtAs consists of both two hole pockets around $\Gamma$ and electron pockets at $K,K'$ \cite{Youn12}. 
The lack of local inversion symmetry together with finite spin-orbit coupling opens the possibility for singlet-triplet mixings. 
Even so, spin-singlet chiral $d$-wave superconductivity has recently been shown to emerge as the leading instability in a fRG calculation, driven mainly by the pockets around $K,K'$ \cite{Fischer14}. However, the weak three-dimensionality was shown to lead to the nodal lines at $K,K'$ crossing a pair of small Fermi surfaces centered around $K,K'$. This results in protected Majorana-Weyl nodes in the bulk and accompanying protected surface states \cite{Fischer14}. Slightly different fRG calculations have on the other hand shown that spin-triplet $f$-wave pairing might instead be the most likely superconducting state thanks to enhanced ferromagnetic fluctuations due to proximity to a van Hove singularity \cite{Wang13}.
The difference in the results in \cite{Fischer14,Wang13} may stem from different choices for the bare interactions and further constraints in the theoretical modeling may be needed in order to arrive at unique theoretical conclusions.

\subsubsection{Na$_x$CoO$_2 \cdot y$H$_2$O }
The water-intercalated cobalt oxide superconductor Na$_x$CoO$_2 \cdot y$H$_2$O consists of thick insulating layers of Na ions and H$_2$O molecules separating superconducting CoO$_2$ layers and has a $T_c$ of about 5~K \cite{Takada03}. This material has many similarities to the high-temperature cuprate superconductors, but also disparities; the Co ions form a triangular lattice, there are multiple bands around the Fermi level, and the system has a filling level far from half-filling. 
Early Knight shift measurements were ambiguous but more recent measurements on single crystals indicate a spin-singlet state \cite{Zheng06}. 
There also exists seemingly conflicting data on the existence of nodes \cite{Fujimoto04, Kanigel04, Oeschler08}. 
Early proposals for the symmetry of the order parameter included, among other orders, both spin-triplet $f$-wave and chiral $d$-wave symmetries. For example, disconnected Fermi surfaces have been suggested to favor spin-triplet $f$-wave superconductivity \cite{Kuroki04}. Considering instead RVB physics and antiferromagnetic exchange interactions the spin-singlet chiral $d$-wave state has been proposed as the leading pairing instability \cite{BaskaranTria, KumarShastry, WangLeeLee, HonerkampTria, Ogata03}.
Most recently, fRG and cluster calculations in a three-orbital model have shown that a chiral $d$-wave state seems to agree best with the current experimental evidence \cite{Kiesel13}. The chiral $d$-wave state was found to arise due to a combined effect of magnetic fluctuations, Fermi surface topology, and the varying orbital characters of the bands.

\subsubsection{$\beta$-$M$NCl }
Another very interesting honeycomb material family is the layered nitrides $\beta$-$M$NCl ($M=$Hf, Zr), which become superconducting with carrier doping by Na or Li intercalation \cite{Yamanaka96, Yamanaka98}. They consist of honeycomb $M$N bilayers alternated with Cl bilayers and exhibit transition temperatures as high as  26~K. Although the pristine materials are band insulators without magnetic ordering, various experimental results, including a weak isotope effect \cite{Tou03, Taguchi07}, have pointed to an unconventional superconducting state. Both NMR \cite{Tou03} and specific heat \cite{Taguchi05,Kasahara09} measurements are consistent with anisotropic spin-singlet pairing with a fully open energy gap, consistent with a chiral $d$-wave state. 
Early band structure calculations pointed to several intriguing similarities between $\beta$-$M$NCl  and the high-temperature cuprate superconductors despite the different lattice structures \cite{Felser99}.
More recently, theoretical results using the fluctuation exchange method (FLEX) on an effective two-band model on the honeycomb lattice have confirmed that chiral $d$-wave superconductivity, mediated by spin-fluctuations, is possible at relative high temperatures in these materials \cite{Kuroki10}. Also variational Monte Carlo studies of an ionic-Hubbard model have shown that a chiral $d$-wave state can emerge \cite{Watanabe13}. 

\subsubsection{$\kappa$-(BEDT-TTF)$_2$X }
The quasi-two-dimensional organic salts $\kappa$-(BEDT-TTF)$_2$$X$, were X is an inorganic monovalent anion, are arranged on an anisotropic triangular lattice, or equivalently a square lattice with isotropic nearest neighbor but one-directional next nearest neighbor hopping \cite{Powell06}. 
Many of these organic salts are either antiferromagnets, Mott insulators, or superconductors \cite{Williams91}, with the phase determined by pressure \cite{Lefebvre00} and frustration. They thus have strong similarities with the high-temperature cuprate superconductors, but with the distinction that they are situated on an anisotropic triangular lattice \cite{McKenzie97}. Experimental results seem ambiguous with regards to if the superconducting state is fully gapped or contains nodes, see e.g.~\cite{Powell04, Powell06} and references therein.
The interplay between antiferromagnetism, the Mott transition, and $d$-wave superconductivity has been studied by different theoretical methods \cite{Kyung06, Sahebsara06, Grover10}. Also chiral $d$-wave superconductivity has been proposed for lattice structures close to the isotropic triangular lattice, when the nearest and next nearest neighbor hopping amplitudes are of similar size \cite{Ogata03}.

\subsubsection{MoS$_2$ }
The transition metal dichalcogenides are layered semiconductors which enable easy exfoliation to two-dimensional layers \cite{Wang12}. Recently thin flakes of MoS$_2$ was found to become superconducting upon doping, with superconductivity forming a dome structure as function of doping \cite{Taniguchi12, Ye12}.
In monolayer MoS$_2$ two layers of S atoms sit in a hexagonal lattice stacked in an eclipsed fashion, whereas the Mo atoms sit in-between in cages formed by six S atoms. Seen from above this forms a honeycomb lattice with Mo and S at the two sites, see e.g.~\cite{Helveg00, Cao12}. The Fermi surface is disconnected and centered around the $K,K'$ points.
A recent theoretical study has explored both electron- and phonon-driven superconductivity and suggested a spin-triplet $f$-wave state, where the order parameter has different signs on the two Fermi surfaces, due to strong short-range repulsion \cite{Roldan13}. Another very recent study \cite{Law14} has proposed a spin-singlet $p+ip$-state on each Fermi surface, which is in agreement with the chiral $d$-wave state projected onto the Fermi surfaces \cite{Linder09}.

\subsubsection{In$_3$Cu$_2$VO$_9$ }
Yet another material with an effective honeycomb structure which has recently drawn attention as a possible host for chiral $d$-wave superconducting state is In$_3$Cu$_2$VO$_9$. Here a singly occupied Cu orbitals forms a spin $S = 1/2$ honeycomb lattice. Experimentally, the ground state has been identified to likely be a N\'{e}el antiferromagnet; no three-dimensional ordering has been found, but strong two-dimensional magnetic correlations are present even down to low temperatures \cite{Moller08, Yan12}. Theory results for the strong-coupling limit $t$-($t'$)-$J$ model have proposed that a chiral $d$-wave state will likely emerge upon doping this magnetic ground state. The methods used include both renormalized and slave-boson mean-field theory \cite{Wu13}, as well as a recently developed variational approach using Grassman tensor product states \cite{Gu13}.

\subsubsection{(111) Bilayer SrIrO$_3$ }
Transition-metal oxides involving $4d$ or $5d$ electrons offers a very fertile ground to study electron correlations. For example, the iridates Li$_2$IrO$_3$ and Na$_2$IrO$_3$, where the Ir atoms form a honeycomb lattice, have been shown to be described by a Kitaev model \cite{Kitaev06} with an admixture of Heisenberg coupling \cite{Jackeli09, Chaloupka10,Singh12,Witczak14,Reuther14}. For dominating Kitaev coupling the ground state is a spin liquid, which has been shown to support a time-reversal invariant $p$-wave superconducting state at finite doping \cite{You12, Hyart12,Scherer14}. But if the Heisenberg term is allowed to dominate, the spin-singlet chiral $d$-wave state would instead be expected to appear with doping. 
Recent advances has also opened up the possibility of artificial transition-metal oxide heterostructures \cite{Zubko11, Hwang12}. For (111) bilayers of perovskite $AB$O$_3$ transition-metal oxides, the transition-metal ions form a buckled honeycomb lattice, which has been proposed as a platform to explore a multitude of quantum effects \cite{Xiao11,Yang11bilayer, Ruegg11}. Most interestingly, 
the antiferromagnetic Heisenberg interaction has been found to dominate over the Kitaev interaction in (111) bilayers of SrIrO$_3$ \cite{Okamoto13}. Due to the strong electron interactions such bilayers can thus effectively be described by a $t$-$J$ model. Carrier doping, by for example partially substituting Ir by Ru or Os (hole doping) or Sr by La (electron doping), has recently been theoretically shown to stabilize a chiral $d$-wave superconducting state \cite{Okamoto13,Okamoto13b}.

\subsubsection{MgB$_2$ and similar materials }
We also mention here that, in terms of superconductors, MgB$_2$ is perhaps the most well-known superconducting honeycomb layered system with $T_c = 39$~K \cite{Nagamatsu01}. 
The B atoms in MgB$_2$ form honeycomb layers separated by triangular Mg layers. Due to a large charge transfer, the Fermi surface is located inside the bonding $\sigma$-band of the honeycomb lattice \cite{Kortus01}. 
Despite the high transition temperature, superconductivity has been established to be phonon-driven and have an $s$-wave symmetry, however, with an unusual two-gap structure \cite{Liu01,Choi02}.
Closely related to MgB$_2$ are the ternary silicides $M$AlSi ($M=$ alkaline-earth atoms), which are also superconducting, albeit at lower temperatures. They have a lattice structure where $M$ occupy the Mg sites and Al and Si are randomly distributed on the B sites \cite{Imai02}.

% These are as far as I can tell only models with no found superconducting state so I have not included them (needed to draw the line somewhere):
%Kagome lattice at 1/6 hole doping: PRB 85, 144402 (2012): Model with chiral d-wave SC
% Kitaev-Heisenberg model: chiral d-wave SC Hyart et al.

% -------------------------------------------------- %
% SUMMARY
% -------------------------------------------------- %
\section{Summary and perspective}
\label{sec:summary}
% Summary existence:
In this review we have strived to summarize the growing body of theoretical work which shows that a spin-singlet $d_{x^2-y^2}\pm id_{xy}$-wave, or chiral $d$-wave, superconducting state can be achieved in doped graphene. Methods ranging from mean-field theory of an effective Hamiltonian capturing effective resonance valence bond correlations \cite{Black-Schaffer07}, to weak-coupling \cite{Nandkishore11} and functional renormalization group \cite{Honerkamp08, Wang11, Kiesel12} calculations have all demonstrated a chiral $d$-wave superconducting state appearing in doped graphene. Doping levels reaching the van Hove singularity at $1/4$ electron or hole doping have been shown to be especially promising for this unconventional superconducting state to be the ground state.

% Summary properties:
We have also reviewed the properties of the chiral $d$-wave superconducting state in doped graphene. This state breaks both time-reversal and parity symmetries and is fully gapped for finite doping levels. Further, the quasiparticle bulk band structure has a non-trivial topology represented by a Chern or Skyrmion number $\mathcal = \pm 2$ \cite{Volovik89,Volovikbook92,Volovik97}. This directly gives rise to two co-propagating, i.e.~chiral, edge states crossing the bulk energy gap and carrying a spontaneous current \cite{Volovik97,Black-Schaffer12PRL}, as well as generates quantized spin and thermal Hall conductances \cite{Senthil99}. Moreover, the two chiral states $d_{x^2-y^2}+ id_{xy}$ and $d_{x^2-y^2}- id_{xy}$ are in general degenerate and thus domain walls, with four chiral modes crossing the bulk band gap  \cite{Volovik97}, will likely be present in the superconducting state. 

% THUS:
Theoretically, there is thus little doubt that graphene is a very promising place to look for chiral $d$-wave superconductivity. In particular this holds true for doping reaching the van Hove singularity.
Of course, even if intrinsic superconductivity is found in heavily doped graphene, it could still arise due to phonon-mediated interactions and be of conventional nature. 
Conveniently, the very specific and often exotic properties of chiral superconductors provides a number of particular observations, which should allow for making an unambiguous identification of a chiral pairing phase.

However, what is less certain, due to an uncertainly in the model parameters, is whether chiral superconducting pairing will occur at any reasonable temperatures. When one uses the currently best available input for the theoretical models, the pairing scales can theoretically reach a few Kelvins, but only in a parameter window of limited extent. Hence no strong predictions can be made. 
Interestingly, some work has shown that the chiral $d$-wave state in graphene can be enhanced by proximity effect to external superconductors \cite{Black-Schaffer09,Black-Schaffer13vortex}. This might thus offer a promising alternative approach to an experimental discovery of chiral $d$-wave superconductivity in graphene.

A possibly even more severe obstacle for experimental realization is posed by the yet largely unknown deviations of real graphene from the simple theoretical models currently used. A standard problem for unconventional superconductivity is impurities, which, of course, are also present in graphene. We have described a few promising theoretical studies on the impact of impurities on chiral pairing on the honeycomb lattice \cite{Black-Schaffer12PRL, Lothman14}, but is has to be awaited if these expectations are fulfilled by an experimental system. 
Moreover, in most theoretical studies, the doping necessary to achieve superconductivity is assumed to be homogeneous and treated in a rigid band picture. This may not be the most realistic assumption, as the dopants may either cause significant disorder or additional band structure features when they are not disordered. This might lead to strong deviations from the results found so far. Some promising experimental work on achieving high doping levels in graphene exist, see e.g.~\cite{McChesney10, Efetov10}, but further theoretical and experimental work is clearly needed to understand these issues more systematically.

There exist thus several experimental challenges for realizing chiral $d$-wave superconductivity in heavily doped graphene. Nevertheless, the significant amount of existing theoretical work shows that this can be a reachable goal and it offers the exciting prospect of realizing the first spin-singlet $d$-wave chiral superconductor, and that in a widely known and easy accessible material.
We have in this review also provided a partial list of known superconductors which are actively proposed to be candidates for hosting a spin-singlet chiral $d$-wave superconducting state. The physics behind superconductivity in these materials is in many cases similar to that proposed for graphene. Studying chiral $d$-wave superconductivity in graphene thus offers a template for chiral $d$-wave superconductivity in many different materials, of which undoubtedly some will be experimentally confirmed to be chiral $d$-wave superconductors in the future. 

%
% -------------------------------------------------- %
% ACKNOWLEDGEMENTS
% -------------------------------------------------- %
\ack
We would like to thank  A.~V.~Balatsky, L.~Boeri, M.~Fogelstr\"om, K.~Le Hur, J.~Linder, T.~L\"othman, M.~Scherer, A.~P.~Schnyder, R.~Thomale, A.-M.~S.~Tremblay, and W.~Wu for valuable discussions. This work was supported by the Swedish Research Council (VR) and the German Research Foundation (DFG).

For figures with copyright permission from the American Physical Society: Readers may view, browse, and/or download material for temporary copying purposes only, provided these uses are for noncommercial personal purposes. Except as provided by law, this material may not be further reproduced, distributed, transmitted, modified, adapted, performed, displayed, published, or sold in whole or part, without prior written permission from the American Physical Society.

\section*{References}
%\bibliographystyle{myunsrt}
%\bibliography{graphene}

\begin{thebibliography}{100}

\bibitem{Novoselov04}
K.~S. Novoselov, A.~K. Geim, S.~V. Morozov, D.~Jiang, Y.~Zhang, S.~V. Dubonos,
  I.~V. Grigorieva, and A.~A. Firsov.
\newblock {\em Science}, 306:666, 2004.

\bibitem{CastroNeto09}
A.~H. Castro~Neto, F.~Guinea, N.~M.~R. Peres, K.~S. Novoselov, and A.~K. Geim.
\newblock {\em Rev.\ Mod.\ Phys.}, 81:109, 2009.

\bibitem{KotovRMP12}
V.~N. Kotov, B.~Uchoa, V.~M. Pereira, F.~Guinea, and A.~H. Castro~Neto.
\newblock {\em Rev. Mod. Phys.}, 84:1067--1125, 2012.

\bibitem{Tchougreeff92}
A.~L. Tchougreeff and R.~Hoffmann.
\newblock {\em J.\ Phys.\ Chem.}, 96:8993, 1992.

\bibitem{Sorella92}
S.~{Sorella} and E.~{Tosatti}.
\newblock {\em EPL}, 19:699, 1992.

\bibitem{Khevshchenko01}
D.~V. Khveshchenko.
\newblock {\em Phys. Rev. Lett.}, 87:246802, 2001.

\bibitem{Herbut06}
I.~F. Herbut.
\newblock {\em Phys. Rev. Lett.}, 97:146401, 2006.

\bibitem{Hou_Chamon07}
C.-Y. Hou, C.~Chamon, and Ch. Mudry.
\newblock {\em Phys. Rev. Lett.}, 98:186809, 2007.

\bibitem{Honerkamp08}
C.~Honerkamp.
\newblock {\em Phys.\ Rev.\ Lett.}, 100:146404, 2008.

\bibitem{Raghu08}
S.~{Raghu}, X.-L. {Qi}, C.~{Honerkamp}, and S.-C. {Zhang}.
\newblock {\em Phys. Rev. Lett.}, 100:156401, 2008.

\bibitem{Liu_Li09}
G.-Z. Liu, W.~Li, and G.~Cheng.
\newblock {\em Phys. Rev. B}, 79:205429, 2009.

\bibitem{Drut09b}
J.~E. Drut and T.~A. L\"{a}hde.
\newblock {\em Phys.\ Rev.\ Lett.}, 102:026802, 2009.

\bibitem{Herbut09}
I.~F. Herbut, V.~Juri\v{c}i\'{c}, and B.~Roy.
\newblock {\em Phys.\ Rev.\ B}, 79:085116, 2009.

\bibitem{Gamayun10}
O.~V. Gamayun, E.~V. Gorbar, and V.~P. Gusynin.
\newblock {\em Phys. Rev. B}, 81:075429, 2010.

\bibitem{Meng10}
Z.~Y. Meng, T.~C. Lang, S.~Wessel, F.~F. Assaad, and A.~Muramatsu.
\newblock {\em Nature}, 464:847, 2010.

\bibitem{Ulybyshev13}
M.~V. Ulybyshev, P.~V. Buividovich, M.~I. Katsnelson, and M.~I. Polikarpov.
\newblock {\em Phys. Rev. Lett.}, 111:056801, 2013.

\bibitem{Sorella12}
S.~{Sorella}, Y.~{Otsuka}, and S.~{Yunoki}.
\newblock {\em Sci. Rep.}, 2:992, 2012.

\bibitem{Feldman09}
B.~E. Feldman, J.~Martin, and A.~Yacoby.
\newblock {\em Nature Phys.}, 5:889--893, 2009.

\bibitem{Weitz10}
R.~T. Weitz, M.~T. Allen, B.~E. Feldman, J.~Martin, and A.~Yacoby.
\newblock {\em Science}, 330:812--816, 2010.

\bibitem{Mayorov11}
A.~S. Mayorov, D.~C. Elias, M.~Mucha-Kruczynski, R.~V. Gorbachev,
  T.~Tudorovskiy, A.~Zhukov, S.~V. Morozov, M.~I. Katsnelson, V.~I. FalÕko,
  A.~K. Geim, and K.~S. Novoselov.
\newblock {\em Science}, 333:860--863, 2011.

\bibitem{Bao12}
W.~Bao, J.~Velasco, F.~Zhang, L.~Jing, B.~Standley, D.~Smirnov, M.~Bockrath,
  A.~H. MacDonald, and C.~N. Lau.
\newblock {\em Proc. Natl. Acad. Sci. U.S.A.}, 109:10802--10805, 2012.

\bibitem{Velasco12}
J.~Velasco, W.~Jing, Y.~Lee, P.~Kratz, V.~Aji, M.~Bockrath, C.~N. Lau,
  C.~Varma, R.~Stillwell, D.~Smirnov, F.~Zhang, J.~Jung, and A.~H. MacDonald.
\newblock {\em Nature Nanotech.}, 7:156--160, 2012.

\bibitem{vanElferenBi}
A.~Veligura, H.~J. van Elferen, N.~Tombros, J.~C. Maan, U.~Zeitler, and B.~J.
  van Wees.
\newblock {\em Phys. Rev. B}, 85:155412, 2012.

\bibitem{vanElferenTri}
H.~J. van Elferen, A.~Veligura, N.~Tombros, E.~V. Kurganova, B.~J. van Wees,
  J.~C. Maan, and U.~Zeitler.
\newblock {\em Phys. Rev. B}, 88:121302, 2013.

\bibitem{FreitagBi}
F.~Freitag, M.~Weiss, R.~Maurand, J.~Trbovic, and C.~Sch\"onenberger.
\newblock {\em Phys. Rev. B}, 87:161402, 2013.

\bibitem{Xue12}
M.~Xue, G.~Chen, H~Yang, Y~Zhu, D.~Wang, J.~He, and T.~Cao.
\newblock {\em J. Am. Chem. Soc.}, 134:6536--6539, 2012.

\bibitem{Profeta12}
G.~Profeta, M.~Calandra, and F.~Mauri.
\newblock {\em Nature Phys.}, 8:131--134, 2012.

\bibitem{Black-Schaffer12PRL}
A.~M. Black-Schaffer.
\newblock {\em Phys. Rev. Lett.}, 109:197001, 2012.

\bibitem{Nayak08}
C.~Nayak, S.~H. Simon, A.~Stern, M.~Freedman, and S.~Das~Sarma.
\newblock {\em Rev. Mod. Phys.}, 80:1083--1159, 2008.

\bibitem{Serban10}
I.~Serban, B.~B\'eri, A.~R. Akhmerov, and C.~W.~J. Beenakker.
\newblock {\em Phys. Rev. Lett.}, 104:147001, 2010.

\bibitem{KallinChiral}
C.~{Kallin}.
\newblock {\em Rep. Prog. Phys.}, 75:042501, 2012.

\bibitem{MacKenzieSruo}
A.~P. {MacKenzie} and Y.~{Maeno}.
\newblock {\em Rev. Mod. Phys.}, 75:657--712, 2003.

\bibitem{VollhardtWoelfle}
D.~{Vollhardt} and P.~{W\"{o}lfle}.
\newblock {\em The superfluid phases of Helium-3}.
\newblock Taylor \& Francis, 1990.

\bibitem{MooreRead}
G.~{Moore} and N.~{Read}.
\newblock {\em Nucl. Phys. B}, 360:362--396, 1991.

\bibitem{Sigrist91}
M.~Sigrist and K.~Ueda.
\newblock {\em Rev. Mod. Phys.}, 63:239--311, 1991.

\bibitem{Raghu10}
S.~Raghu, S.~A. Kivelson, and D.~J. Scalapino.
\newblock {\em Phys. Rev. B}, 81:224505, 2010.

\bibitem{Kohn65}
W.~Kohn and J.~M. Luttinger.
\newblock {\em Phys. Rev. Lett.}, 15:524--526, 1965.

\bibitem{Nandkishore14}
R.~Nandkishore, R.~Thomale, and A.~V. Chubukov.
\newblock {\em Phys. Rev. B}, 89:144501, 2014.

\bibitem{Black-Schaffer07}
A.~M. Black-Schaffer and S.~Doniach.
\newblock {\em Phys.\ Rev.\ B}, 75:134512, 2007.

\bibitem{Wu13}
W.~Wu, M.~M. Scherer, C.~Honerkamp, and K.~Le~Hur.
\newblock {\em Phys. Rev. B}, 87:094521, 2013.

\bibitem{Pathak10}
S.~Pathak, V.~B. Shenoy, and G.~Baskaran.
\newblock {\em Phys. Rev. B}, 81:085431, 2010.

\bibitem{Ma11}
T.~Ma, Z.~Huang, F.~Hu, and H.-Q. Lin.
\newblock {\em Phys. Rev. B}, 84:121410, 2011.

\bibitem{Nandkishore11}
R.~Nandkishore, L.~S. Levitov, and A.~V. Chubukov.
\newblock {\em Nature Phys.}, 8:158, 2012.

\bibitem{Wang11}
W.-S. Wang, Y.-Y. Xiang, Q.-H. Wang, F.~Wang, F.~Yang, and D.-H. Lee.
\newblock {\em Phys. Rev. B}, 85:035414, 2012.

\bibitem{Kiesel12}
M.~L. Kiesel, C.~Platt, W.~Hanke, D.~A. Abanin, and R.~Thomale.
\newblock {\em Phys. Rev. B}, 86:020507, 2012.

\bibitem{Gonzalez08}
J.~Gonz\'alez.
\newblock {\em Phys. Rev. B}, 78:205431, 2008.

\bibitem{Fogelstrom97}
M.~Fogelstr\"om, D.~Rainer, and J.~A. Sauls.
\newblock {\em Phys. Rev. Lett.}, 79:281--284, 1997.

\bibitem{Covington97}
M.~Covington, M.~Aprili, E.~Paraoanu, L.~H. Greene, F.~Xu, J.~Zhu, and C.~A.
  Mirkin.
\newblock {\em Phys. Rev. Lett.}, 79:277--280, 1997.

\bibitem{Balatsky97}
A.~V. Balatsky.
\newblock {\em Phys. Rev. Lett.}, 80:1972--1975, 1998.

\bibitem{Krishana97}
K.~Krishana, N.~P. Ong, Q.~Li, G.~D. Gu, and N.~Koshizuka.
\newblock {\em Science}, 277:83--85, 1997.

\bibitem{Elhalel07}
G.~Elhalel, R.~Beck, G.~Leibovitch, and G.~Deutscher.
\newblock {\em Phys. Rev. Lett.}, 98:137002, 2007.

\bibitem{Maeno03}
A.~P. Mackenzie and Y.~Maeno.
\newblock {\em Rev. Mod. Phys.}, 75:657--712, 2003.

\bibitem{Kallin12}
C.~Kallin.
\newblock {\em Rep. Prog. Phys.}, 75:042501, 2012.

\bibitem{Fulde64}
P.~Fulde and R.~A. Ferrell.
\newblock {\em Phys. Rev.}, 135:A550--A563, 1964.

\bibitem{Larkin64}
A.~I. Larkin and Y.~N. Ovchinnikov.
\newblock {\em Zh. Eksp. Teor. Fiz}, 47:1136, 1964.

\bibitem{Roy10}
B.~Roy and I.~F. Herbut.
\newblock {\em Phys. Rev. B}, 82:035429, 2010.

\bibitem{Murray13}
J.~M. Murray and O.~Vafek.
\newblock {\em Phys. Rev. B}, 89:205119, 2014.

\bibitem{Shibayama00}
Y.~Shibayama, H.~Sato, T.~Enoki, and M.~Endo.
\newblock {\em Phys. Rev. Lett.}, 84:1744--1747, 2000.

\bibitem{Enoki09}
T.~Enoki and K.~Takai.
\newblock {\em Solid State Commun.}, 149:1144 -- 1150, 2009.

\bibitem{Tao11}
C.~Tao, L.~Jiao, O.~V. Yazyev, Y.-C. Chen, J.~Feng, X.~Zhang, R.~B. Capaz,
  J.~M. Tour, A.~Zettl, S.~G. Louie, H.~Dai, and M.~F. Crommie.
\newblock {\em Nature Phys.}, 7:616--620, 2011.

\bibitem{Esquinazi03}
P.~Esquinazi, D.~Spemann, R.~H\"ohne, A.~Setzer, K.-H. Han, and T.~Butz.
\newblock {\em Phys. Rev. Lett.}, 91:227201, 2003.

\bibitem{Cervenka09}
J.~Cervenka, M.~I. Katsnelson, and C.~F.~J. Flipse.
\newblock {\em Nature Phys.}, 5:840, 2009.

\bibitem{Sepioni10}
M.~Sepioni, R.~R. Nair, S.~Rablen, J.~Narayanan, F.~Tuna, R.~Winpenny, A.~K.
  Geim, and I.~V. Grigorieva.
\newblock {\em Phys. Rev. Lett.}, 105:207205, 2010.

\bibitem{Elias11}
D.~C. Elias, R.~V. Gorbachev, S.~V. Mayorov, A. S. and-Morozov, A.~A. Zhukov,
  P.~Blake, L.~A. Ponomarenko, I.~V. Grigorieva, K.~S. Novoselov, F.~Guinea,
  and A.~K. Geim.
\newblock {\em Nature Phys.}, 7:701, 2011.

\bibitem{Wehling11U}
T.~O. Wehling, E.~\ifmmode \mbox{\c{S}}\else \c{S}\fi{}a\ifmmode
  \mbox{\c{s}}\else \c{s}\fi{}\ifmmode \imath \else \i
  \fi{}o\ifmmode~\breve{g}\else \u{g}\fi{}lu, C.~Friedrich, A.~I. Lichtenstein,
  M.~I. Katsnelson, and S.~Bl\"ugel.
\newblock {\em Phys. Rev. Lett.}, 106:236805, 2011.

\bibitem{Drut09}
J.~E. Drut and T.~A. L\"{a}hde.
\newblock {\em Phys.\ Rev.\ B}, 79:165425, 2009.

\bibitem{Hirsch85}
J.~E. Hirsch.
\newblock {\em Phys. Rev. Lett.}, 54:1317--1320, 1985.

\bibitem{GrosJoyntRice87}
C.~Gros, R.~Joynt, and T.~M. Rice.
\newblock {\em Phys. Rev. B}, 36:381--393, 1987.

\bibitem{ZhangRice88}
F.~C. Zhang and T.~M. Rice.
\newblock {\em Phys. Rev. B}, 37:3759--3761, 1988.

\bibitem{Choy96}
T.~C. Choy and B.~A. McKinnon.
\newblock {\em Phys. Rev. B}, 52:14539--14543, 1995.

\bibitem{Zhang88}
F.~C. Zhang, C.~Gros, T.~M. Rice, and H.~Shiba.
\newblock {\em Supercond. Sci. Tech.}, 1:36, 1988.

\bibitem{Lederer89}
P.~Lederer, D.~Poilblanc, and T.~M. Rice.
\newblock {\em Phys. Rev. Lett.}, 63:1519--1522, 1989.

\bibitem{Anderson04}
P.~W. Anderson, P.~A. Lee, M.~Randeria, T.~M. Rice, N.~Trivedi, and F.~C.
  Zhang.
\newblock {\em J. Phys.: Condens. Matter}, 16:R755, 2004.

\bibitem{Edegger07}
B.~Edegger, V.~N. Muthukumar, and C.~Gros.
\newblock {\em Advances in Physics}, 56:927--1033, 2007.

\bibitem{Ogata08}
M.~Ogata and H.~Fukuyama.
\newblock {\em Rep. Prog. Phys.}, 71:036501, 2008.

\bibitem{Vollhardt84}
D.~Vollhardt.
\newblock {\em Rev. Mod. Phys.}, 56:99--120, 1984.

\bibitem{Anderson73}
P.~W. Anderson.
\newblock {\em Mater. Res. Bull.}, 8:153 -- 160, 1973.

\bibitem{Anderson87}
P.~W. Anderson.
\newblock {\em Science}, 235:1196, 1987.

\bibitem{Paulingbook}
L.~Pauling.
\newblock {\em Nature of the Chemical Bond}.
\newblock Cornell University Press, New York, 1960.

\bibitem{CastroNeto09view}
A.~H. Castro~Neto.
\newblock {\em Physics}, 2:30, 2009.

\bibitem{Baskaran02}
G.~Baskaran.
\newblock {\em Phys.\ Rev.\ B}, 65:212505, 2002.

\bibitem{Uchoa07}
B.~Uchoa and A.~H. Castro~Neto.
\newblock {\em Phys.\ Rev.\ Lett.}, 98:146801, 2007.

\bibitem{Linder09}
J.~Linder, A.~M. Black-Schaffer, T.~Yokoyama, S.~Doniach, and A.~Sudb\o{}.
\newblock {\em Phys. Rev. B}, 80:094522, 2009.

\bibitem{Jiang08}
Y.~Jiang, D.-X. Yao, E.~W. Carlson, H.-D. Chen, and J.~P. Hu.
\newblock {\em Phys. Rev. B}, 77:235420, 2008.

\bibitem{Lothman14}
T.~L{\"o}thman and A.~M. {Black-Schaffer}.
\newblock {\em ArXiv:1402.3195 (unpublished)}, 2014.

\bibitem{Kuznetsova05}
V.~Kuznetsova and B.~Barzykin.
\newblock {\em Europhys.\ Lett.}, 72:437, 2005.

\bibitem{Tran11}
M.-T. Tran and K.-S. Kim.
\newblock {\em Phys. Rev. B}, 83:125416, 2011.

\bibitem{Black-Schaffer14tJ}
A.~M. {Black-Schaffer}, W.~Wu, and K.~{Le Hur}.
\newblock {\em Phys. Rev. B}, 90:054521, 2014.

\bibitem{Black-Schaffer12}
A.~M. Black-Schaffer.
\newblock {\em Phys. Rev. Lett.}, 109:197001, 2012.

\bibitem{Gu13}
Z.-C. Gu, H.-C. Jiang, D.~N. Sheng, H.~Yao, L.~Balents, and X.-G. Wen.
\newblock {\em Phys. Rev. B}, 88:155112, 2013.

\bibitem{JiangMesaros14}
S.~{Jiang}, A.~{Mesaros}, and Y.~{Ran}.
\newblock {\em Phys. Rev. X}, 4:031040, 2014.

\bibitem{Scalapino}
D.~J. Scalapino.
\newblock {\em J. Low Temp. Phys.}, 117:179--188, 1999.

\bibitem{Onari}
S.~Onari, R.~Arita, K.~Kuroki, and H.~Aoki.
\newblock {\em Phys. Rev. B}, 70:094523, 2004.

\bibitem{Metzner12}
W.~Metzner, M.~Salmhofer, C.~Honerkamp, V.~Meden, and K.~Sch\"onhammer.
\newblock {\em Rev. Mod. Phys.}, 84:299--352, 2012.

\bibitem{Platt13}
C.~Platt, W.~Hanke, and R.~Thomale.
\newblock {\em Adv. Phys.}, 62:453--562, 2013.

\bibitem{Solyom79}
J.~Solyom.
\newblock {\em Adv. Phys.}, 28:201--303, 1979.

\bibitem{Zanchi98}
D.~{Zanchi} and H.~J. {Schulz}.
\newblock {\em EPL}, 44:235--241, 1998.

\bibitem{Halboth00}
C.~J. {Halboth} and W.~{Metzner}.
\newblock {\em Phys. Rev. B}, 61:7364--7377, 2000.

\bibitem{Honerkamp01}
C.~{Honerkamp}, M.~{Salmhofer}, N.~{Furukawa}, and T.~M. {Rice}.
\newblock {\em Phys. Rev. B}, 63:035109, 2001.

\bibitem{Husemann09}
C.~Husemann and M.~Salmhofer.
\newblock {\em Phys. Rev. B}, 79:195125, 2009.

\bibitem{Maier13}
S.~A. Maier and C.~Honerkamp.
\newblock {\em Phys. Rev. B}, 86:134404, 2012.

\bibitem{Honerkamp01b}
C.~Honerkamp and M.~Salmhofer.
\newblock {\em Phys. Rev. B}, 64:184516, 2001.

\bibitem{Nandkishore12}
R.~{Nandkishore} and A.~V. {Chubukov}.
\newblock {\em Phys. Rev. B}, 86:115426, 2012.

\bibitem{Nandkishore12b}
R.~Nandkishore, G.-W. Chern, and A.~V. Chubukov.
\newblock {\em Phys. Rev. Lett.}, 108:227204, 2012.

\bibitem{Katanin04}
A.~A. Katanin and A.~P. Kampf.
\newblock {\em Phys. Rev. Lett.}, 93:106406, 2004.

\bibitem{Feldman08}
J.~Feldman and M.~Salmhofer.
\newblock {\em Rev. Math. Phys.}, 20:275--334, 2008.

\bibitem{Scherer12a}
M.~M. Scherer, S.~Uebelacker, and C.~Honerkamp.
\newblock {\em Phys. Rev. B}, 85:235408, 2012.

\bibitem{Scherer12b}
M.~M. Scherer, S.~Uebelacker, D.~D. Scherer, and C.~Honerkamp.
\newblock {\em Phys. Rev. B}, 86:155415, 2012.

\bibitem{Laughlin98}
R.~B. Laughlin.
\newblock {\em Phys. Rev. Lett.}, 80:5188--5191, 1998.

\bibitem{Balatsky98}
A.~V. Balatsky.
\newblock {\em Phys. Rev. Lett.}, 80:1972--1975, 1998.

\bibitem{Gustafsson13}
D.~Gustafsson, D.~Golubev, M.~Fogelstr\"{o}m, T.~Claeson, S.~Kubatkin,
  T.~Bauch, and F.~Lombardi.
\newblock {\em Nat. Nanotechnol.}, 8:25, 2013.

\bibitem{AltlandZirnbauer97}
A.~Altland and M.~R. Zirnbauer.
\newblock {\em Phys. Rev. B}, 55:1142--1161, 1997.

\bibitem{Schnyder08}
A.~P. Schnyder, S.~Ryu, A.~Furusaki, and A.~W.~W. Ludwig.
\newblock {\em Phys. Rev. B}, 78:195125, 2008.

\bibitem{Sato10}
M.~Sato, Y.~Takahashi, and S.~Fujimoto.
\newblock {\em Phys. Rev. B}, 82:134521, 2010.

\bibitem{TKNN}
D.~J. Thouless, M.~Kohmoto, M.~P. Nightingale, and M.~den Nijs.
\newblock {\em Phys. Rev. Lett.}, 49:405--408, 1982.

\bibitem{Volovik89}
G.~E. Volovik and V.~M. Yakovenko.
\newblock {\em J. Phys.: Condens. Matter}, 1:5263, 1989.

\bibitem{Volovikbook92}
G.~E. Volovik.
\newblock {\em Exotic Properties of Superfluid $^3$He}.
\newblock World Scientific, Singapore, 1992.

\bibitem{Volovik97}
G.~E. Volovik.
\newblock {\em JETP Lett.}, 66:522--527, 1997.

\bibitem{Read00}
N.~Read and D.~Green.
\newblock {\em Phys. Rev. B}, 61:10267--10297, 2000.

\bibitem{Raghu10bands}
S.~Raghu, A.~Kapitulnik, and S.~A. Kivelson.
\newblock {\em Phys. Rev. Lett.}, 105:136401, 2010.

\bibitem{Senthil99}
T.~Senthil, J.~B. Marston, and M.~P.~A. Fisher.
\newblock {\em Phys. Rev. B}, 60:4245--4254, 1999.

\bibitem{Horovitz03}
B.~Horovitz and A.~Golub.
\newblock {\em Phys. Rev. B}, 68:214503, 2003.

\bibitem{Hasan10}
M.~Z. Hasan and C.~L. Kane.
\newblock {\em Rev. Mod. Phys.}, 82:3045--3067, 2010.

\bibitem{Jackiw76}
R.~Jackiw and C.~Rebbi.
\newblock {\em Phys. Rev. D}, 13:3398--3409, 1976.

\bibitem{Halperin82}
B.~I. Halperin.
\newblock {\em Phys. Rev. B}, 25:2185--2190, 1982.

\bibitem{Qi11}
X.-L. Qi and S.-C. Zhang.
\newblock {\em Rev. Mod. Phys.}, 83:1057--1110, 2011.

\bibitem{Hu94}
C.-R. Hu.
\newblock {\em Phys. Rev. Lett.}, 72:1526--1529, 1994.

\bibitem{Kashiwaya00}
S.~Kashiwaya and Y.~Tanaka.
\newblock {\em Rep.\ Prog.\ Phys.}, 63:1641, 2000.

\bibitem{Braunecker05}
B.~Braunecker, P.~A. Lee, and Z.~Wang.
\newblock {\em Phys. Rev. Lett.}, 95:017004, 2005.

\bibitem{Caroli64}
C.~Caroli, P.~G.~De Gennes, and J.~Matricon.
\newblock {\em Phys. Lett.}, 9:307 -- 309, 1964.

\bibitem{Kopnin91}
N.~B. Kopnin and M.~M. Salomaa.
\newblock {\em Phys. Rev. B}, 44:9667--9677, 1991.

\bibitem{Volovik99}
G.~E. Volovik.
\newblock {\em JETP Lett.}, 70:609--614, 1999.

\bibitem{Ryu10}
S.~Ryu, A.~P. Schnyder, A.~Furusaki, and A.~W.~W. Ludwig.
\newblock {\em New Journal of Physics}, 12:065010, 2010.

\bibitem{Teo10}
Jeffrey C.~Y. Teo and C.~L. Kane.
\newblock {\em Phys. Rev. B}, 82:115120, 2010.

\bibitem{Black-Schaffer09}
A.~M. Black-Schaffer and S.~Doniach.
\newblock {\em Phys.\ Rev.\ B}, 79:064502, 2009.

\bibitem{Lu13}
H.-Y. Lu, S.~Chen, Y.~Xu, L.-Q. Zhang, D.~Wang, and W.-S. Wang.
\newblock {\em Phys. Rev. B}, 88:085416, 2013.

\bibitem{Balatsky06}
A.~V. Balatsky, I.~Vekhter, and J.-X. Zhu.
\newblock {\em Rev. Mod. Phys.}, 78:373--433, 2006.

\bibitem{Anderson59}
P.~W. Anderson.
\newblock {\em J. Phys. Chem. Solids}, 11:26 -- 30, 1959.

\bibitem{Graf00}
M.~J. Graf, A.~V. Balatsky, and J.~A. Sauls.
\newblock {\em Phys. Rev. B}, 61:3255--3258, 2000.

\bibitem{Black-Schaffer13cuprate}
A.~M. Black-Schaffer, D.~S. Golubev, T.~Bauch, F.~Lombardi, and
  M.~Fogelstr\"{o}m.
\newblock {\em Phys. Rev. Lett.}, 110:197001, 2013.

\bibitem{Pellegrino10}
F.~M.~D. Pellegrino, G.~G.~N. Angilella, and R.~Pucci.
\newblock {\em Eur. Phys. J. B}, 76:469--473, 2010.

\bibitem{McChesney10}
J.~L. McChesney, A.~Bostwick, T.~Ohta, T.~Seyller, K.~Horn, J.~Gonz\'alez, and
  E.~Rotenberg.
\newblock {\em Phys. Rev. Lett.}, 104:136803, 2010.

\bibitem{Efetov10}
D.~K. Efetov and P.~Kim.
\newblock {\em Phys. Rev. Lett.}, 105:256805, 2010.

\bibitem{Heersche07}
H.~B. Heersche, P.~Jarillo-Herrero, J.~B. Oostinga, L.~M.~K. Vandersypen, and
  A.~F. Morpurgo.
\newblock {\em Nature}, 446:56, 2007.

\bibitem{Shailos07}
A.~Shailos, W.~Nativel, A.~Kasumov, C.~Collet, M.~Ferrier, S.~Gu\'{e}ron,
  R.~Deblock, and H.~Bouchiat.
\newblock {\em Europhys.\ Lett.}, 79:57008, 2007.

\bibitem{Du08}
X.~Du, I.~Skachko, and E.~Y. Andrei.
\newblock {\em Phys.\ Rev.\ B}, 77:184507, 2008.

\bibitem{Black-Schaffer10}
A.~M. Black-Schaffer and S.~Doniach.
\newblock {\em Phys. Rev. B}, 81:014517, 2010.

\bibitem{Volovik88}
G.~E. Volovik.
\newblock {\em J. Phys. C}, 21:L215, 1988.

\bibitem{Salomaa89}
M.~M. Salomaa and G.~E. Volovik.
\newblock {\em J. Phys.: Condens. Matter}, 1:277, 1989.

\bibitem{Sauls09}
J.~A. Sauls and M.~Eschrig.
\newblock {\em New J. Phys.}, 11:075008, 2009.

\bibitem{Black-Schaffer13vortex}
A.~M. Black-Schaffer.
\newblock {\em Phys. Rev. B}, 88:104506, 2013.

\bibitem{Sun09}
K.~Sun, H.~Yao, E.~Fradkin, and S.~A. Kivelson.
\newblock {\em Phys. Rev. Lett.}, 103:046811, 2009.

\bibitem{Vafek14}
O.~Vafek, J.~M. Murray, and V.~Cvetkovic.
\newblock {\em Phys. Rev. Lett.}, 112:147002, 2014.

\bibitem{Milovanovic12}
J.~Vu\ifmmode \check{c}\else \v{c}\fi{}i\ifmmode \check{c}\else
  \v{c}\fi{}evi\ifmmode~\acute{c}\else \'{c}\fi{}, M.~O. Goerbig, and M.~V.
  Milovanovi\ifmmode~\acute{c}\else \'{c}\fi{}.
\newblock {\em Phys. Rev. B}, 86:214505, 2012.

\bibitem{Roy13}
Bitan Roy.
\newblock {\em Phys. Rev. B}, 88:075415, 2013.

\bibitem{Liu13}
F.~Liu, C.-C. Liu, K.~Wu, F.~Yang, and Y.~Yao.
\newblock {\em Phys. Rev. Lett.}, 111:066804, 2013.

\bibitem{Hannay65}
N.~B. Hannay, T.~H. Geballe, B.~T. Matthias, K.~Andres, P.~Schmidt, and
  D.~MacNair.
\newblock {\em Phys.\ Rev.\ Lett.}, 14:225, 1965.

\bibitem{Belash87}
I.~T. Belash, A.~D. Bronnikov, O.~V. Zharikov, and A.~V. Palnichenko.
\newblock {\em Solid State Commun.}, 64:1445, 1987.

\bibitem{Weller05}
T.~E. Weller, M.~Ellerby, S.~S. Saxena, R.~P. Smith, and N.~T. Skipper.
\newblock {\em Nature Phys.}, 1:39, 2005.

\bibitem{Csanyi05}
G.~Csanyi, P.~B. Littlewood, A.~H. Nevidomskyy, C.~J. Pickard, and B.~D.
  Simons.
\newblock {\em Nature Phys.}, 1:42, 2005.

\bibitem{Kopelevich00}
Y.~Kopelevich, P.~Esquinazi, J.~H.~S. Torres, and S.~Moehlecke.
\newblock {\em J. Low Temp.\ Phys.}, 119:691, 2000.

\bibitem{daSilva01}
R.~R. da~Silva, J.~H.~S. Torres, and Y.~Kopelevich.
\newblock {\em Phys.\ Rev.\ Lett.}, 87:147001, 2001.

\bibitem{Moehlecke04}
S.~Moehlecke, Y.~Kopelevich, and M.~B. Maple.
\newblock {\em Phys.\ Rev.\ B}, 69:134519, 2004.

\bibitem{Scheike12}
T.~Scheike, W.~B\"{o}hlmann, P.~Esquinazi, J.~Barzola-Quiquia, A.~Ballestar,
  and A.~Setzer.
\newblock {\em Adv. Mater.}, 24:5826--5831, 2012.

\bibitem{LeeFeng}
T.~K. Lee and S.~Feng.
\newblock {\em Phys. Rev. B}, 41:11110--11113, 1990.

\bibitem{Koretsune02}
T.~Koretsune and M.~Ogata.
\newblock {\em Phys. Rev. Lett.}, 89:116401, 2002.

\bibitem{BaskaranTria}
G.~{Baskaran}.
\newblock {\em Phys. Rev. Lett.}, 91:097003, 2003.

\bibitem{KumarShastry}
B.~{Kumar} and B.~S. {Shastry}.
\newblock {\em Phys. Rev. B}, 68:104508, 2003.

\bibitem{WangLeeLee}
Q.-H. Wang, D.-H. Lee, and P.~A. Lee.
\newblock {\em Phys. Rev. B}, 69:092504, 2004.

\bibitem{HonerkampTria}
C.~Honerkamp.
\newblock {\em Phys. Rev. B}, 68:104510, 2003.

\bibitem{Nishikubo11}
Y~Nishikubo, K.~Kudo, and M.~Nohara.
\newblock {\em J. Phys. Soc. Jpn}, 80:055002, 2011.

\bibitem{Biswas13}
P.~K. Biswas, H.~Luetkens, T.~Neupert, T.~St\"urzer, C.~Baines, G.~Pascua,
  A.~P. Schnyder, M.~H. Fischer, J.~Goryo, M.~R. Lees, H.~Maeter,
  F.~Br\"uckner, H.-H. Klauss, M.~Nicklas, P.~J. Baker, A.~D. Hillier,
  M.~Sigrist, A.~Amato, and D.~Johrendt.
\newblock {\em Phys. Rev. B}, 87:180503, 2013.

\bibitem{Bruckner13}
F.~{Br{\"u}ckner}, R.~{Sarkar}, M.~{G{\"u}nther}, H.~{K{\"u}hne},
  H.~{Luetkens}, T.~{Neupert}, A.~P. {Reyes}, P.~L. {Kuhns}, P.~K. {Biswas},
  T.~{St{\"u}rzer}, D.~{Johrendt}, and H.-H. {Klauss}.
\newblock {\em ArXiv:1312.6166 (unpublished)}, 2013.

\bibitem{Goryo12}
J.~Goryo, M.~H. Fischer, and M.~Sigrist.
\newblock {\em Phys. Rev. B}, 86:100507, 2012.

\bibitem{Youn12}
S.~J. Youn, M.~H. Fischer, S.~H. Rhim, M.~Sigrist, and D.~F. Agterberg.
\newblock {\em Phys. Rev. B}, 85:220505, 2012.

\bibitem{Fischer14}
M.~H. Fischer, T.~Neupert, C.~Platt, A.~P. Schnyder, W.~Hanke, J.~Goryo,
  R.~Thomale, and M.~Sigrist.
\newblock {\em Phys. Rev. B}, 89:020509, 2014.

\bibitem{Wang13}
W.-S. {Wang}, Y.~{Yang}, and Q.-H. {Wang}.
\newblock {\em ArXiv:1312.3071 (unpublished)}, 2013.

\bibitem{Takada03}
K.~Takada, H.~Sakurai, E.~Takayama-Muromachi, F.~Izumi, R.~A. Dilanian, and
  T.~Sasaki.
\newblock {\em Nature}, 422:53--55, 2003.

\bibitem{Zheng06}
G.~Zheng, K.~Matano, D.~P. Chen, and C.~T. Lin.
\newblock {\em Phys. Rev. B}, 73:180503, 2006.

\bibitem{Fujimoto04}
T.~Fujimoto, G.-Q. Zheng, Y.~Kitaoka, R.~L. Meng, J.~Cmaidalka, and C.~W. Chu.
\newblock {\em Phys. Rev. Lett.}, 92:047004, 2004.

\bibitem{Kanigel04}
A.~Kanigel, A.~Keren, L.~Patlagan, K.~B. Chashka, P.~King, and A.~Amato.
\newblock {\em Phys. Rev. Lett.}, 92:257007, 2004.

\bibitem{Oeschler08}
N.~Oeschler, R.~A. Fisher, N.~E. Phillips, J.~E. Gordon, M.-L. Foo, and R.~J.
  Cava.
\newblock {\em Phys. Rev. B}, 78:054528, 2008.

\bibitem{Kuroki04}
K.~Kuroki, Y.~Tanaka, and R.~Arita.
\newblock {\em Phys. Rev. Lett.}, 93:077001, 2004.

\bibitem{Ogata03}
M.~Ogata.
\newblock {\em J. Phys. Soc. Jpn}, 72:1839--1842, 2003.

\bibitem{Kiesel13}
M.~L. Kiesel, C.~Platt, W.~Hanke, and R.~Thomale.
\newblock {\em Phys. Rev. Lett.}, 111:097001, 2013.

\bibitem{Yamanaka96}
S.~Yamanaka, H.~Kawaji, K.~Hotehama, and M.~Ohashi.
\newblock {\em Advanced Materials}, 8:771--774, 1996.

\bibitem{Yamanaka98}
S.~Yamanaka, K.~Hotehama, and H.~Kawaji.
\newblock {\em Nature}, 392:580--582, 1998.

\bibitem{Tou03}
H.~Tou, Y.~Maniwa, and S.~Yamanaka.
\newblock {\em Phys. Rev. B}, 67:100509, 2003.

\bibitem{Taguchi07}
Y.~Taguchi, T.~Kawabata, T.~Takano, A.~Kitora, K.~Kato, M.~Takata, and
  Y.~Iwasa.
\newblock {\em Phys. Rev. B}, 76:064508, 2007.

\bibitem{Taguchi05}
Y.~Taguchi, M.~Hisakabe, and Y.~Iwasa.
\newblock {\em Phys. Rev. Lett.}, 94:217002, 2005.

\bibitem{Kasahara09}
Y.~Kasahara, T.~Kishiume, T.~Takano, K.~Kobayashi, E.~Matsuoka, H.~Onodera,
  K.~Kuroki, Y.~Taguchi, and Y.~Iwasa.
\newblock {\em Phys. Rev. Lett.}, 103:077004, 2009.

\bibitem{Felser99}
C.~Felser and R.~Seshadri.
\newblock {\em J. Mater. Chem.}, 9:459--464, 1999.

\bibitem{Kuroki10}
K.~Kuroki.
\newblock {\em Phys. Rev. B}, 81:104502, 2010.

\bibitem{Watanabe13}
T.~Watanabe and S.~Ishihara.
\newblock {\em J.\ Phys.\ Soc.\ Jpn}, 82:034704, 2013.

\bibitem{Powell06}
B.~J. Powell and R.~H. McKenzie.
\newblock {\em J. Phys.: Condens. Matter}, 18:R827, 2006.

\bibitem{Williams91}
J.~M. Williams, A.~J. Schultz, U.~Geiser, K.~D. Carlson, A.~M. Kini, H.~H.
  Wang, W.-K. Kwok, M.-H. Whangbo, and J.~E. Schriber.
\newblock {\em Science}, 252:1501--1508, 1991.

\bibitem{Lefebvre00}
S.~Lefebvre, P.~Wzietek, S.~Brown, C.~Bourbonnais, D.~J\'erome, C.~M\'ezi\`ere,
  M.~Fourmigu\'e, and P.~Batail.
\newblock {\em Phys. Rev. Lett.}, 85:5420--5423, 2000.

\bibitem{McKenzie97}
R.~H. McKenzie.
\newblock {\em Science}, 278:820--821, 1997.

\bibitem{Powell04}
B.~J. Powell and R.~H. McKenzie.
\newblock {\em Phys. Rev. B}, 69:024519, 2004.

\bibitem{Kyung06}
B.~Kyung and A.-M.~S. Tremblay.
\newblock {\em Phys. Rev. Lett.}, 97:046402, 2006.

\bibitem{Sahebsara06}
P.~Sahebsara and D.~S\'en\'echal.
\newblock {\em Phys. Rev. Lett.}, 97:257004, 2006.

\bibitem{Grover10}
T.~Grover, N.~Trivedi, T.~Senthil, and P.~A. Lee.
\newblock {\em Phys. Rev. B}, 81:245121, 2010.

\bibitem{Wang12}
Q.~H. Wang, K.~Kalantar-Zadeh, A.~Kis, J.~N. Coleman, and M.~S. Strano.
\newblock {\em Nat. Nanotech.}, 7:699--712, 2012.

\bibitem{Taniguchi12}
K.~Taniguchi, A.~Matsumoto, H.~Shimotani, and H.~Takagi.
\newblock {\em Appl. Phys. Lett.}, 101:042603, 2012.

\bibitem{Ye12}
J.~T. Ye, Y.~J. Zhang, R.~Akashi, M.~S. Bahramy, R.~Arita, and Y.~Iwasa.
\newblock {\em Science}, 338:1193--1196, 2012.

\bibitem{Helveg00}
S.~Helveg, J.~V. Lauritsen, E.~L\ae{}gsgaard, I.~Stensgaard, J.~K. N\o{}rskov,
  B.~S. Clausen, H.~Tops\o{}e, and F.~Besenbacher.
\newblock {\em Phys. Rev. Lett.}, 84:951--954, 2000.

\bibitem{Cao12}
T.~Cao, G.~Wang, W.~Han, H.~Ye, C.~Zhu, J.~Shi, Q.~Niu, P.~Tan, E.~Wang,
  B.~Liu, and J.~Feng.
\newblock {\em Nat. Commun.}, 3:887, 2012.

\bibitem{Roldan13}
R.~Rold\'an, E.~Cappelluti, and F.~Guinea.
\newblock {\em Phys. Rev. B}, 88:054515, 2013.

\bibitem{Law14}
N.~F.~Q. {Yuan}, K.~F. {Mak}, and K.~T. {Law}.
\newblock {\em Phys. Rev. Lett.}, 113:097001, 2014.


\bibitem{Moller08}
A.~M\"oller, U.~L\"ow, T.~Taetz, M.~Kriener, G.~Andr\'e, F.~Damay, O.~Heyer,
  M.~Braden, and J.~A. Mydosh.
\newblock {\em Phys. Rev. B}, 78:024420, 2008.

\bibitem{Yan12}
Y.~J. Yan, Z.~Y. Li, T.~Zhang, X.~G. Luo, G.~J. Ye, Z.~J. Xiang, P.~Cheng,
  L.~J. Zou, and X.~H. Chen.
\newblock {\em Phys. Rev. B}, 85:085102, 2012.

\bibitem{Kitaev06}
A.~Kitaev.
\newblock {\em Ann. Phys. (N. Y.)}, 321:2 -- 111, 2006.

\bibitem{Jackeli09}
G.~Jackeli and G.~Khaliullin.
\newblock {\em Phys. Rev. Lett.}, 102:017205, 2009.

\bibitem{Chaloupka10}
J.~Chaloupka, G.~Jackeli, and G.~Khaliullin.
\newblock {\em Phys. Rev. Lett.}, 105:027204, 2010.

\bibitem{Singh12}
Yogesh Singh, S.~Manni, J.~Reuther, T.~Berlijn, R.~Thomale, W.~Ku, S.~Trebst,
  and P.~Gegenwart.
\newblock {\em Phys. Rev. Lett.}, 108:127203, 2012.

\bibitem{Witczak14}
William Witczak-Krempa, Gang Chen, Yong~Baek Kim, and Leon Balents.
\newblock {\em Annu. Rev. Condens. Matter Phys.}, 5:57--82, 2014.

\bibitem{Reuther14}
J.~{Reuther}, R.~{Thomale}, and S.~{Rachel}.
\newblock {\em Phys. Rev. B}, 90:100405(R), 2014.

\bibitem{You12}
Yi-Zhuang You, Itamar Kimchi, and Ashvin Vishwanath.
\newblock {\em Phys. Rev. B}, 86:085145, 2012.

\bibitem{Hyart12}
Timo Hyart, Anthony~R. Wright, Giniyat Khaliullin, and Bernd Rosenow.
\newblock {\em Phys. Rev. B}, 85:140510, 2012.

\bibitem{Scherer14}
D.~D. {Scherer}, M.~M. {Scherer}, G.~{Khaliullin}, C.~{Honerkamp}, and
  B.~{Rosenow}.
\newblock {\em Phys. Rev. B}, 90:045135, 2014.

\bibitem{Zubko11}
P.~Zubko, S.~Gariglio, M.~Gabay, P.~Ghosez, and J.-M. Triscone.
\newblock {\em Annual Review of Condensed Matter Physics}, 2:141--165, 2011.

\bibitem{Hwang12}
H.~Y. Hwang, Y.~Iwasa, M.~Kawasaki, B.~Keimer, N.~Nagaosa, and Y.~Tokura.
\newblock {\em Nature Mater.}, 11:103--113, 2012.

\bibitem{Xiao11}
Xiao D., W.~Zhu, Y.~Ran, N.~Nagaosa, and S.~Okamoto.
\newblock {\em Nat. Commun.}, 2:596, 2011.

\bibitem{Yang11bilayer}
Kai-Yu Yang, Wenguang Zhu, Di~Xiao, Satoshi Okamoto, Ziqiang Wang, and Ying
  Ran.
\newblock {\em Phys. Rev. B}, 84:201104, 2011.

\bibitem{Ruegg11}
A.~R\"uegg and G.~A. Fiete.
\newblock {\em Phys. Rev. B}, 84:201103, 2011.

\bibitem{Okamoto13}
S.~Okamoto.
\newblock {\em Phys. Rev. Lett.}, 110:066403, 2013.

\bibitem{Okamoto13b}
S.~Okamoto.
\newblock {\em Phys. Rev. B}, 87:064508, 2013.

\bibitem{Nagamatsu01}
J.~Nagamatsu, N.~Nakagawa, T.~Muranaka, Y.~Zenitani, and J.~Akimitsu.
\newblock {\em Nature}, 410:63--64, 2001.

\bibitem{Kortus01}
J.~Kortus, I.~I. Mazin, K.~D. Belashchenko, V.~P. Antropov, and L.~L. Boyer.
\newblock {\em Phys. Rev. Lett.}, 86:4656--4659, 2001.

\bibitem{Liu01}
A.~Y. Liu, I.~I. Mazin, and J.~Kortus.
\newblock {\em Phys. Rev. Lett.}, 87:087005, 2001.

\bibitem{Choi02}
H.~J. Choi, D.~Roundy, H.~Sun, M.~L. Cohen, and S.~G. Louie.
\newblock {\em Nature}, 418:758--760, 2002.

\bibitem{Imai02}
M.~Imai, K.~Nishida, T.~Kimura, and H.~Abe.
\newblock {\em Appl. Phys. Lett.}, 80:1019--1021, 2002.

\end{thebibliography}

\end{document}